\RequirePackage{fix-cm}
\documentclass[twocolumn,epjc3]{svjour3}  
\usepackage{microtype} 
\smartqed  
\RequirePackage{graphicx}
\usepackage{xcolor}
\usepackage{subfigure}
\usepackage{siunitx}
\usepackage{amsmath}
\usepackage{amssymb}
\usepackage{tabto}
\usepackage{svg}
\usepackage{subfloat}
\usepackage{subcaption}
\usepackage[switch]{lineno}
\usepackage{booktabs} 
\linenumbers

\newcommand{\utca}{\unit{$\mu$}TCA\ }
\newcommand{\sigmavis}{\ensuremath{\sigma_{\text{vis}}}\ }

\newcommand{\fbinv}{\unit{fb\ensuremath{^{-1}}}\ }

\newcolumntype{P}[1]{>{\raggedright\arraybackslash}m{#1}}
\newcolumntype{C}[1]{>{\centering\arraybackslash}m{#1}}

\makeatletter
\renewcommand\institute@and{\unskip; \institute@mark}
\makeatother
\RequirePackage[colorlinks,citecolor=black,urlcolor=black,linkcolor=black]{hyperref}
\usepackage{orcidlink}

\journalname{Eur. Phys. J. C}
\begin{document}
\title{Design and performance of the Fast Beam Condition Monitor for luminosity and background measurement at the CMS Experiment in LHC Run 3}


\author{The CMS BRIL Collaboration 
         \and Eliana Acurio\,\orcidlink{0000-0002-9630-3342}\thanksref{epn}
         \and Ying An\,\orcidlink{0000-0003-1299-1879}\thanksref{desy}
         \and Iakov Andreev\,\orcidlink{0009-0002-5926-9664}\thanksref{desy}
         \and Tomas Atehortua\,\orcidlink{0000-0002-2647-8104 }\thanksref{vanderbilt}
         \and Georg Auzinger\,\orcidlink{0000-0001-7077-8262}\thanksref{cern}
         \and Hamed Bakhshiansohi\,\orcidlink{0000-0001-5741-3357}\thanksref{isfahan,desy}
         \and Gergely Bálint\,\orcidlink{0009-0000-7778-3531}\thanksref{elte}
         \and Luis Ignacio Banos\,\orcidlink{0000-0001-6195-3102}\thanksref{desy}
         \and Juan Pedro Barajas Ibarria\,\orcidlink{0009-0009-1952-0907}\thanksref{sonora}
         \and Hugo Alberto Becerril\,\orcidlink{0000-0001-5387-712X}\thanksref{desy}
         \and Jose Feliciano Benitez\,\orcidlink{0000-0002-2633-6712}\thanksref{sonora}
         \and Philip H.~Butler\,\orcidlink{0000-0001-9878-2140}\thanksref{newzealand}
         \and Alan Campbell\,\orcidlink{0000-0003-4439-5748}\thanksref{desy}
         \and Edgar Fernando Carrera Jarrín\,\orcidlink{0000-0002-0857-8507}\thanksref{usfq}
         \and Alfredo Castaneda Hernandez\,\orcidlink{0000-0003-4766-1546}\thanksref{sonora}
         \and Marco Costa\,\orcidlink{0000-0003-0156-0790}\thanksref{infn-torino}
         \and Antonio Cota Rodriguez\,\orcidlink{0000-0001-8026-6236}\thanksref{sonora}
         \and Luis Enrique Cuevas Picos\,\orcidlink{0009-0007-2135-9986 }\thanksref{sonora}
         \and Anne Dabrowski\,\orcidlink{0000-0003-2570-9676}\thanksref{cern}
         \and Vladyslav Danilov\thanksref{desy}
         \and Andr\'es Guillermo Delannoy Sotomayor\,\orcidlink{0000-0003-1252-6213}\thanksref{tennessee}
         \and Hedwin Aaron Encinas Acosta\,\orcidlink{0000-0002-8239-6648}\thanksref{sonora}
         \and Krisztián Farkas\,\orcidlink{0000-0003-1740-6974}\thanksref{elte}
         \and Anna Fehérkuti\,\orcidlink{0000-0002-5043-2958}\thanksref{elte,wigner}
         \and Mahmoud Moussa Abdelkhalek Gadallah\,\orcidlink{0000-0002-8305-6661}\thanksref{elte,assiut}
         \and Luis Gabriel Gallegos Mariñez\thanksref{sonora}
         \and Angela Giraldi\,\orcidlink{0000-0003-4423-2631}\thanksref{desy}
         \and Alfredo Gurrola\,\orcidlink{0000-0002-2793-4052}\thanksref{vanderbilt}
         \and Moritz Guthoff\,\orcidlink{0000-0002-3974-589X}\thanksref{desy}
         \and Mykyta Haranko\,\orcidlink{0000-0002-9376-9235}\thanksref{cern}
         \and Maria Hempel\thanksref{desy}
         \and Hans Henschel\thanksref{desy}
        \and Shilpi Jain\,\orcidlink{0000-0003-1770-5309}\thanksref{minnesota,tata}
         \and Ármin Kadlecsik\,\orcidlink{0000-0001-5559-0106}\thanksref{elte,wigner}
         \and Jan Kaplon\,\orcidlink{0000-0002-7528-7128}\thanksref{cern}
         \and Ádám Kardos\,\orcidlink{0000-0001-5592-4567}\thanksref{elte,debrecen}
         \and Olena Karacheban\,\orcidlink{0000-0002-2785-3762}\thanksref{email,cern,rutgers,btu} 
         \and Nimmitha Karunarathna\,\orcidlink{0000-0002-3412-0508}\thanksref{tennessee}
         \and Joscha Ahauser\,\orcidlink{0000-0002-4781-5704}\thanksref{desy,kit}
         \and Georgios Krintiras\,\orcidlink{0000-0002-0380-7577}\thanksref{kansas}  
         \and Leen Lambers\,\orcidlink{0000-0001-6937-5167}\thanksref{btu} 
         \and Jessica Leonard\,\orcidlink{0000-0003-1761-8221}\thanksref{desy}
         \and Mois\'es David L\'eon Coello\,\orcidlink{0000-0002-3761-911X}\thanksref{elte,sonora} 
         \and Maksim Jenihhin\,\orcidlink{0000-0001-8165-9592}\thanksref{taltech}
         \and Willard Johns\,\orcidlink{0000-0001-5291-8903}\thanksref{vanderbilt}
         \and Markus Klute\,\orcidlink{0000-0002-0869-5631}\thanksref{kit}
         \and Wolfgang Lange\thanksref{desy}
         \and Kuan-Yu Lin\,\orcidlink{0000-0002-2269-3632}\thanksref{desy}
         \and Wei Heng Liu\,\orcidlink{0009-0009-1785-4831}\thanksref{cern,oxford}
         \and Wolfgang Lohnmann\,\orcidlink{0000-0002-8705-0857}\thanksref{desy,btu,cern}
         \and Arkady Lokhovitskiy\,\orcidlink{0000-0002-4016-0039}\thanksref{debrecen,newzealand}
         \and Robert Loos\thanksref{cern}
         \and Péter Major\,\orcidlink{0000-0002-5476-0414}\thanksref{maryland,elte}
         \and Sophie Mallows\thanksref{kit,cern}
         \and Jeremiah Mans\,\orcidlink{0000-0003-2840-1087}\thanksref{minnesota}
         \and Daniel Marlow\,\orcidlink{0000-0002-6395-1079}\thanksref{princeton}
         \and Badder Marzocchi\,\orcidlink{0000-0001-6687-6214}\thanksref{minnesota}
         \and John-Erik Mei{\ss}ner\thanksref{desy}
         \and Philip Jason Meltzer\thanksref{rutgers}
         \and Stefano Mersi\,\orcidlink{0000-0003-2155-6692}\thanksref{cern}
         \and Andreas Bernhard Meyer\,\orcidlink{0000-0001-8532-2356}\thanksref{desy} 
         \and Dmitri Mihhailov\thanksref{taltech}
         \and Valeria Monti\,\orcidlink{0000-0001-7774-8963}\thanksref{infn-torino}
         \and Michele Mormile\,\orcidlink{0000-0003-0456-7250}\thanksref{desy,kit}
         \and Javier Alberto Murillo Quijada\,\orcidlink{0000-0003-4933-2092}\thanksref{sonora}
         \and Vlodymyr Myronenko\,\orcidlink{0000-0002-3984-4732}\thanksref{desy}
         \and Cristina Oropeza Barrera\,\orcidlink{0000-0001-9724-0016}\thanksref{iberoamericana}
         \and Younes Otarid\,\orcidlink{0009-0009-4180-9264}\thanksref{cern}
         \and Christopher Palmer\,\orcidlink{0000-0002-5801-5737}\thanksref{maryland} 
         \and Jose David Ochoa Flores\,\orcidlink{0009-0001-1395-4159}\thanksref{usfq} 
         \and Michelangelo Pari\,\orcidlink{0000-0002-1852-9549}\thanksref{cern} 
         \and Gabriella Pásztor\,\orcidlink{0000-0003-0707-9762}\thanksref{email,elte}
         \and Marek Penno\,\orcidlink{0009-0005-6820-8021}\thanksref{desy}
         \and Fabio Lucas Pereira Carneiro\,\orcidlink{0009-0001-6182-0112}\thanksref{cern,btu}
         \and Elena Popova\,\orcidlink{0000-0001-7556-8969}\thanksref{maryland} 
         \and Attila József Rádl\,\orcidlink{0000-0001-8810-0388}\thanksref{elte,wigner}
         \and Evan Altair Ranken\,\orcidlink{0000-0001-7472-5029}\thanksref{email,desy}
         \and Beatriz Ribeiro-Lopes\,\orcidlink{0000-0003-0823-447X}\thanksref{desy,ghent}
         \and Francesco Romeo\,\orcidlink{0000-0002-1297-6065}\thanksref{vanderbilt}
         \and Carlos Romero\,\orcidlink{0009-0007-2597-875X}\thanksref{northwestern}
         \and Jonas R\"ubenach\thanksref{desy}
         \and Alexander Ruede\,\orcidlink{0000-0003-2294-7150}\thanksref{cern}       
         \and Vladimir Ryjov\thanksref{cern}
         \and Santeri Saariokari\,\orcidlink{0000-0002-6798-2454}\thanksref{helsinki,cern}
         \and Alessia Saggio\,\orcidlink{0000-0002-7385-3317}\thanksref{desy}  
         \and Valerie Scheurer\,\orcidlink{0000-0003-0664-7780}\thanksref{desy,saopaolo}
         \and Vahid Sedighzadeh Dalavi\,\orcidlink{0000-0002-8975-687X}\thanksref{isfahan}
         \and Mohammad Sedghi\thanksref{isfahan}
         \and Tatiana Selezneva\,\orcidlink{0009-0000-8339-5172}\thanksref{kit}
         \and Konstantin Sharko\,\orcidlink{0000-0002-7614-5236}\thanksref{desy}
         \and Alexey Shevelev\,\orcidlink{0000-0003-4600-0228}\thanksref{maryland}
         \and Konstantin Shibin\,\orcidlink{0000-0002-8041-1779}\thanksref{taltech}
         \and Sunil Somalwar\,\orcidlink{0000-0002-8856-7401}\thanksref{rutgers}
         \and Rafael Sosa\,\orcidlink{0000-0002-2240-6699}\thanksref{desy}
         \and Stefan Spanier\,\orcidlink{0000-0002-7049-4646}\thanksref{tennessee}
         \and David Stickland\,\orcidlink{0000-0003-4702-8820}\thanksref{princeton}
         \and Robert Stone\,\orcidlink{0000-0001-6229-695X}\thanksref{rutgers}
         \and Andromachi Tsirou\,\orcidlink{0000-0003-2713-1724 }\thanksref{cern,athens}
         \and Oleksii Turkot\,\orcidlink{0000-0001-5352-7744}\thanksref{desy}
         \and Lizardo Valencia Palomo\,\orcidlink{0000-0002-8736-440X}\thanksref{sonora}
         \and Cristina Vazquez Velez\thanksref{cern} 
         \and Roger Rusack\,\orcidlink{0000-0002-7633-749X}\thanksref{minnesota}
         \and Paola Tropea\,\orcidlink{0000-0003-1899-2266}\thanksref{cern}
         \and Balázs Ujvári\,\orcidlink{0000-0003-0498-4265}\thanksref{debrecen}
         \and Mayda Velasco\,\orcidlink{0000-0002-1619-3121}\thanksref{northwestern}
         \and Piero Giorgio Verdini\,\orcidlink{0000-0002-0042-9507}\thanksref{infn-pisa}
         \and David Walter\,\orcidlink{0000-0001-8584-9705}\thanksref{desy,mit}
         \and Joanna Wa\'{n}czyk\,\orcidlink{0000-0002-8562-1863}\thanksref{email,cern,epfl}
         \and Jesus Alberto Velazquez Corral\,\orcidlink{0009-0000-0455-237X}\thanksref{kansas} 
         \and Daniel John Wilbern\,\orcidlink{0000-0002-1958-7003}\thanksref{northwestern}
         \and Zhen Xie\,\orcidlink{0000-0002-4411-9615}\thanksref{princeton}
         \and Maryam Zeinali\,\orcidlink{0000-0001-8367-6257}\thanksref{isfahan,tehran}
         \and Wolfram Zeuner\,\orcidlink{0009-0004-8806-0047}\thanksref{cern,kit}
}

\thankstext{email}{Corresponding authors e-mail: jwanczyk@cern.ch, olena.karacheban@cern.ch, gabriella.pasztor@cern.ch, e.ranken@cern.ch}
\thankstext{ghent}{Now at Ghent University, Ghent, Belgium}
\thankstext{saopaolo}{Now at 
Universidade Estadual Paulista, São Paulo, Brazil}
\thankstext{assiut}{Also at Physics Department, Faculty of Science, Assiut University, Assiut, Egypt} 
\thankstext{athens}{Also at National and Kapodistrian University of Athens, Athens, Greece}
\thankstext{wigner}{Also at HUN-REN Wigner Research Centre for Physics, Budapest, Hungary}
\thankstext{tata}{Now at Tata Institute of Fundamental Research, Mumbai, India} 
\thankstext{tehran}{Also at Sharif University of Technology, Tehran, Iran}
\thankstext{epfl}{Also at Ecole Polytechnique F\'ed\'erale Lausanne, Lausanne, Switzerland}
\thankstext{oxford}{Now at University of Oxford, Oxford, United Kingdom}
\thankstext{mit}{Now at Massachusetts Institute of Technology, Cambridge, Massachusetts, USA}


\institute{Escuela Politecnica Nacional, Quito, Ecuador \label{epn}
\and Universidad San Francisco de Quito, Quito, Ecuador \label{usfq}
\and Tallinn University of Technology, Tallinn, Estonia \label{taltech}
\and Helsinki Institute of Physics (HIP), Helsinki, Finland \label{helsinki}
\and Brandenburg University of Technology, Cottbus, Germany \label{btu}
\and Deutsches Elektronen-Synchrotron, Hamburg, Germany\label{desy} 
\and Institut f\"ur Experimentelle Teilchenphysik, Karlsruhe, Germany \label{kit}
\and MTA-ELTE Lend\"ulet CMS Particle and Nuclear Physics Group, E\"otv\"os Lor\'and University,
Budapest, Hungary \label{elte}
\and HUN-REN Wigner Research Centre for Physics, Budapest, Hungary \label{wigner}
\and University of Debrecen (HU), Debrecen, Hungary \label{debrecen}
\and Isfahan University of Technology, Isfahan, Iran \label{isfahan}
\and Istituto Nazionale di Fisica Nucleare (INFN), Sezione di Pisa, Pisa, Italy \label{infn-pisa}
\and INFN Sezione di Torino, Universit\'a di Torino, Torino, Italy \label{infn-torino}
\and University of Sonora (UNISON), Hermosillo, Mexico \label{sonora}
\and Universidad Iberoamericana, Mexico City, Mexico \label{iberoamericana}
\and University of Canterbury, Christchurch, New Zealand\label{newzealand}
\and CERN, European Organization for Nuclear Research, Geneva, Switzerland \label{cern}
\and Northwestern University, Evanston, Illinois, USA \label{northwestern}
\and The University of Kansas, Lawrence, Kansas, USA \label{kansas} 
\and University of Maryland, College Park, Maryland, USA \label{maryland}
\and University of Minnesota, Minneapolis, Minnesota, USA \label{minnesota}
\and Rutgers, The State University of New Jersey, Piscataway, New Jersey, USA \label{rutgers}
\and Princeton University, Princeton, New Jersey, USA \label{princeton}
\and University of Tennessee, Knoxville, Tennessee, USA \label{tennessee}
\and Vanderbilt University, Nashville, Tennessee, USA \label{vanderbilt}
}


\date{Received: date / Accepted: date}

\nolinenumbers

\maketitle

\begin{abstract}

The Fast Beam Condition Monitor (BCM1F) has been used at the CMS Experiment since the first LHC circulating beams in 2008. Originally meant as a beam-induced background monitor for fast beam losses detection, it showed a potential also for luminosity measurements in 2012 running, and has been used for luminosity measurements since the beginning of Run 2 data taking in 2015 as a part of the Beam Radiation, Instrumentation and Luminosity (BRIL) system. Over the years, the system has undergone various upgrades to the sensors, the front-end and back-end electronics, providing improvements in the precision of the measurements, that remain valid in the higher pileup conditions of LHC Run 3 (2022-2026). Based on the experience of all BCM1F Run 2 upgrades, the detector was completely rebuilt prior to LHC Run 3 using AC-coupled silicon-pad diodes and active cooling. This latest detector version exhibits excellent linearity with instantaneous luminosity and achieves nanosecond-level timing precision, enabling improved systematic corrections for luminosity and background measurements. 
This paper presents a detailed overview of the detector system for LHC Run 3, including the selection and qualification of sensors as well as a summary of the  readout system. It also outlines the processing and calibration strategy for luminosity data, discussing operational hurdles and comparing BCM1F measurements to other CMS luminosity measurements to assess the system's performance as a luminometer. Lastly, the implications for the design of a future luminosity detector to be used in the envisioned HL-LHC upgrade are discussed.
   
\keywords{CMS \and BCM1F detector \and Luminometer \and Luminosity measurement \and Beam induced background \and Beam instrumentation }
\end{abstract}

\tableofcontents

\section{Introduction}

The Fast Beam Condition Monitor (BCM1F) has been in place since the beginning of LHC operations in 2008~\cite{run1-bcm1f}. The original design utilised single-crystalline diamond sensors, and was initially intended for beam-induced background (BIB) measurement~\cite{6153979}. During LHC Run 1, its performance was studied, and its potential for luminosity measurement was explored~\cite{bcm1f-jessica}. A major upgrade was then performed in preparation for LHC Run 2~\cite{bcm1f-LS1}, introducing a dedicated printed circuit board (PCB), improved diamond sensors with two-pad metallisation~\cite{HEMPEL201665}, and a specialised fast application-specific integrated circuit (ASIC)~\cite{asic}. The front-end timing performance enabled measurements of the per-bunch luminosity as well as the incoming beam-induced background. Considerable effort was then invested in studying the detector’s performance as a luminometer~\cite{bcm1f-LS1_2,olena-thesis}, because precise luminosity measurement is of critical importance for the physics programme of the CMS experiment. 
To provide uninterrupted, reliable, real-time luminosity measurements to CMS and the LHC, redundant luminometers are required. BCM1F along with the Pixel Luminosity Telescope (PLT)~\cite{PLT} are two independent CMS luminometers.  Together with the CMS Forward Hadron Calorimeter \cite{Mans:1481837}, which is equipped with a dedicated luminosity readout, they provide the primary online, per-bunch luminosity measurements for CMS under the responsibility of the Beam Radiation, Instrumentation and Luminosity (BRIL) system.

In the early years of Run 2, BCM1F incorporated multiple redundant sensor types: single- and poly-crystalline diamond produced via Chemical Vapour Deposition (sCVD and pCVD), and silicon, in order to evaluate their performance for BIB as well as luminosity measurements under nominal beam conditions~\cite{run2_bcm1f}. Originally diamond sensors were preferred, as they do not require cooling, and are therefore not affected by periods of cooling service interventions during LHC operation nor the costs and maintenance of such a cooling system. In comparison, silicon sensors require cooling to ensure longevity and reduced leakage current under increased radiation. During the first test of silicon sensors in Run 2, no cooling was integrated yet into the BCM1F mechanics. Despite this, silicon demonstrated superior characteristics in terms of signal-to-noise ratio and linear response~\cite{run2_bcm1f}, but showed build-up of leakage current, as expected under radiation exposure without cooling. By contrast, the performance of diamond sensors was limited by charge trapping, which caused the response to depend on a combination of the total collision rate and the periods without collisions (for example, the gaps in the LHC filling pattern) during which de-trapping of shallow defects could occur.
Both the non-linearity as a function of the total collision rate and its dependence on radiation-induced damage affected the accuracy of the luminosity measurement~\cite{olena-thesis,run2_bcm1f,florianK}. These effects were difficult to fully correct, as calibration adjustments were challenging under rapidly changing luminosity conditions. In addition, for single-crystal diamond sensors in particular, occasional large breakthrough currents were observed after irradiation, leading to high-voltage trips.

\begin{figure}[!htb]
   \centering
   \includegraphics*[width=.99\columnwidth]{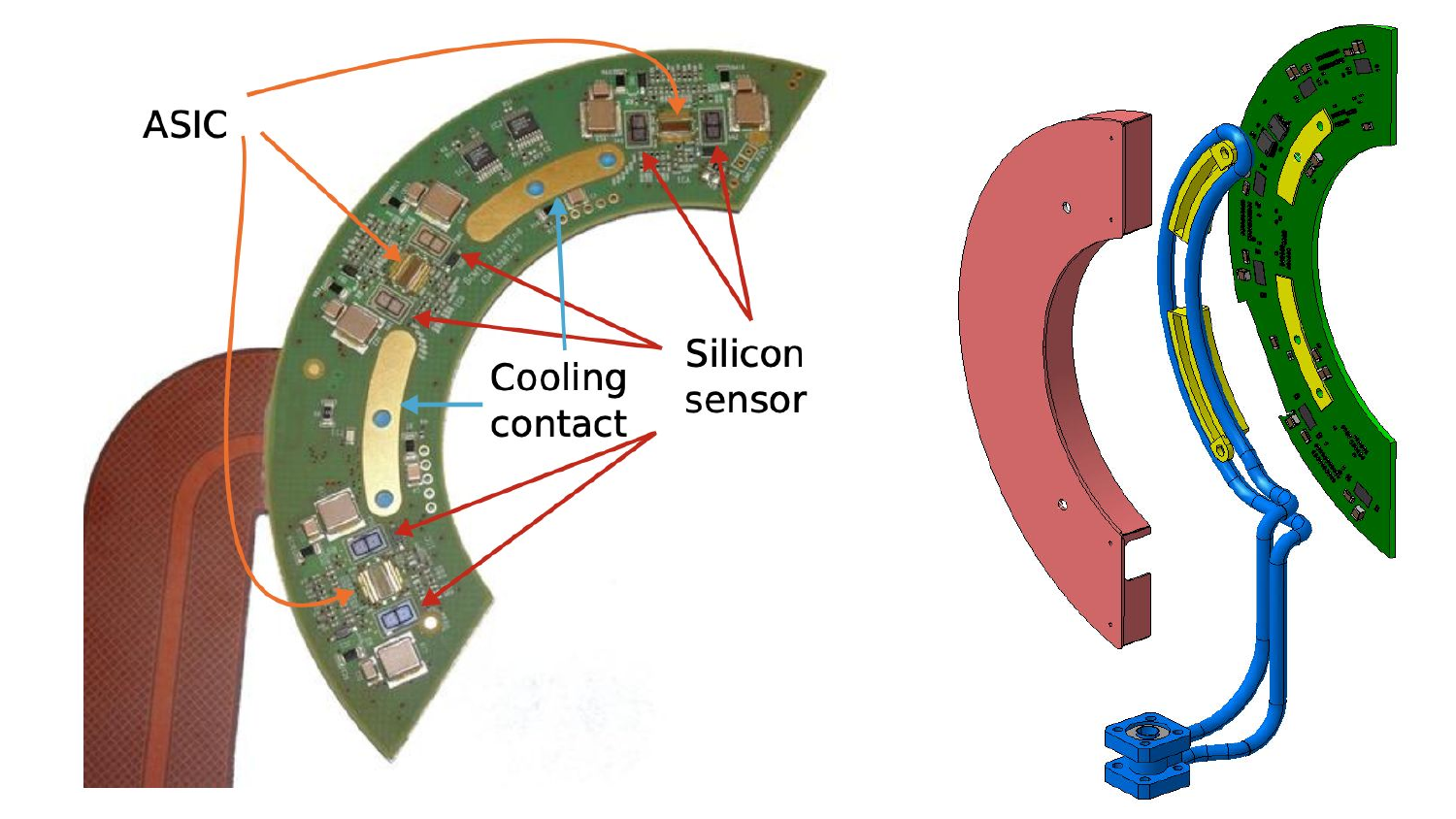}
   \caption{Photo of the Run 3 BCM1F C-shape (left) and drawing of the C-shape with cooling pipe and protection cover, as required in the final assembly (right)}
   \label{fig:cshape}
\end{figure}

In preparation for the demanding conditions of increased instantaneous and integrated luminosity planned for LHC Run 3, BCM1F underwent a comprehensive upgrade during the preceding Long Shutdown 2 (LS2) period, which took place between 2019 and 2022, with the goal not only of providing reliable online luminosity to the LHC and CMS, but also precision luminosity measurement needed to exploit fully the LHC luminosity for the CMS physics program.  Based on lessons learned during Run 2, a key advancement for the Run 3 system was the adoption of radiation-resistant sensors based on acceptor-doped (p-type) silicon~\cite{Jonas_thesis,Joanna_thesis}.  An innovative three-dimensional (3D) printed titanium active cooling loop was integrated into the design, to actively cool the silicon sensors. The complete BCM1F detector consists of four C-shaped PCBs with added contacts for the cooling pipe, hereafter referred to as C-shapes. Two C-shapes form a ring around the beam pipe, placed at $\pm$1.83\,m away from the interaction point (IP) on both ends. Sensors and ASICs are placed at a radius of about 7\,cm. Figure~\ref{fig:cshape} (left) shows a photograph of the C-shape PCB with three ASICs, each processing data from two two-pad sensors (four silicon pads in total), which  are further described in Section~\ref{sec:sensors_and_beamTest}. One C-shape consists of 12 independent channels, adding up to a total of 48 BCM1F channels.
Figure~\ref{fig:cshape} (right)  shows a drawing  of the C-shape PCB with the cooling pipe and the protection cover, which all assembled together form the C-shape ready for installation.

In the following sections, a comprehensive description of the design and operation of the upgraded BCM1F detector implemented for LHC Run 3 is presented. Detailed results on sensor characterization and performance, test-beam measurements, and operational experience within CMS during Run 3 are discussed.
\section{Detector system description}
\label{sec:detector_description}

The design of BCM1F was shaped by the stringent installation constraints within the CMS detector around the LHC beam pipe, including radiation hardness, high-voltage stability, minimal size, and a limited number of connections. This design combines flexible and rigid PCB elements. Folding of the flexible sections allowed the assembly to fit within the carriage structure for insertion into CMS. To allow signal transmission over long distances, the signal is converted from electrical to optical close to the detector, and the optical signal is transmitted to the CMS service cavern, where data processing takes place. This section describes the mechanical design of BCM1F, the motivation for the installation location, the means of signal transmission, and the back-end electronics used for signal processing.

\subsection{Mechanical design}

\begin{figure}[!htb]
   \centering
   \includegraphics*[width=.99\columnwidth]{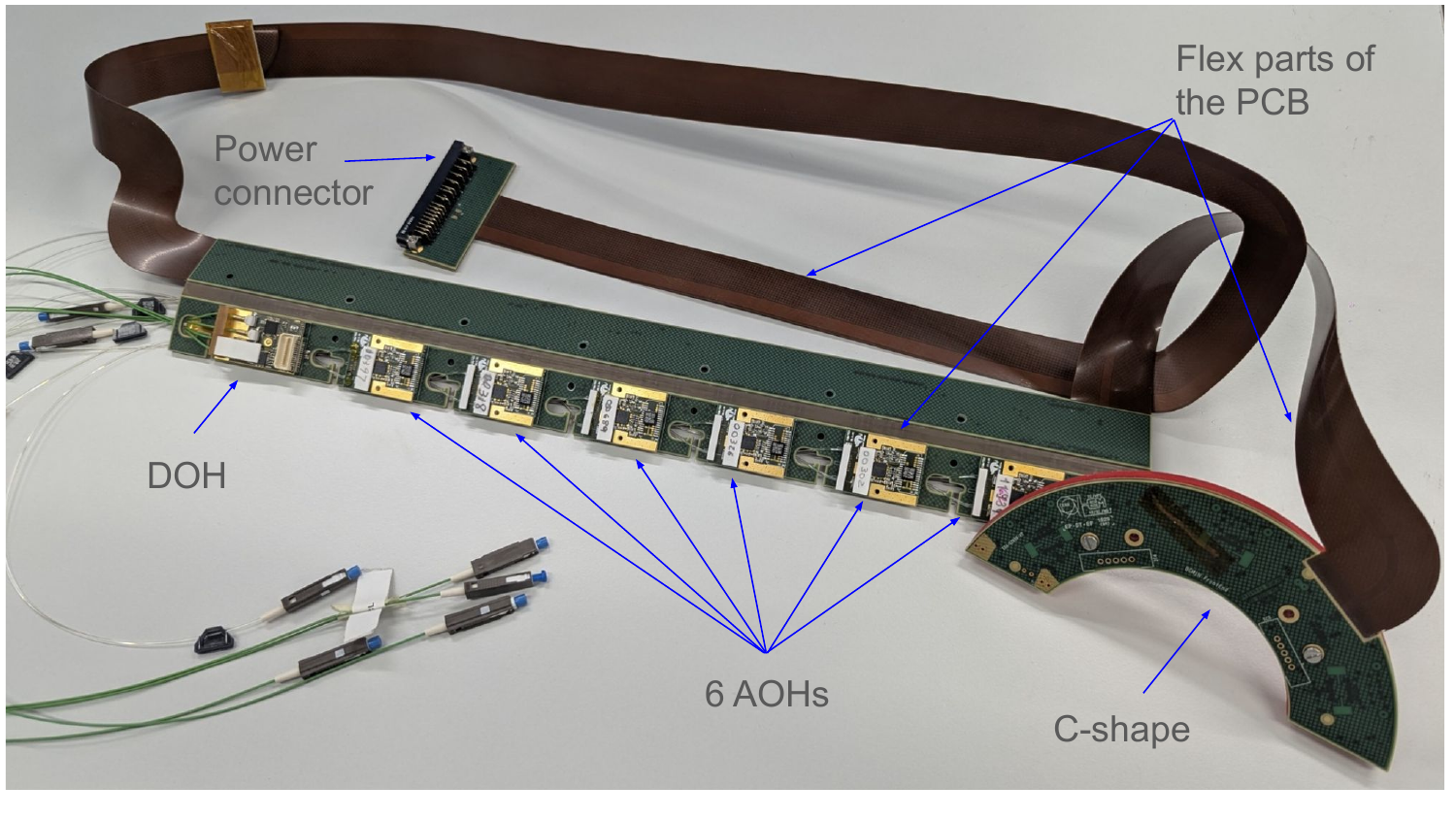}
   \caption{Photo of the full BCM1F front end with the C-shape, the optomother board hosting the AOHs and DOH, the power connector and the two flex sections.}
   \label{fig:cshape_photo}
\end{figure}

The BCM1F employs a custom hybrid PCB combining rigid sections with integrated flexible portions, terminated with a single power connector at the end of the flexible section (Fig.~\ref{fig:cshape_photo}). 


The rigid PCB section, or C-shape, contains the sensors, ASICs, and passive components. A flexible section links the C-shape to the optical board, which comprises two rigid halves joined by a foldable section and carries six analogue–optical hybrids (AOH) and one digital–optical hybrid (DOH)~\cite{DOH}.

\begin{figure}[!htb]
   \centering
   \includegraphics*[width=.99\columnwidth]{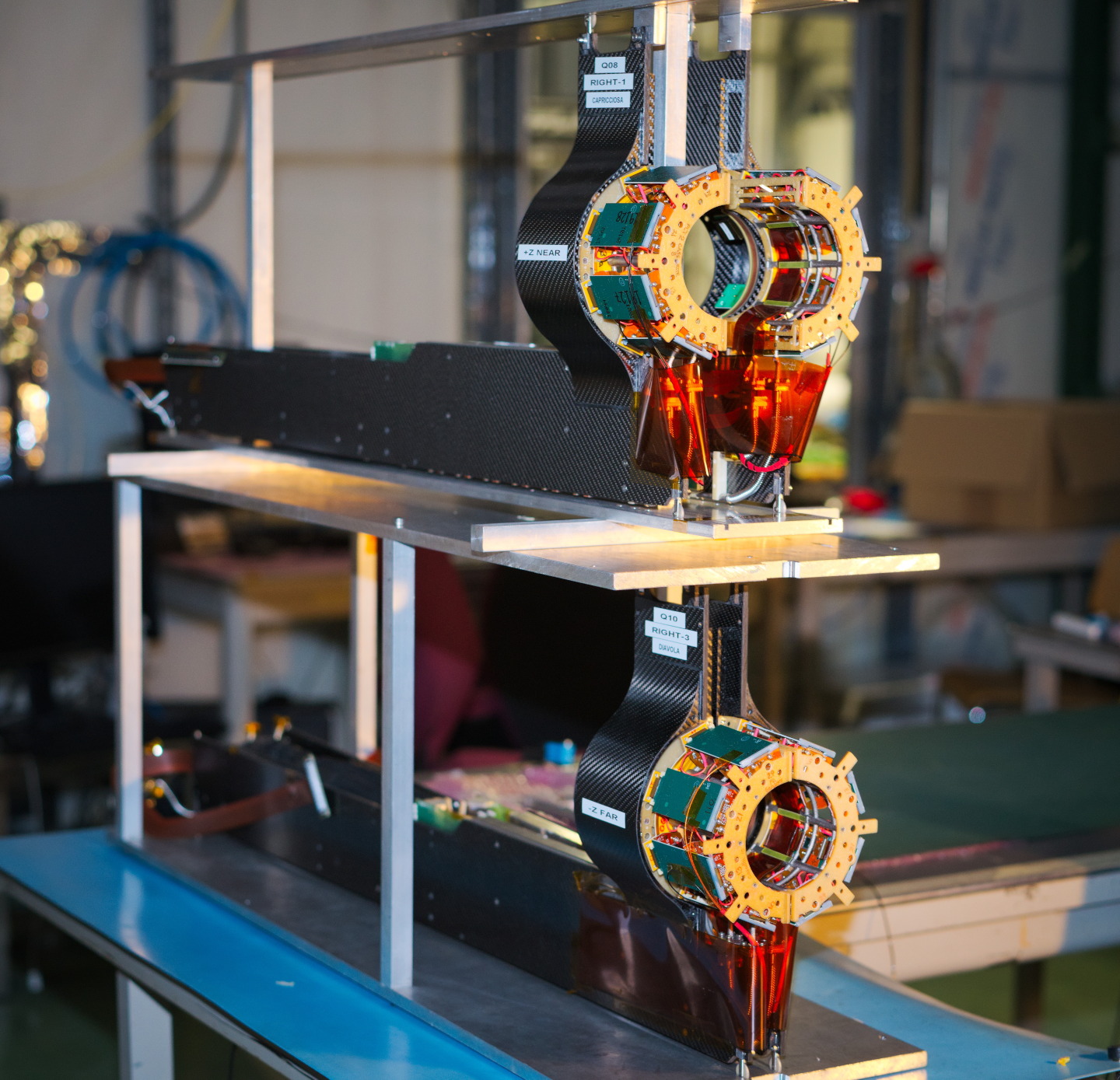}
   \caption{Photo of the BCM1F and PLT carriages ready for installation in CMS. The PLT is visible in front, with the BCM1F mounted behind it}
   \label{fig:carriages_photo}
\end{figure}

For installation inside CMS, the BCM1F is integrated in a carriage together with the Pixel Luminosity Telescope (PLT) as well as Beam Conditions Monitor for Losses (BCML). Four carriages mounted in the transport structure and ready for installation are shown in Fig.~\ref{fig:carriages_photo}.

A cooling loop, connected to the CMS Strip Tracker cooling plant, circulates liquid C\(_6\)F\(_{14}\). The surface of the cooling loop is machined flat and polished to ensure optimal thermal contact. The PCB is secured to the cooling loop using screws to provide a mechanically stable and thermally efficient interface. The loop cools the AOH diodes to maintain a stable temperature, since the AOH signal output varies with temperature. The C-shape containing the sensors is cooled via thermal contacts on the PCB, with two cooling blocks positioned in the loop of the cooling pipe, as shown in Fig.~\ref{fig:cshape}. 
During Run 2, this configuration maintained the C-shape at an operational temperature of approximately –16~$^\circ$C, corresponding to a cooling-plant set point of –20~$^\circ$C. For Run 3, the cooling-plant temperature was further lowered to –25~$^\circ$C, keeping the C-shape at around -18~$^\circ$C.

\subsection{Installation location} 
\label{sec:location}

The time of flight for relativistic particles between the CMS interaction point and the BCM1F detector, positioned 1.83\,m from the IP, is 6.25\,ns (1/4 of the bunch crossing interval). Due to this ``golden" location, beam-induced background accompanying an incoming LHC bunch produces hits in the detector 12.5\,ns before the collision products, making signal and background maximally separated within the LHC bunch spacing of 25\,ns.

\subsection{Signal creation and transmission}
\label{sec:signalamp}

The electron-hole pairs, drifting in the sensor after a charged particle crossing, induce the signal at the ASIC inputs. This signal is amplified using a dedicated 4-channel charge-sensitive amplifier with an input-charge-independent rise time of 7\,ns, typical 10\,ns full width at half maximum (FWHM) and recovery to baseline within 25\,ns, allowing the separation of detector hits in subsequent bunch crossings (BXs). Each BX along the LHC orbit is labelled by a bunch crossing identification number (BCID) between 0 and 3563.

\begin{figure}
    \centering
    \includegraphics[width=\columnwidth]{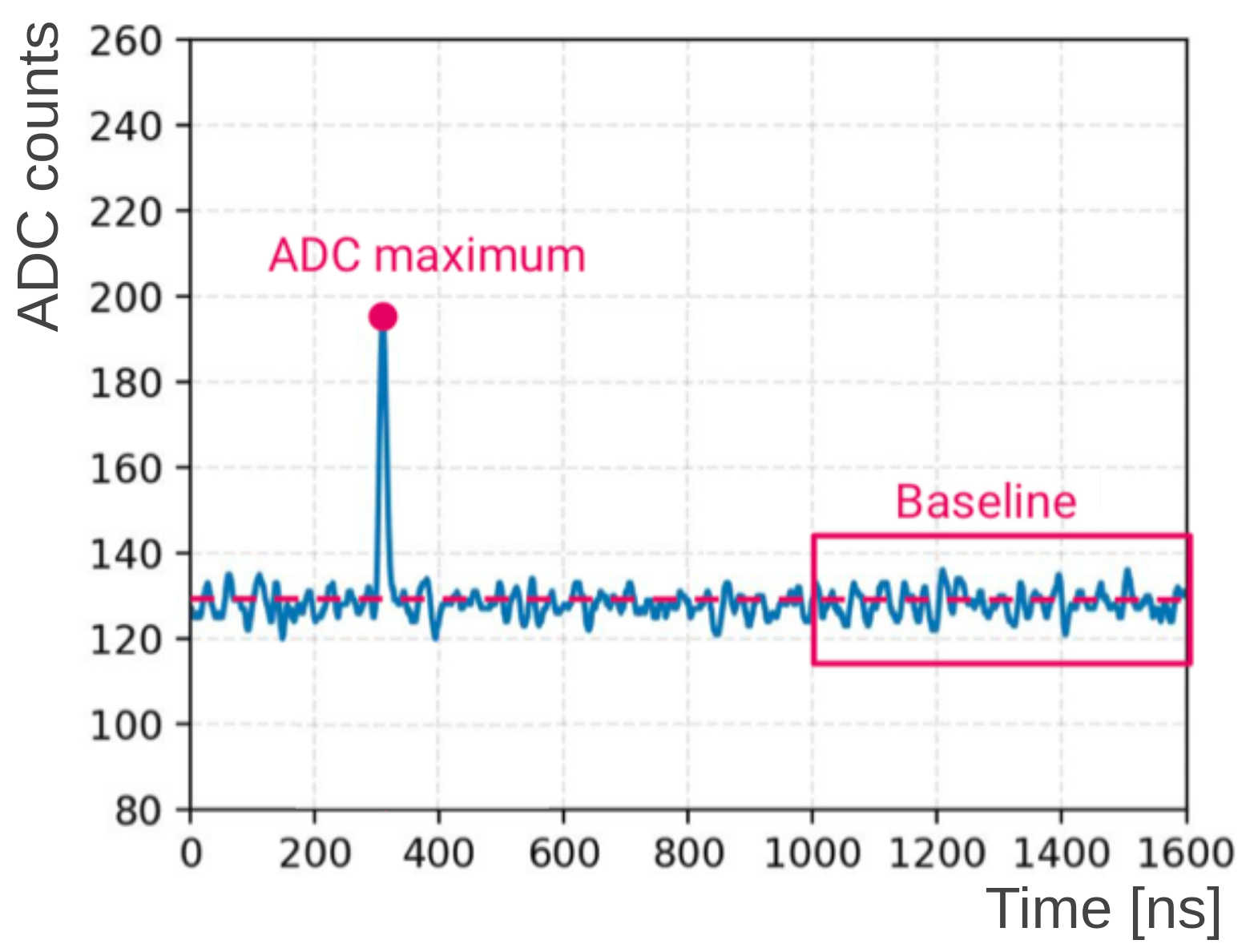}
    \caption{Example of a signal digitized with the VME ADC (see Sec~\ref{sec:vme_backend}). The ADC pedestal baseline is shown with a dashed line at about 130 counts. A particle hit from a lab source is visible at around 300\,ns, with signal to noise ratio above 10.}
    \label{fig:BCM1F_signal_shape}
\end{figure}

An example of the BCM1F signal after  amplification, as sampled by an analogue to digital converter (ADC), is shown in Fig.~\ref{fig:BCM1F_signal_shape}. In this example the ADC baseline fluctuates around 130 ADC counts and reaches about 195 ADC counts when a particle hit is detected. Signal-to-noise ratio is above 10. Signal amplitude is defined as the difference between the maximum of the peak and the mean value of the baseline. Further details of the back-end electronics used to obtain Fig.~\ref{fig:BCM1F_signal_shape} are described in Sec.~\ref{sec:backend_and_opt}.

Electrical pulses from the front-end ASICs are converted into optical signals on the BCM1F optomother board, shown in  Fig.~\ref{fig:cshape_photo}, using  AOH boards that have been designed and built for the phase-0 CMS tracker~\cite{AOH}. 

Through 12-channel single mode optical fibres of about 150\,m in length, the optical signals with a wavelength of 1310\,nm are transmitted from the CMS experimental cavern to the back-end electronics located in the more accessible CMS service cavern. 

\subsection{Back-end electronics} \label{sec:backend_and_opt}
\begin{figure}
    \centering
    \includegraphics[width=\columnwidth]{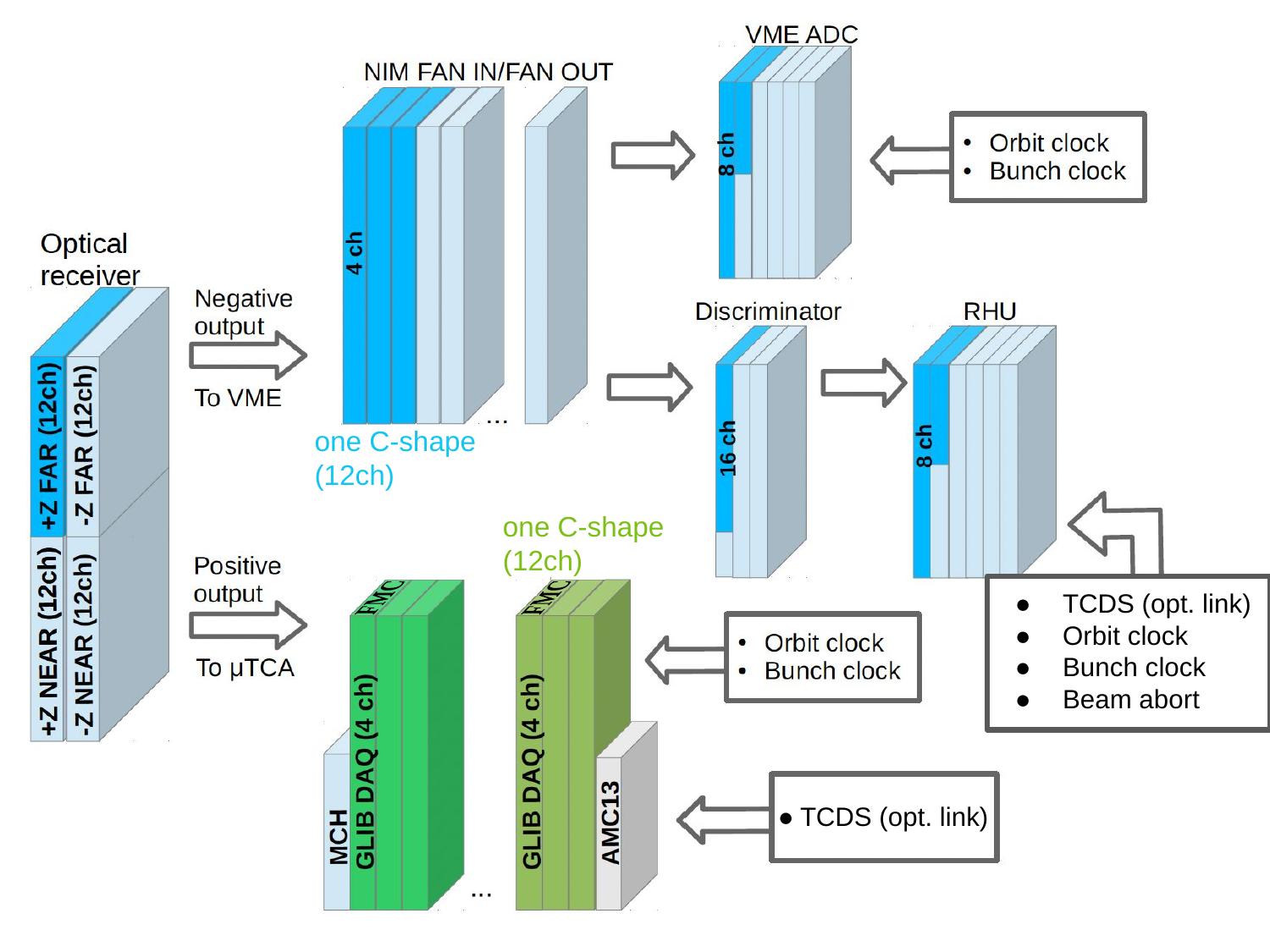}
    \caption{Schematic view of the BCM1F back end~\cite{olena-thesis}. For illustration purposes, boards used for the readout of one C-shape (12 channels) are shown in bold colour}
    \label{fig:backend}
\end{figure}

In the CMS service cavern, optical receivers convert the optical signal back to electrical signals, and additionally produce an inverted copy of the signal. The two copies of the signal  are then forwarded for processing by two back-end readout chains based on different standards: a VME system described in Sec.~\ref{sec:vme_backend}, and a $\mu$TCA system developed as an upgrade to the VME-based back end and described in Sec.~\ref{sec:utca_backend}. Figure~\ref{fig:backend} shows a sketch of the full BCM1F back end. Both back-end systems were operated during Run 3 to ensure redundancy and reliability. This dual operation offered complementary methods for signal identification and monitoring, enhancing the understanding of systematic effects in the readout chain and providing redundancy in online luminosity measurements. Both systems construct synchronised, real-time orbit-occupancy histograms to measure per-bunch-crossing luminosity and beam-induced background.

For testing and monitoring both readout chains, the front-end ASIC generates test pulses of an amplitude approximately equivalent to the signal of a minimum ionizing particle (MIP). Injection of the test pulse is initiated at the ASIC input via an external trigger. Upon activation, a fixed amount of charge is injected into the preamplifier, and the signal follows the whole amplification and readout chain. Monitoring of the test pulse amplitude makes it possible to disentangle sensor ageing effects from the ageing of the other electronics components in the front-end readout, as both are exposed to the harsh radiation environment of CMS.

\subsubsection{VME-based back end}\label{sec:vme_backend}

The BCM1F VME back-end system~\cite{bcm1f-jessica,olena-thesis,TWEPP} processes BCM1F signals received from the negative output of optical receivers, as illustrated on the upper part of Fig.~\ref{fig:backend}. 
Linear FAN IN/FAN OUT boards are used to duplicate the signal to distribute it further. One signal path leads to a VME analogue-to-digital converter (ADC, CAEN V1721), which can record samples of raw data used for commissioning and offline  stability monitoring. The second signal path is sent to a fixed threshold discriminator (CAEN v895) to distinguish hits from electronic noise. The discriminated signal is subsequently routed to a custom-developed Real-Time Histogramming Unit~\cite{Penno:154451} (RHU), whose function is to project the 
arrival of signal into time bins along an LHC orbit to form an occupancy histogram.

The RHU aggregates the discriminated signals  as a function of the relative time delay with respect to the LHC orbit clock signal (corresponding to one full LHC orbit (a single turn) of 89\,$\mu$s), using a time bin width of 6.25\,ns. This yields 4 bins per 25\,ns bunch crossing and a total of 14\,256 occupancy bins.
 
After a time aggregation period of 4096 LHC orbits, the histogram is closed and made available for readout.
This period, known in CMS as a \textit{Lumi Nibble} (NB), is broadcast centrally to all CMS luminometers by the Timing and Control Distribution System (TCDS) \cite{CMS_run3_development}. The broadcast ensures a common, synchronized integration period across all CMS online luminosity systems. Double buffering in the RHU prevents any loss of data.

The hardware is configured such that the collision products of the first bunch crossing (BCID~0) of the orbit fall into the third bin of the histogram and consequently the BIB hits fall in the first bin.

\begin{figure}
    \centering
    \includegraphics[width=\columnwidth]{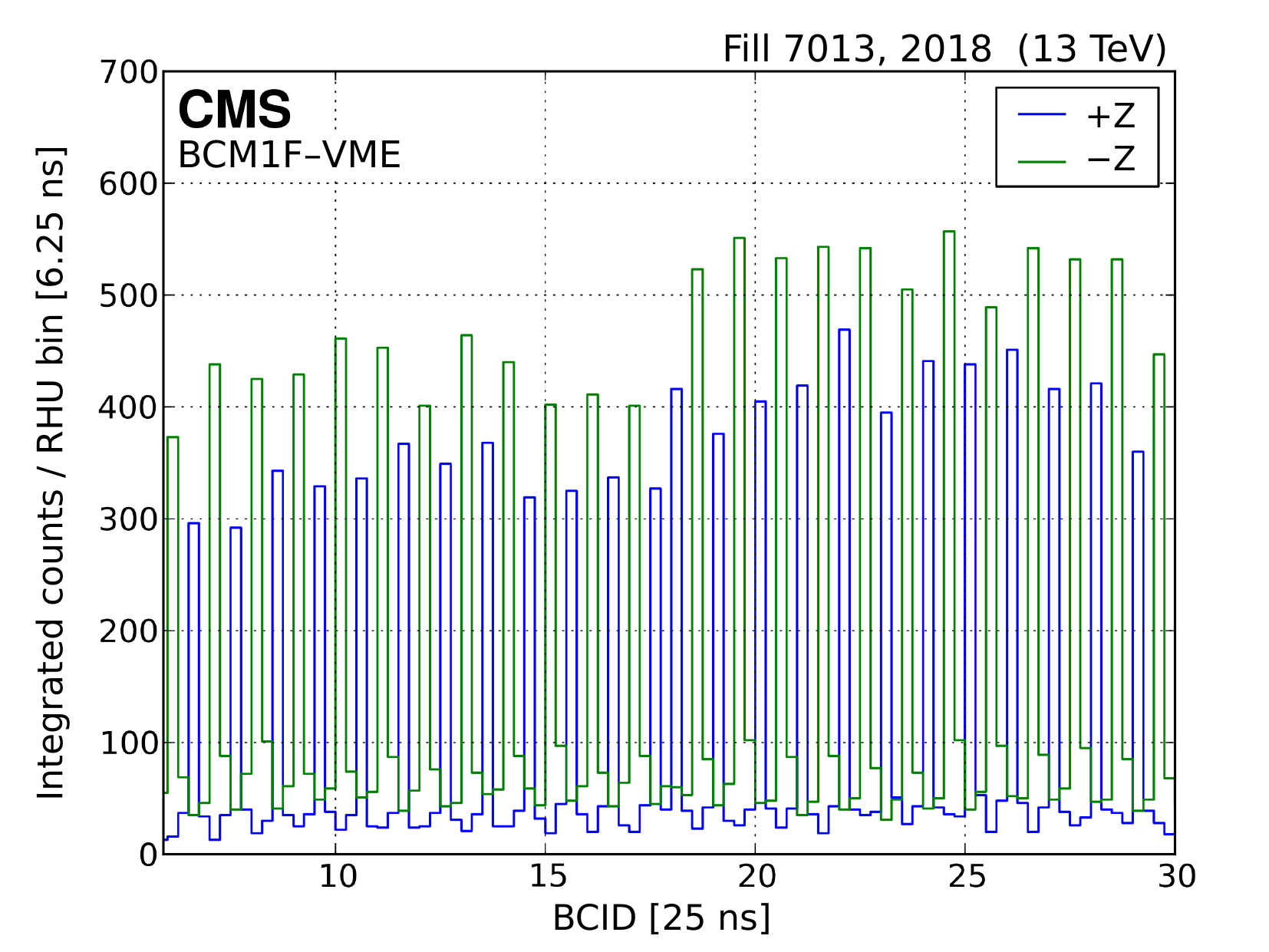}
    \caption{
BCM1F data recorded during a typical fill in LHC Run 2, demonstrating the timing difference in arrival of hits resulting from BIB at the two BCM1F detector locations. One bunch crossing corresponds to four RHU bins of 6.25\,ns each. Bunch crossings of BCID 6–17 were filled only for Beam 2, and those of BCID 18–29 were filled only for Beam 1. The incoming Beam-1-induced background is measured by the BCM1F first on the +Z end, arriving 12.5\,ns later at the -Z end, and vice versa for Beam 2. Data recorded before the beams were brought into collision is shown, to prevent a large contribution from afterglow hits due to activation of the detector material}
    \label{fig:BIB_unpaired_bunches}
\end{figure}

To illustrate the BCM1F timing structure and ability to measure BIB independently for two beams, the data of fill number 7013, recorded before the beams were brought into collision is shown for BCIDs 6 to 29 in Fig.~\ref{fig:BIB_unpaired_bunches}. The sum of the rate from all channels on the -Z and +Z detector locations are shown. One bunch crossing is represented by four bins of 6.25\,ns each. Bunch crossings 6 to 17 were only filled for beam~2, and 18 to 29 were filled for beam~1. In CMS, beam~1 enters on the +Z end and exits on the -Z end, while beam~2 enters on the -Z end and exits on the +Z end. Hence the first (second) set of 12 bunches leave hits due to beam induced background in the -Z (+Z) detectors and 12.5\,ns later in the +Z (-Z) detectors. No collision products are visible in this plot as the beams were passing each other without colliding. Normally, collision products appear in the same bin as the outgoing beam background.

\subsubsection{$\mu$TCA-based back end}\label{sec:utca_backend}

The \utca back end was developed as an upgrade to the VME-based system, with signal processing, amplitude monitoring, and histogramming all implemented on the on-board FPGA. 


The \utca back-end system consists of a NAT MicroTCA Hub (MCH) module with three Gigabit Link Interface Boards~\cite{GLIB_2013} (GLIB) each connected to a four-channel FPGA Mezzanine daughter card (FMC125 produced by 4DSP).
Each FMC performs 8-bit analogue-to-digital conversion at sampling rates of up to 1.25\,GS/s per channel. The inputs are AC-coupled, and a phase-locked loop (PLL) synchronises the system with the LHC bunch crossing clock at 40.08\,\unit{MHz}. The ADC is also configured to lock to a frequency of 30 times the LHC bunch clock, enabling a synchronised sampling of 1.204\,GS/s per channel.
Peak detection and histogramming are performed on the Xilinx Virtex-6 FPGA hosted on the GLIB. This consolidates the functionality of both the VME ADC and RHU modules, supporting real-time hit identification, occupancy and amplitude histogram aggregation, and raw ADC data recording. 
The TCDS signals are decoded by a double-width Advanced Mezzanine Card (AMC13)~\cite{amc13} after which they are read by the MCH, which manages system control and communication.
Communication between software and firmware is achieved through the IPbus protocol~\cite{Larrea_2015} with IP addresses assigned to each GLIB. To maintain the sample structure, the data are streamed into a first-in-first-out (FIFO) queue that transmits all 30 samples per bunch crossing. 

Two orbit and clock sources are available: CMS TCDS signals from the AMC13, and orbit and bunch clock signals derived directly from the LHC RF clock via analogue inputs on the FMC125. In the nominal configuration, the LHC clock and orbit signals are used, which allows for synchronization with the LHC beam for all beam modes. The CMS TCDS signal is used only to define the 4-NB integration period for the \utca histogramming, the CMS run number and the LHC fill identifiers, enabling the \utca system to be synchronous to the other CMS luminometers. The \utca histogramming module provides a timing binning resolution of six bins per 25\,ns bunch crossing.
 
 The \utca signal processing continuously receives samples from the FMC125 in a real-time pipeline, without any triggering mechanism or dead-time. Hit identification is handled entirely in the FPGA using a peak-finding algorithm described below. The algorithm is deterministic and based on a fixed number of clock cycles.

\subsubsection*{Derivative-based peak finding algorithm for real-time signal processing}\label{sec:peakFinder}

\begin{figure}[!htb]
   \centering
   \includegraphics*[width=.99
   \columnwidth]{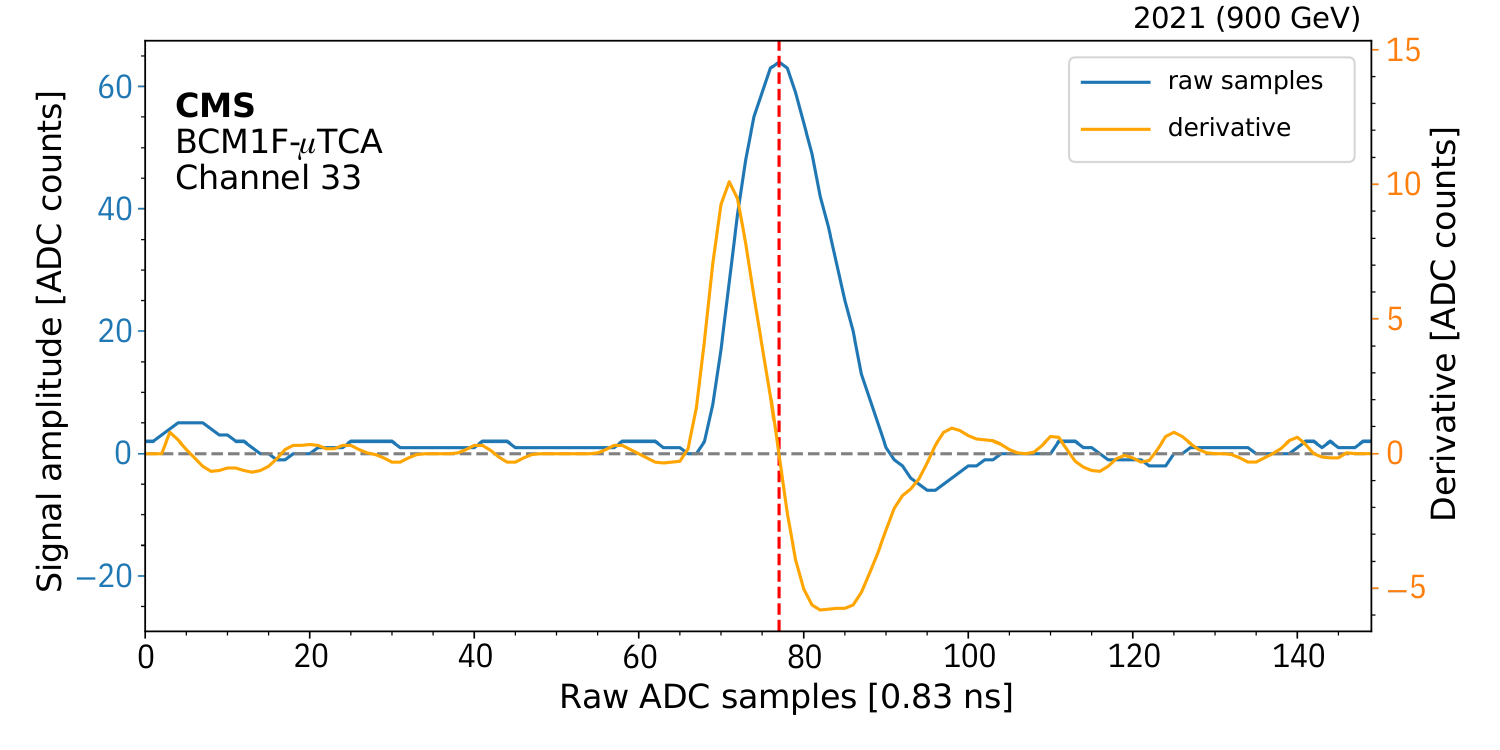}
   \caption{BCM1F \utca  sampled hit signal (blue), overlaid with the corresponding derivative (orange). The detected peak at the derivative zero-crossing is marked with dashed red line~\cite{Joanna_thesis}}
   \label{fig:raw_utca}
\end{figure}

The \utca real-time signal-processing chain uses a peak-finding algorithm based on identifying the zero crossing of the derivative of the signal. 
An example is shown in Fig.~\ref{fig:raw_utca}, which displays a typical single particle hit signal (in blue) and its derivative (in orange). The location of the peak can be identified by the location of the zero crossing of the derivative, and is used to identify the hits and their corresponding arrival times. The arrival time calculation benefits from the property of the ASIC shaping, that the amplifier peaking time is constant with signal amplitude~\cite{asic}. The integral of the positive side of the derivative signal is used as a monitor of the signal amplitude and relative gain stability of the system. 

This derivative-based method has the added advantage of resolving hits that occur close together in time — resulting in overlapping analogue pulses — as long as the derivative of the pulse shape crosses zero (from positive to negative) for each contributing hit. This feature enhances the reliability of hit detection and the precision of the corresponding arrival time measurement, even under high-pileup conditions. 
The likelihood of double hits remains extremely low (0.01\% at a pileup of around 65).
Moreover, the approach mitigates low-frequency fluctuations such as baseline shifts.
However, a simple derivative threshold can mistakenly interpret high-frequency noise with steep gradients as genuine signals. 

To address this, low-pass filtering techniques are applied to suppress these unwanted components. 
The filter window size is carefully chosen considering the signal bandwidth and the required level of noise reduction, ensuring linear differentiation.
The implementation and performance of this method were studied with simulations, as detailed in Ref.~\cite{alex-thesis}. 
In practice, the derivative is computed using a smooth, noise-robust differentiator (SNRD), with a window size of $N=7$ samples:
\begin{equation}
   \mathrm{SNRD} = \frac{5(x_1-x_{-1}) + 4(x_2-x_{-2}) +x_3-x_{-3}}{32h},
\end{equation}\label{eq:snrd} 
where $x_i$ represents the sample with the index $i$ around the sample where the SNRD is calculated, and $h$ is the distance (25/30 ns) between samples. 

To optimize peak detection performance, three key algorithm parameters must be carefully adjusted:
\paragraph{1. Derivative Threshold (DdT),}
which is based on the gradient of the fast-rising edge of the signal pulse produced by particle hits. Since signal and noise produce different derivatives at the fast-rising edge, this method effectively discriminates against low-amplitude, Gaussian-like electronic noise.
This is because signal pulses are shaped by the ASIC with a constant peaking time. A rough upper limit on the derivative threshold can be estimated using the test pulse, but the final setting should be done with the beam and tuned based on the signal amplitude of each channel.

The algorithm distinguishes pulses using a two-step process based on the signal derivative. First, it detects when the derivative exceeds a predefined threshold, marking the potential start of a pulse. It then locates the subsequent zero-crossing of the derivative, which typically corresponds to the pulse peak, as shown in Fig.~\ref{fig:raw_utca} with a red dashed line. In the case of overlapping input pulses from multiple hits close in time, only a single signal maximum may appear due to the merged signals. To resolve these, the algorithm  searches for multiple threshold crossings of the derivative followed by their respective zero-crossings. This enables the resolution of the contributing hits and their corresponding relative arrival time.
A relevant example of hit detection from the differentiated signal is shown in Sec.~\ref{sec:uTCA_commissioning}.

\paragraph{2. Amplitude Threshold (DaT),} is defined as the minimum cut on the measured pulse amplitude. It serves as a secondary filter, triggered only after the derivative threshold (DdT) is crossed. It aims to reject small, fast pulses that are unlikely to result from a particle hit.
The pulse amplitude is computed as the integral over the samples between the derivative level ascent and level descent zero-crossings. This method makes it insensitive to baseline shifts, as it does not rely on explicit baseline subtraction.

\paragraph{3. Peak isolation or Time over the threshold (ToT):}
verifies the validity of the detected peak by analysing surrounding raw samples to ensure it represents a true signal, and in case of overlapping pulses, defines separation between them. Specifically, it checks that a defined number of preceding and succeeding points have a smaller derivative value than the peak itself. 

Due to differences in sensor performance and signal paths, these parameters must be optimized separately for each channel. Additionally, after irradiation, the parameters must be regularly monitored and adjusted to account for detector signal degradation.

\begin{figure}[!htb]
   \centering
   \includegraphics*[width=.8\columnwidth]{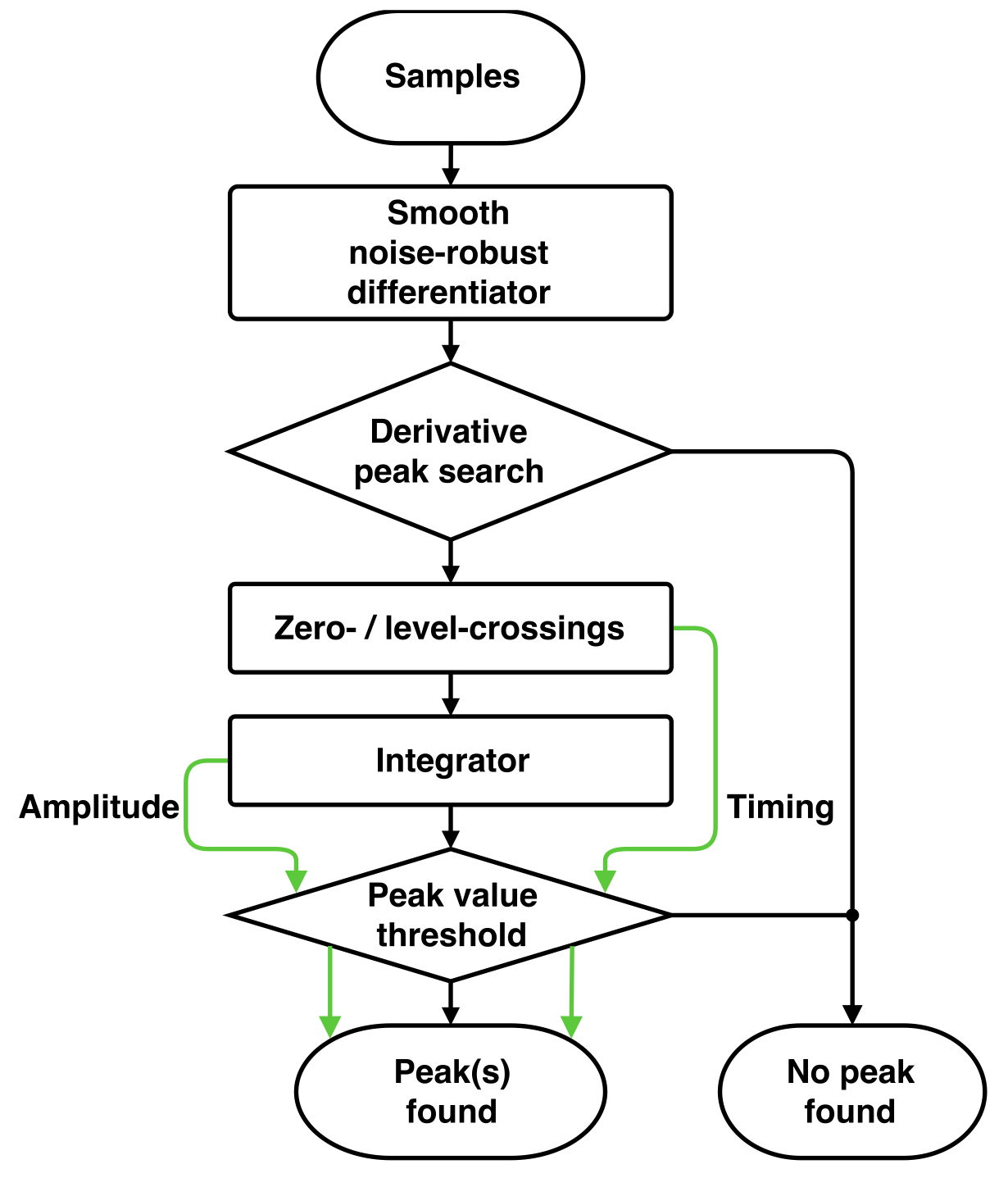}
   \caption{Peak finding algorithm flow chart~\cite{alex-thesis} as implemented in the FPGA of the \utca GLIB}
   \label{fig:peakFindingAlgo}
\end{figure}

A schematic representation of the complete data flow within the peak-finding algorithm is shown in Fig.~\ref{fig:peakFindingAlgo}.

By combining an amplitude cut-off with a derivative threshold, signals originating from electronic noise are effectively rejected. Since noise typically produces low-amplitude fluctuations with shallow gradients, their maximum derivative remains well below the threshold set for genuine particle hits.
\section{Sensor characterization}
\label{sec:sensors_and_beamTest}

Silicon sensors (Sec.~\ref{sec:sensors_design}) used in the BCM1F detector assembly for LHC Run 3 were tested in the laboratory for stable response and basic properties (Sec.~\ref{sec:ivcv}). Only sensors that passed the initial characterization were used to construct the detector. A prototype detector was studied using an electron beam at the DESY test beam facility, where detailed studies of sensor response and validation of the guard ring grounding scheme were performed on a BCM1F prototype detector (Sec.~\ref{sec:LS2_test_beam}).  

\subsection{BCM1F Run 3 silicon sensor design}
\label{sec:sensors_design}

\begin{figure}
    \centering
    \includegraphics[width=\columnwidth]{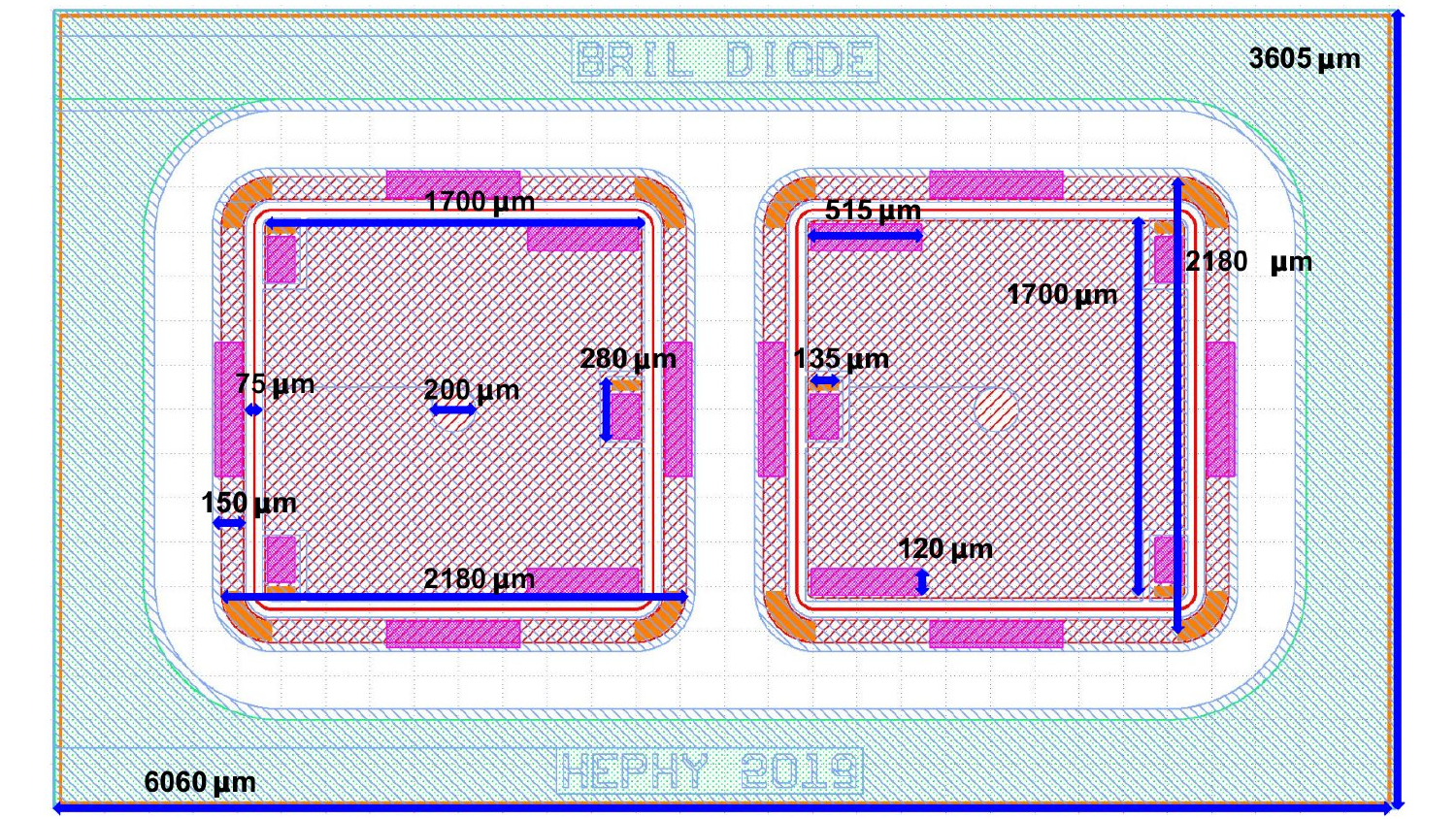}
    \caption{ Two-pad silicon sensor design. Each pad has three identical DC-pads, two AC-pads and is surrounded by the independent guard ring. Areas with direct contact to the bulk of the sensor are shown in orange. Pads shown in pink are dedicated for wire-bonding. The thin lines around each pad show the location of the p-stop implant between the pad and the guard ring. Areas without metallization are indicated in white. }
    \label{fig:2pad_design}
\end{figure}

\begin{figure}[!htb]
   \centering
   \includegraphics*[width=.99\columnwidth]{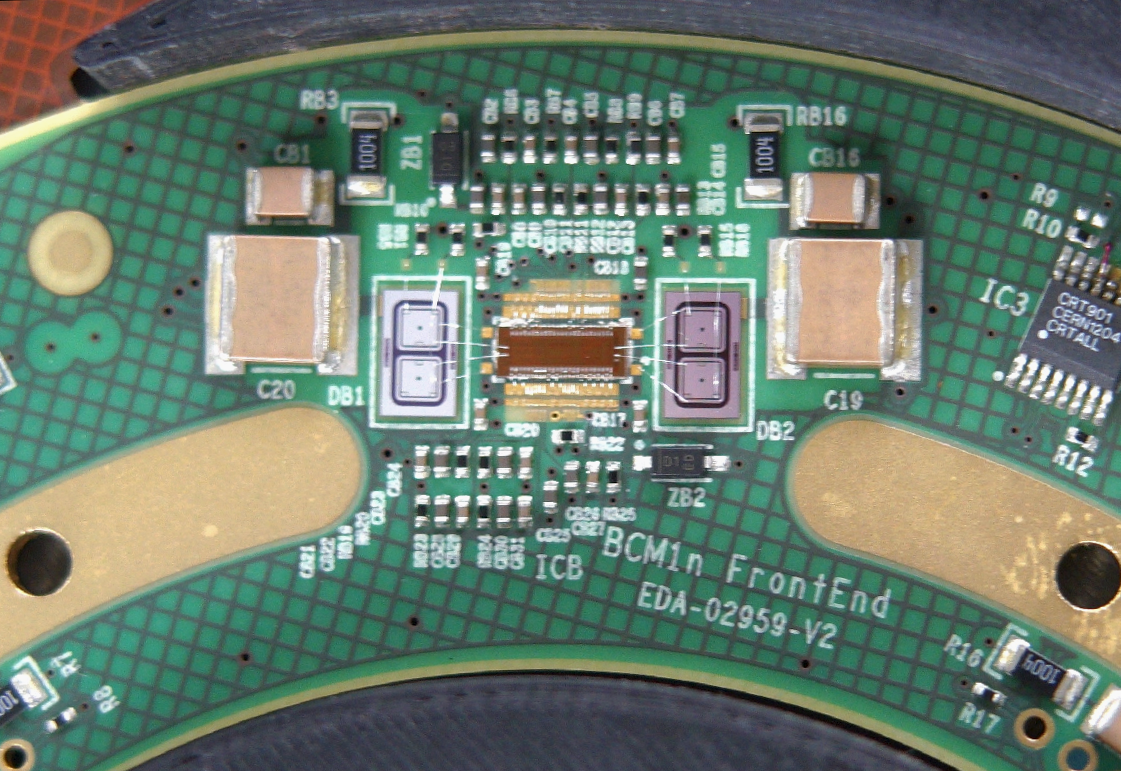}
   \caption{Photo of the BCM1F C-shape segment with wire bonds between the ASIC inputs and the two-pad silicon sensors visible }
   \label{fig:sensers_ASIC_photo}
\end{figure}

The BCM1F detector uses two-pad 290\,$\mu$m thick silicon sensors, which were produced on the half-moons of the CMS outer tracker sensors wafers. The sensor design is shown in Fig.~\ref{fig:2pad_design}, with a pad metallization size of $1.7\times 1.7$\,mm$^2$. For the possibility of bonding from each side of the sensor, each sensor pad has three identical DC-pads (with direct contact to the bulk of the sensor) and two AC-pads (with a SiO$_2$ passivation layer separating the bulk of the sensor and metallization providing capacitive coupling). The wire bonding is done in such a way that one of the DC-pads per pad is grounded through a resistor, and one of the AC-pads is bonded to the ASIC~\cite{asic} input, as shown in Fig.~\ref{fig:sensers_ASIC_photo}. Because the ASIC was originally designed for diamond sensors and cannot compensate for a constant input current, AC coupling was chosen to connect the sensor output to the amplifier. This becomes even more important with irradiation, as leakage current increases. Each sensor pad is surrounded by an independent guard ring, which is connected to the ASIC ground with a wire bond. Grounding of the guard ring suppresses small signals at the edges of the sensor pads, as confirmed by beam test results described in Sec.~\ref{sec:LS2_test_beam}. Independent guard rings for each pad also suppress small signals originating from the area between the two pads on the shared edge.

\subsection{Silicon: IV and CV measurements}
\label{sec:ivcv}

\begin{figure}
    \centering
    \includegraphics*[width=0.49\columnwidth]{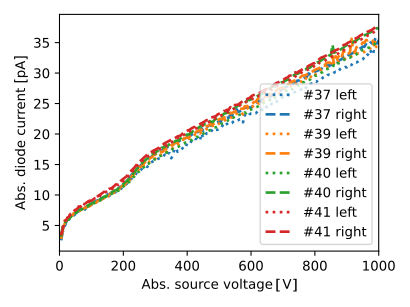}
    \includegraphics*[width=0.49\columnwidth]{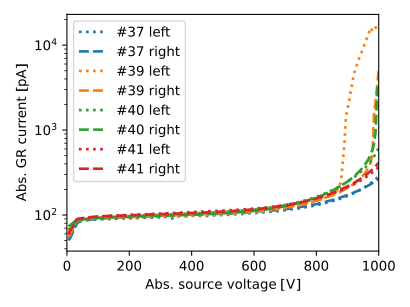}
    \caption{Measured I-V characteristics of silicon diode sensors (left) and guard rings (right)~\cite{Jonas_thesis}. Numbers in the legend are arbitrary sensor identifiers with marker left or right corresponding to the two pads of each sensor.  }
    \label{fig:iv}
\end{figure}

Prior to the assembly of the detector,  current--voltage (I-V) scans were performed on the silicon diode as well as the guard ring. For each case, the leakage current is measured as a function of the applied bias voltage from 5\,V to 1000\,V, with the upper limit well above the operating range of the sensors (normally 300-450\,V). The results are shown in Fig.~\ref{fig:iv}. It is seen that the diode leakage current per sensor remains below 0.05\,nA  up to 1000\,V. The guard ring, meanwhile, begins to demonstrate a sharp increase in current at a bias of around 900\,V, but maintains low current ($\approx$0.1\,nA) at typical operating bias voltages. 

\begin{figure}
    \centering
    \includegraphics*[width=0.6\columnwidth]{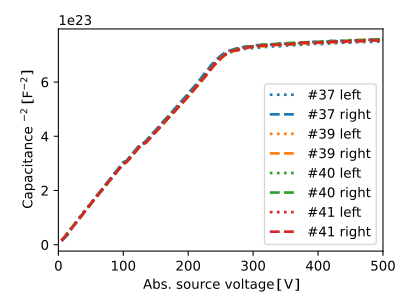}
    \caption{Measured C-V characteristics of the silicon diode sensors~\cite{Jonas_thesis}, with $\frac{1}{C^{-2}}\propto V $  until full depletion, then reaches a plateau. Numbers in the legend are arbitrary sensor identifiers with marker left or right corresponding to the two pads of each sensor. }
    \label{fig:cv}
\end{figure}

Secondly, capacitance--voltage (C-V) scans are performed to determine the full depletion voltage of the diode. The results, shown in Fig.~\ref{fig:cv}, demonstrate a nearly linear increase of the inverse square of the capacitance, $\frac{1}{C^{2}}$, until a certain bias voltage is reached, after which it plateaus. The bias at which this change occurs, identified as the full depletion voltage, was found to be around 260\,V for most sensors.

Resistance and capacitance  were also measured between the AC 
and DC 
connections, to probe for possible defects in the silicon dioxide insulation layer. The mean AC-DC coupling capacitance of the sensors was found to be 360\,pF, while the resistance between AC and DC connections was well above the desired 100\,M$\Omega$.

\subsection{Detector performance measured in a test-beam}
\label{sec:LS2_test_beam} 

A prototype of the BCM1F detector equipped with a front-end ASIC and two two-pad silicon sensors was tested at the DESY II test-beam area~\cite{Younes_thesis}, with one sensor grounded and the second having a floating guard ring. An electron beam was used with an energy of 5.4\,GeV.

For the reconstruction of the particle trajectories, the EUDET pixel beam telescope~\cite{EUDET} was used.
Three telescope planes were positioned before and three planes after the BCM1F detector prototype, with the detector centred along the beam axis. Data readout was triggered by Hamamatsu Photo-Multiplier Tubes (PMTs) positioned in front of and behind the telescope. For each trigger a Trigger Logic Unit number was assigned, which allowed the matching of each reconstructed track with the signal generated in the BCM1F sensors. 

\begin{figure}
\centering
\includegraphics[width=\columnwidth]{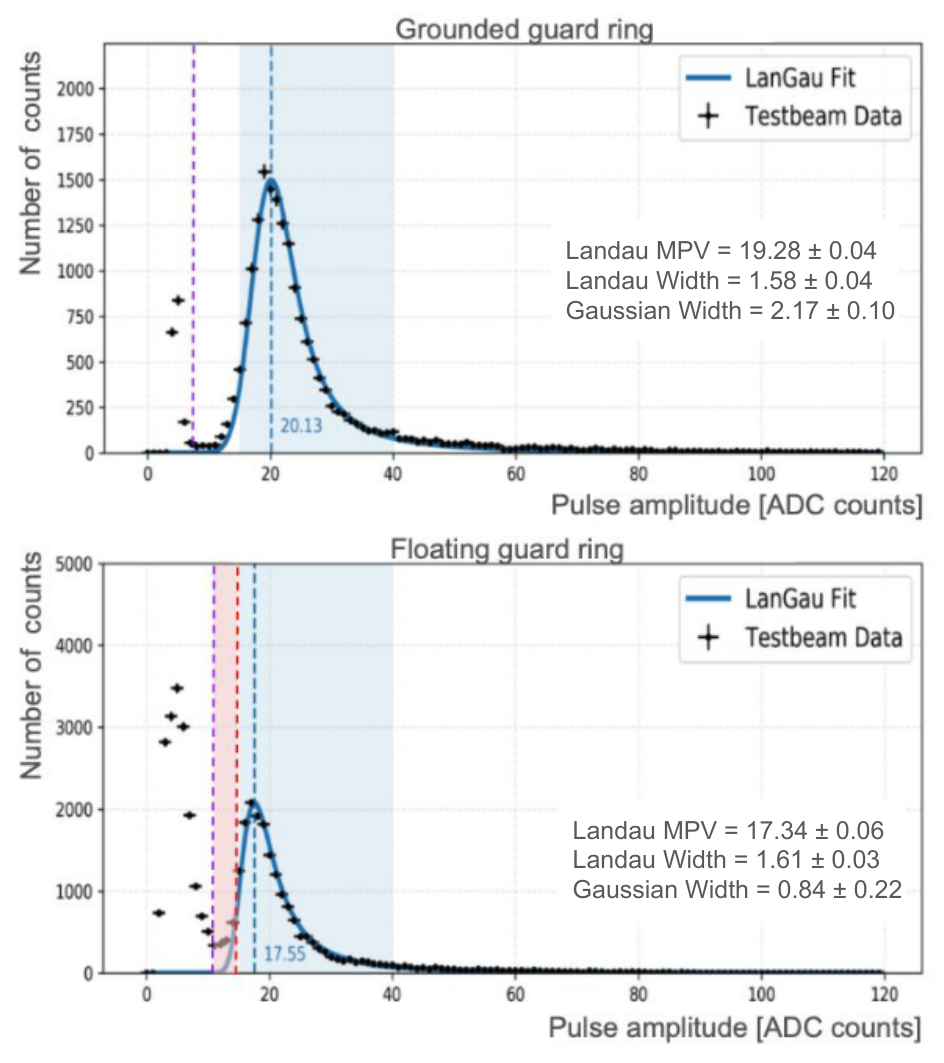}
\caption{Amplitude spectra for sensors with grounded guard ring (top) and floating guard ring (bottom). The violet line indicates the minimum ADC threshold to cut most of the noise. The blue band indicates the fit range for the signal, and the peak, denoted hereafter as most probable value (MPV), is marked by the blue dashed line. The red band shows the transitions region only visible for the sensor with a floating guard ring. The bias voltage applied to the sensors is 300\,V}
\label{fig:Test_beam_amp}
\end{figure}

An example of a digitized signal from a lab source during commissioning is shown above in Fig.~\ref{fig:BCM1F_signal_shape}. By processing high-statistics signals from the test beam, amplitude spectra for both sensors on the BCM1F test board were obtained. Figure~\ref{fig:Test_beam_amp} shows the amplitude spectra for the sensors with a grounded guard ring (top) and with floating guard ring (bottom). The energy loss of electrons in silicon sensors follows a Landau distribution. Each measured spectrum is fit to a Landau distribution convoluted with a Gaussian in order to account for the electronic noise. The fit region is highlighted with a blue band. Pulses at a very low amplitude correspond to noise. They form the first visible peak, which is called the pedestal. The pedestal can be removed by setting the minimum signal amplitude cut, as indicated in the figure with a violet line. The sensor with a floating guard ring does not show a clean separation between the signal peak and the pedestal. 

\begin{figure}
\centering
\includegraphics[width=\columnwidth]{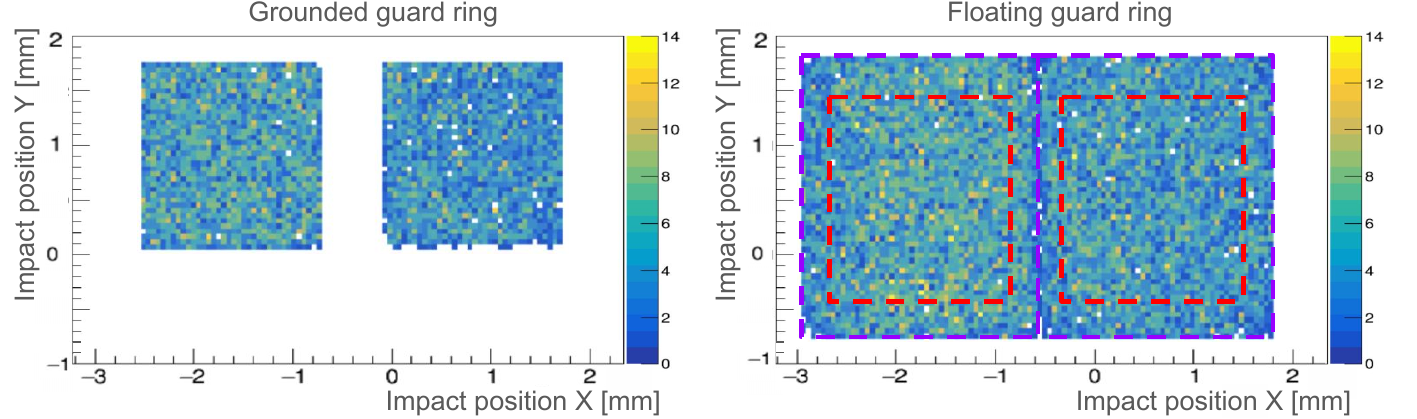}
\caption{Distribution of particle impact positions that correspond to signals above threshold in the sensor with grounded guard ring (left) and that with the floating guard ring (right). In the right plot, the red dashed rectangles indicate the approximate pad area. Hits in the area outside the red dashed rectangles correspond to the low amplitude part of the spectra seen in Fig.~\ref{fig:Test_beam_amp}}
\label{fig:Test_beam_hitmap}
\end{figure}

Tracks reconstructed with the EUDET telescope were used to study the sensor response as a function of the particle impact position and to investigate the origin of the signals in the gap between noise pedestal and Landau peak. Only single-track events were selected to ensure a one-to-one correspondence between signals and impact positions. For hits corresponding to a signal above threshold in the BCM1F sensor, the distributions of particle impact positions in X and Y are shown in  Fig.~\ref{fig:Test_beam_hitmap}. For the sensor with a grounded guard ring, the diode size is observed to be around \mbox{1.9\,$\times$\,1.8\,mm$^2$}, which is consistent with the expected metallisation area. In contrast, for the sensor with floating guard ring, the diode size is observed to be about \mbox{2.6\,$\times$\,2.8\,mm$^2$}, which is much larger than the metallisation area. The latter effect is mainly due to the fact that a floating guard ring extends the sensor electrical field beyond the guard ring edges, yielding a substantially larger active area. When plotting only the impact positions of tracks associated with signals in the red band of the amplitude spectra (Fig.~\ref{fig:Test_beam_amp}, red band), it was confirmed that most of these are associated with the region outside the sensor metallisation (Fig.~\ref{fig:Test_beam_hitmap} frame between the red and violet dashed rectangles). 

The charge collection efficiency (CCE) of the sensor was  studied as a function of the hit position. The sensor was subdivided into 50\,$\mu$m columns. Signal amplitudes, measured in ADC counts, for each of the columns can be converted into energy deposition using the ADC calibration, performed beforehand with known charge injection. The CCE can then be defined as the ratio of the measured charge to the expected charge, calculated for known thickness of the silicon sensor.
The region outside the metallised pad area was found to have much lower CCE, decreasing from about 90\% to about 25\%, while under the metallisation CCE remained stable at 90\% or higher.

Based on the test beam results, it was decided to ground the guard ring for all sensors to maintain a well-defined active area and ensure the separation of the two pads.

\section{Post-installation detector commissioning and operational experience} 
\label{sec:commisioning_ops}

Before installation in CMS, the C-shapes were mounted on the BCM1F/PLT carbon fiber carriage~\cite{PLT} to be inserted into the CMS pixel detector volume. The signals are processed by back-end electronics located in the CMS service cavern (Sec.~\ref{sec:detector_description}).
Commissioning follows the installation. Basic checks can be performed immediately, using test pulses injected into the ASIC inputs to verify the readout chain and determine optimal AOH bias settings. Further commissioning steps, such as timing adjustments, high voltage bias scans, and final threshold adjustments, require beam presence in the LHC.

During operation, detector performance is monitored and adjustments are applied to compensate for radiation-induced efficiency loss. Calibration data are recorded at the start of each LHC fill, including test pulses injected within a range of BCIDs containing no injected bunches (the abort gap discussed in Sec.~\ref{sec:data_reprocessing}). Monitoring of test-pulse amplitudes over time reveals changes in electronics performance, linked mostly to AOH laser output degradation under irradiation. The contribution from ASIC performance degradation is negligible for doses below 100\,MRad~\cite{asic}. 

This section describes BCM1F commissioning, operational procedures, stability measurements, and associated monitoring tools. 

\subsection{Monitoring and optimising the AOH bias settings} \label{sec:AOH_bias}

\begin{figure}
    \centering
    \includegraphics[width=1.0\columnwidth]{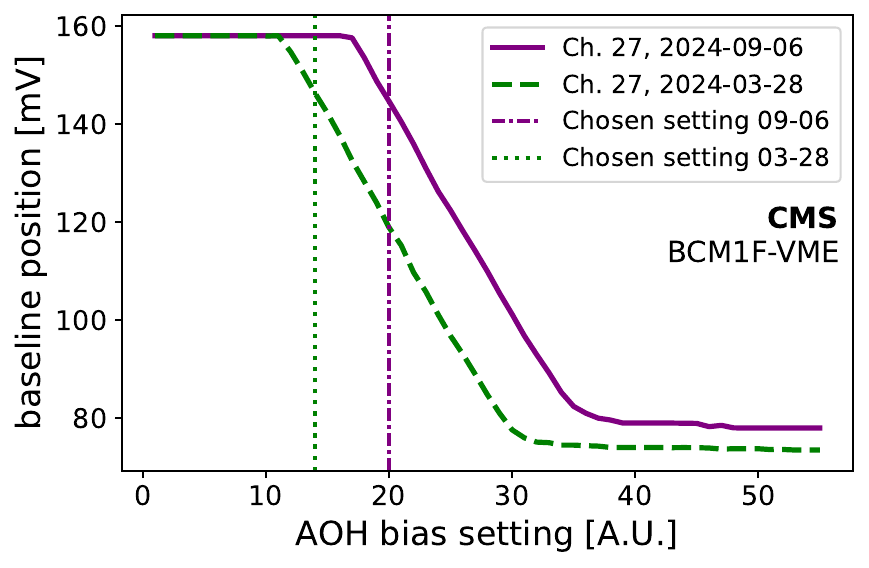}
    \caption{
    Illustration of radiation-induced AOH degradation: The baseline shifts to higher values at a given AOH bias value, or alternatively the operational AOH parameter range shifts to higher values. Data of two AOH bias scans for the same BCM1F channel are shown from March and  September 2024, marking a period during which roughly $90\,\unit{fb}^{-1}$ of integrated luminosity was delivered. The chosen operational setting resulting from each scan is indicated}
    \label{fig:aoh_drift}
\end{figure}

Each AOH is calibrated so that the bias setting enables operation within the full linear response range of the converter. This is determined by scanning bias values with the VME ADC. Test-pulse amplitude and baseline position are measured to identify the linear regime. The scan was first performed immediately after detector installation to verify readout chain functionality. 
Although the AOH components tolerate CMS cavern radiation, the bias threshold of the laser diodes shifts with accumulated dose and operating temperature. Figure~\ref{fig:aoh_drift} shows the results of two AOH bias setting scans performed for the same BCM1F channel in March 2024 (green curve) and in September 2024 (purple curve). The integrated luminosity delivered in this period was $\approx 90\,$fb$^{-1}$ and the corresponding absorbed dose in the location of the AOHs about 50\,\unit{kGy}. Note that the individual AOH slots are not exposed to an equal dose of radiation due to their positioning along the beam axis on the rigid rectangular portion of the PCB (see Fig.~\ref{fig:cshape_photo}). After irradiation, a higher AOH bias setting is required to reach the regime of linear readout response, and the output range is seen to shrink slightly.  The amounts by which performance and ideal bias setting change with respect to the amount of delivered luminosity can vary substantially channel-by-channel.

To maintain optimal bias settings during operation, the calibration procedure runs automatically each week during beam-off periods to determine optimal AOH settings, which can then be updated manually if required to compensate for radiation-induced shifts.

\subsection{VME back-end commissioning and amplitude-based efficiency monitoring during operations} 
\label{sec:rhu_adc}

With the first LHC collisions, the raw signals are recorded by the VME ADCs to confirm the functionality of each channel. Discriminator threshold scans are also performed to tune the inputs for the RHU boards. To maximize the separation between collision signals and beam-induced background in dedicated bins of the RHU, timing adjustment is performed manually with 0.5\,ns delay cables such that signals from beam-induced background fall into the first RHU bin and collision products into the third bin, for each bunch crossing (Sec.~\ref{sec:vme_backend}). 

\begin{figure}
    \centering
    \includegraphics[width=\columnwidth]{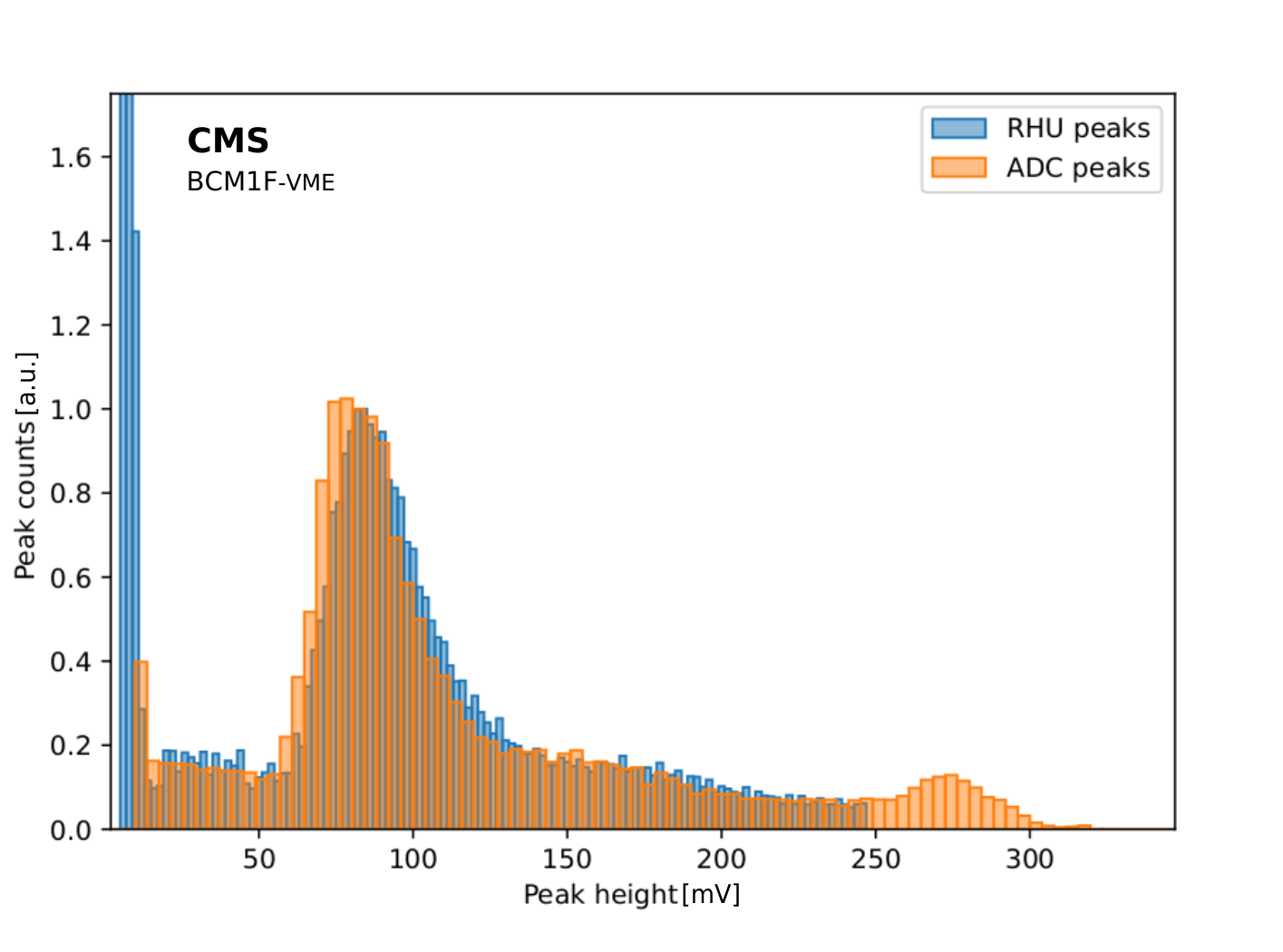}
    \caption{Peak height distribution in a typical BCM1F channel showing the approximate agreement of data taken using the ADCs and RHUs. RHU data was obtained by performing a threshold scan from 6\,mV to 250\,mV, using a step size of 2\,mV. Both datasets were recorded simultaneously, summing $2^{15}$ LHC orbits with 1200 colliding bunches in circulation~\cite{Jonas_thesis}}
    \label{fig:adc_rhu_peaks}
\end{figure}

Using accumulated ADC data and aggregating all measured signal amplitudes, per-channel amplitude spectra and baseline spectra are obtained and used for monitoring. The amplitude spectra are checked to ensure good separation between the noise and the signal peak. 
An example of an  ADC amplitude spectrum is shown in Fig.~\ref{fig:adc_rhu_peaks} (orange histogram). The data shows the noise level of the sensors at low signal amplitude, followed by the pronounced single particle hit signal peak at around 80\,mV. The precise location and proximity to the noise level varies channel-by-channel. 

The VME ADC baseline spectra are checked for stability, to ensure that the baseline fluctuation remains below 4-5\,mV.


To verify that the discriminators and RHU boards are functioning as intended, a scan of discriminator thresholds is performed and compared to the spectra observed by the ADC. During the threshold scan, counts are recorded for discriminator threshold voltages from 6\,mV to 250\,mV in 2\,mV steps. Differentiating the resulting distribution yields the number of counts as a function of signal height (Fig.~\ref{fig:adc_rhu_peaks}, blue histogram), analogous to what is recorded by the VME ADC. 
The relevant ADC data must be taken during a time of stable beams, ideally with at least 400 colliding bunches and for the duration of a few minutes. 


Over time, sensor performance degrades due to radiation damage, shifting the signal peak to lower values. Continuous monitoring is therefore necessary. To compensate for the loss of signal amplitude, higher HV is applied to maintain full depletion when possible.

\begin{figure}
    \centering
    \includegraphics[width=\columnwidth]{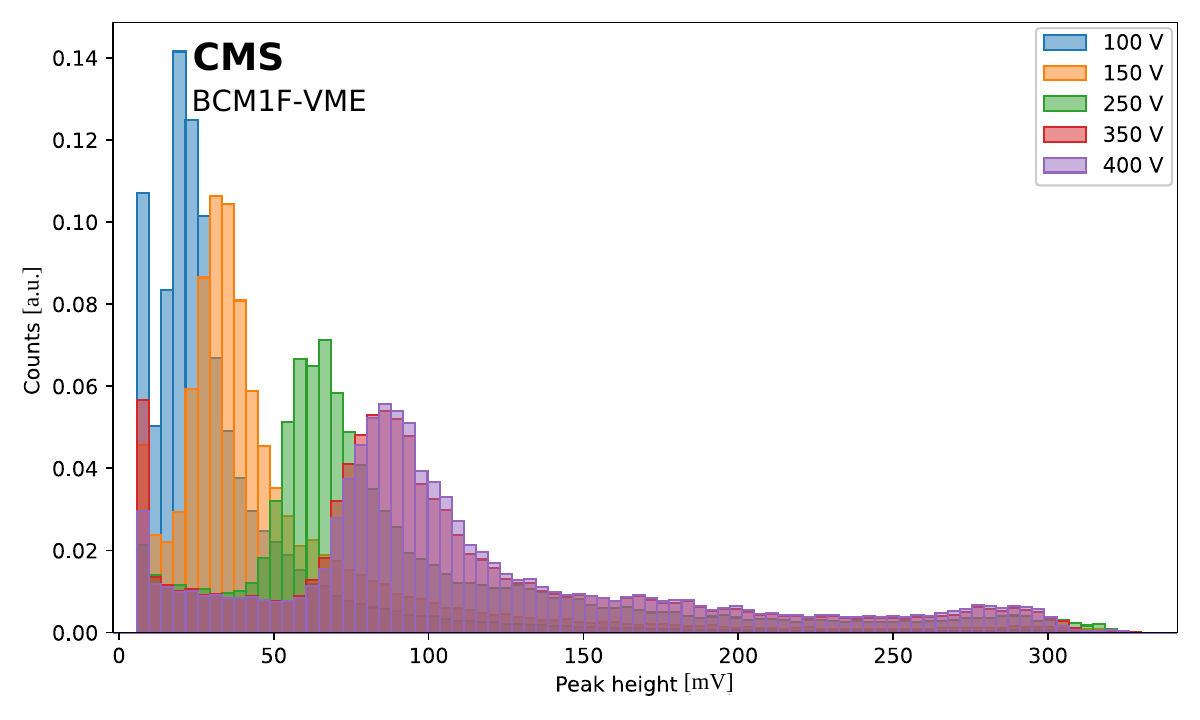}
    \caption{The signal peak height spectrum taken during stable beams with VME ADC back end at different HV settings, showing the dependence of the most probable value on bias voltage after the delivery of $\approx~30$\,\fbinv of integrated luminosity~\cite{Jonas_thesis} 
    }
    \label{fig:hv_adc}
\end{figure}

The VME ADC back end records raw data for about one minute at the beginning of each fill. Processing includes baseline subtraction, histogramming of pulse heights, and a Landau fit to the amplitude spectrum. Histograms are produced separately for particle-induced hits and test pulses during the abort gap. In particular, the amplitude histogram from particle hits serves as a primary diagnostics of the VME back-end performance. An example of the amplitude histogram for several HV settings is shown in Fig.~\ref{fig:hv_adc}, recorded during 2022 data taking after the delivery of about $ 30$\fbinv of integrated luminosity. In the case that the single particle signal peak begins to overlap with the noise level, as seen in the blue histogram at 100\,V bias, the HV must be increased. 

As noted in Sec.~\ref{sec:ivcv}, the HV should be set above the full-depletion voltage of the sensor diodes, identified during pre-installation tests to be below 300\,V. During the initial operation in 2022, no significant gain was seen from increasing the HV above 300\,V. However, as seen in Fig.~\ref{fig:hv_adc}, an HV scan already showed a relative gain in signal strength up to 350\,V following the delivery of $\approx 30\,$fb$^{-1}$. Performing such a scan requires interrupting BCM1F data-taking during stable beams and is performed only every few months, when monitoring data indicate it is required.

The rate at which signal MPVs drift as a function of integrated luminosity is observed to vary widely by channel. Generally, the MPV has been observed to shift by between $0.1$ and 1\,mV per 1\,fb$^{-1}$ of luminosity delivered (equating to roughly 2\,kGy absorbed dose at the sensor location). When an individual channel reaches the point that the discriminator is cutting away $\gtrsim 5\%$ of the single particle hit signal peak, it becomes necessary to either mask the channel or adjust the threshold. The ADC amplitude histogram is used to assess this drift and select proper discriminator thresholds. Good agreement between the RHU and ADC data processing chains is observed (Fig.~\ref{fig:adc_rhu_peaks}). 

\subsection{\utca back-end commissioning and performance monitoring during operations} \label{sec:uTCA_commissioning}

\begin{figure}[h!]
\begin{center}
\includegraphics[width=\columnwidth]{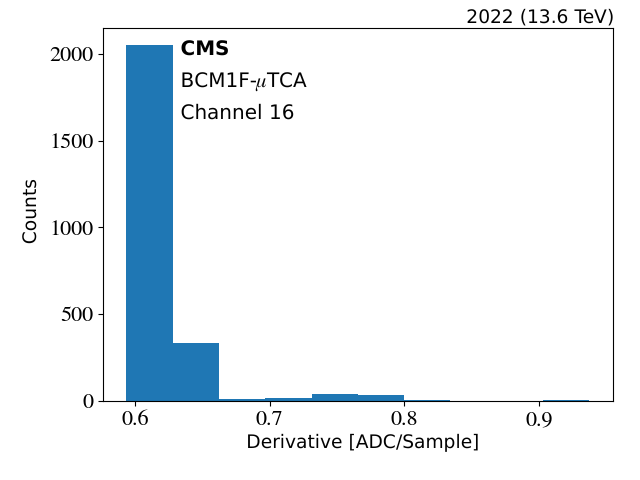}

\includegraphics[width=\columnwidth]{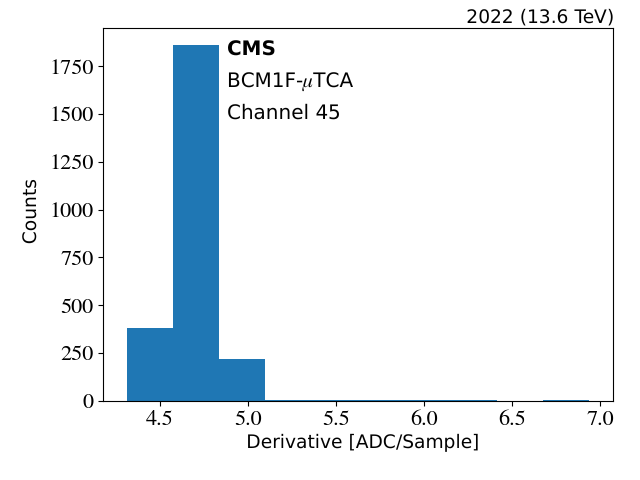} 
  \end{center}
  \caption{
  Distribution of the maximum signal derivative per orbit in the absence of beams is shown for a good BCM1F-\utca channel with stable baseline (top) and a channel with a noisy baseline (bottom) ~\cite{Joanna_thesis}  
  }\label{fig:baseline}
\end{figure}

The \utca back end relies on the firmware peak-finder algorithm, which requires per-channel optimization, as described in Sec.~\ref{sec:peakFinder}. During early commissioning, half of the channels used a simple fixed amplitude threshold, while the remaining channels used a derivative-based threshold, to allow for cross-checks of measured counts with the VME system.
Initial settings were determined with a ModelSim test-bench~\cite{modelsim}, sweeping the derivative (DdT) and amplitude (DaT) thresholds using test pulses and raw data. 
The \utca back-end system enables noise and amplitude monitoring in a single electronics card.  With the use of AC-coupled ADCs, the \utca readout is less sensitive to slow temperature-dependent AOH baseline changes.
The behaviour of the signal baseline in the back end provides insight into the stability of the entire electronic signal path. The signal derivative is shown for two example channels in Fig.~\ref{fig:baseline}: one with a low value (around 0.6 ADC counts per sample) and stable derivative (left), indicating good performance, and another with a steeper rise (around 4.7 ADC counts per sample) and more variable gradient (right), including significant noise contamination.
The maximum detected value, measured in the absence of particle hits, is used as a reference for the minimum derivative threshold. 


\begin{figure}[!htb]
   \centering
   \includegraphics*[width=.9\columnwidth]{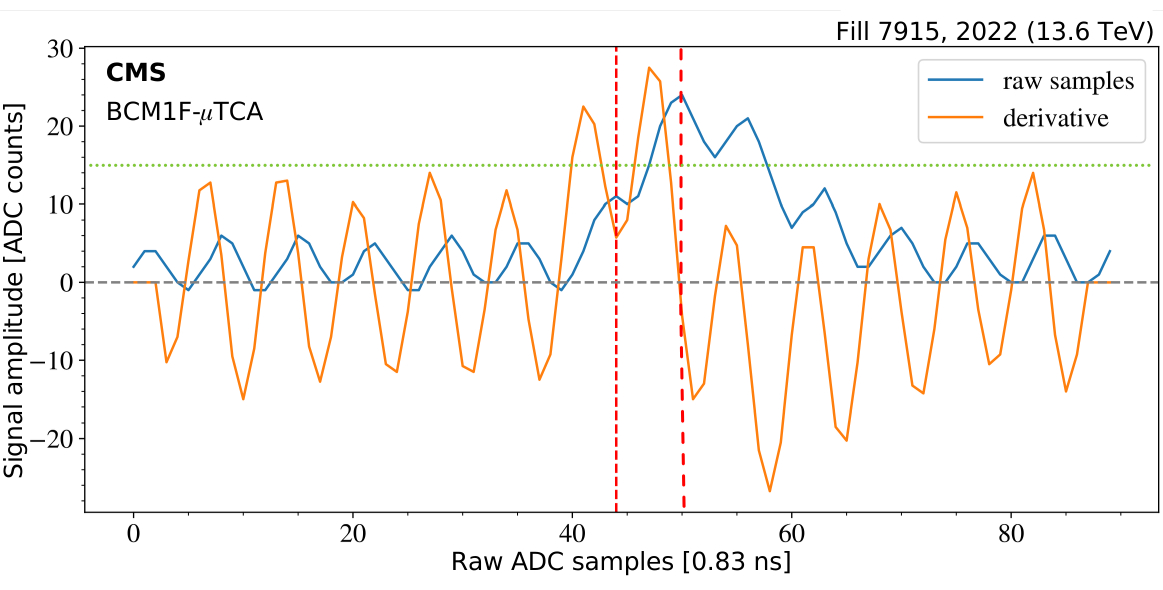}
   \caption{Example of a noisy channel in the presence of a hit: a double pulse is detected (indicated with vertical red dashed lines) corresponding to derivative values exceeding the threshold (horizontal green dashed line)~\cite{Joanna_thesis}}
   \label{fig:doublePulse}
\end{figure}

An example of a good channel with a hit was shown in Fig.~\ref{fig:raw_utca}, where the baseline derivative oscillates around 0. 
An example of derivative threshold (green line) tuning for a problematic channel is shown in Fig.~\ref{fig:doublePulse}. In this case, the raw pulse (blue curve) is distorted by extreme baseline noise, which the peak finder algorithm mistakenly interprets as two stacked pulses.
In this case, the derivative threshold can be set to a higher value to suppress the noise.
Nevertheless, baseline distortion can cause the threshold to be crossed twice, leading to the recognition of two pulses (red vertical dashed lines).
Such extreme noise results in hit rate over counting and the affected channel must be excluded. 
During the Run 3 operations, 6 to 12 channels were affected, depending on the period. 

\begin{figure}[!htb]
   \centering
   \includegraphics*[width=\columnwidth]{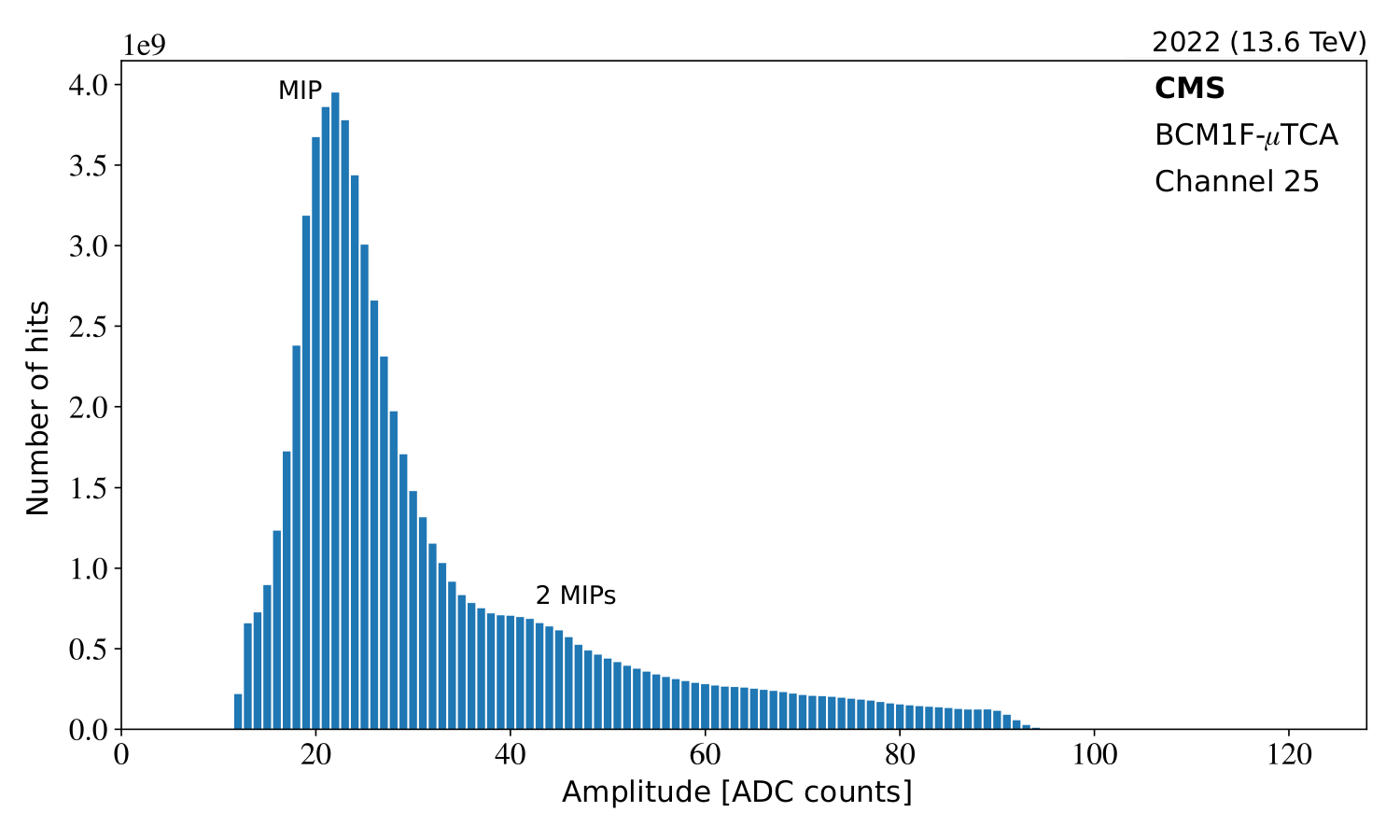}
   \caption{BCM1F-\utca per-channel amplitude spectrum measured for high-energy $pp$ collisions (as calculated online in the FPGA). The distribution is cut at the amplitude threshold of 11 ADC counts~\cite{Joanna_thesis} 
   }
   \label{fig:amplitude_spectra}
\end{figure}

Amplitude thresholds were then tuned using each sensor’s amplitude spectrum to optimize signal-noise separation with colliding beam data. 
Figure~\ref{fig:amplitude_spectra} shows an example of a \utca amplitude spectrum for a well-performing BCM1F channel during 13.6\,TeV pp-collisions, reconstructed using the derivative peak finder. Low-amplitude noise was suppressed through operational thresholds. The spectrum follows a Landau distribution, with its most probable value (MPV) corresponding approximately to the energy loss associated with a single relativistic particle hit in the silicon sensor. Owing to the 10-hour data collection period, a charge deposition peak corresponding to two overlapping particle hits is also visible.

The timing of the full system is crucial for the performance. The main timing differences between BCM1F-\utca channels arise from the separate readout paths of varying lengths on the two ends of CMS. 
To ensure proper synchronization of all channels, timing adjustments are configured in the BCM1F-\utca readout software. Three types of delay relative to the LHC orbit timing are available and implemented per GLIB (serving four channels): in steps of 25\,ns (bunch-crossing granularity); in steps of 3.33\,ns (4 samples) and in steps of 0.83\,ns (single sample delay).

During Run~3, the first circulating beams in the LHC were used to tune the timing for all channels. The resulting configuration was then validated using the first single-bunch collisions, confirming that 90–95\% of the signal was concentrated within a single histogram bin (corresponding to 4.17\,ns).

The amplitude analysis of the BCM1F-\utca\ back end is implemented within the online analysis to provide key performance metrics in real time: the MPV of the test pulse signal in periods with no beams present, the MPV of the amplitude spectrum during collisions, the total hit count during collisions for comparison of per-sensor efficiencies, and their evolution over time.

\begin{figure}[!ht]
\centering
\includegraphics[width=\columnwidth]{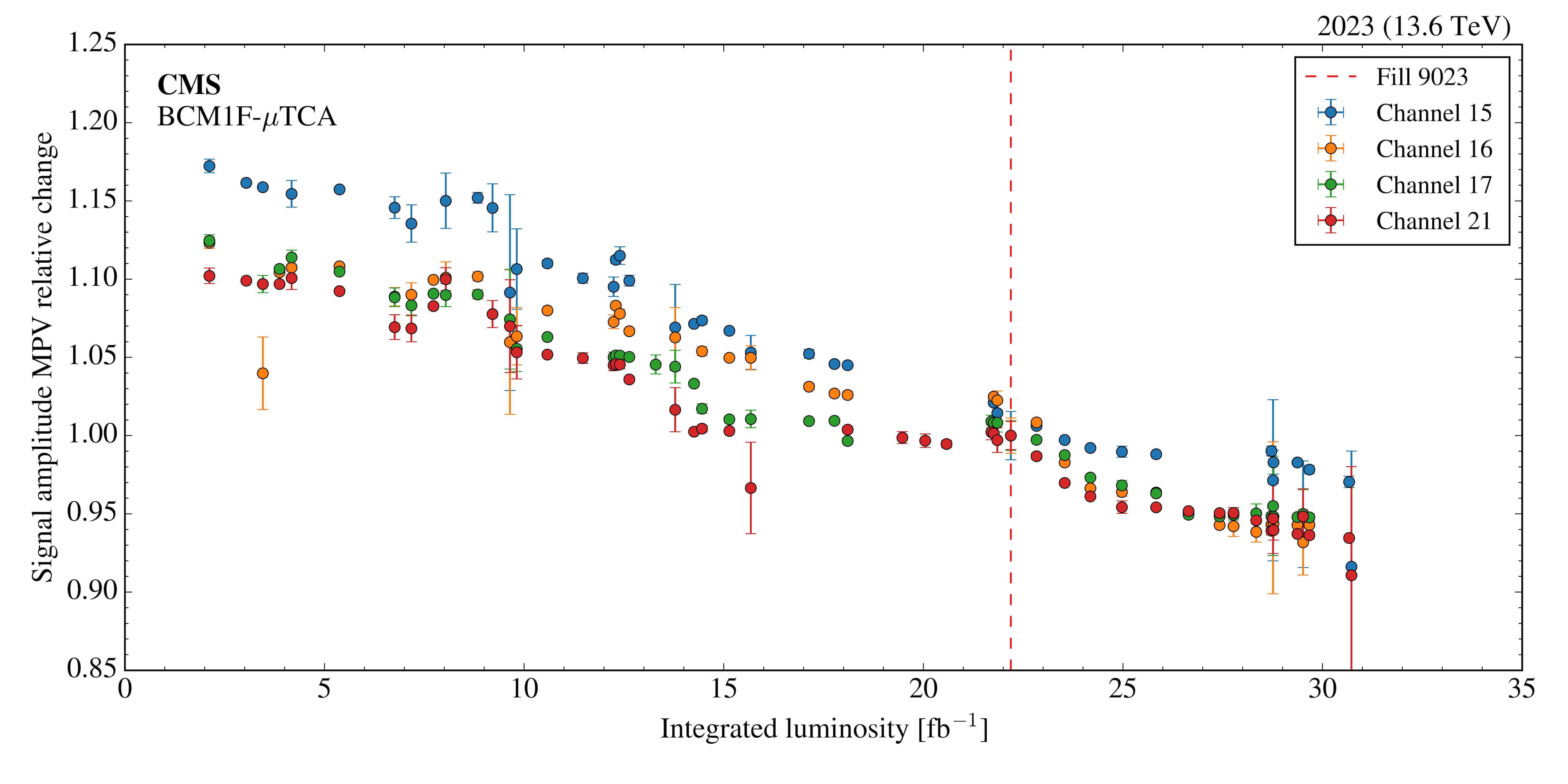}
\includegraphics[width=1.0\columnwidth]
{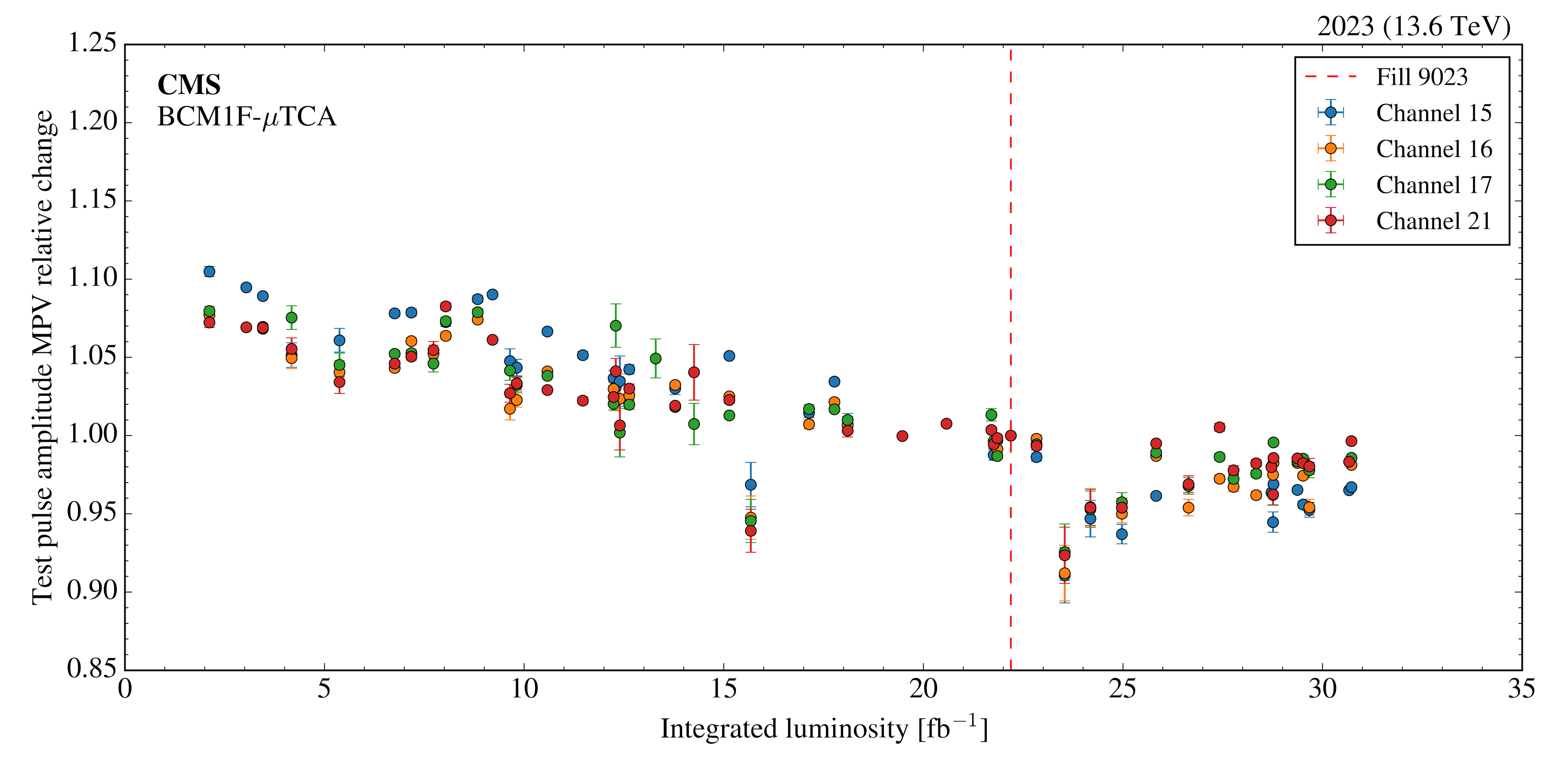}
\caption{%
    The relative change in the MPV of the signal amplitude spectra (top) and the test pulse spectra (bottom) for four BCM1F-\utca channels during 2023 data-taking. 
    The values are normalized to the MPV measured during the absolute luminosity calibration period, indicated by the red dashed line
    }
\label{fig:bcm1f:mpv-change}
\end{figure}

\begin{figure}[!ht]
\centering
\includegraphics[width=\columnwidth]{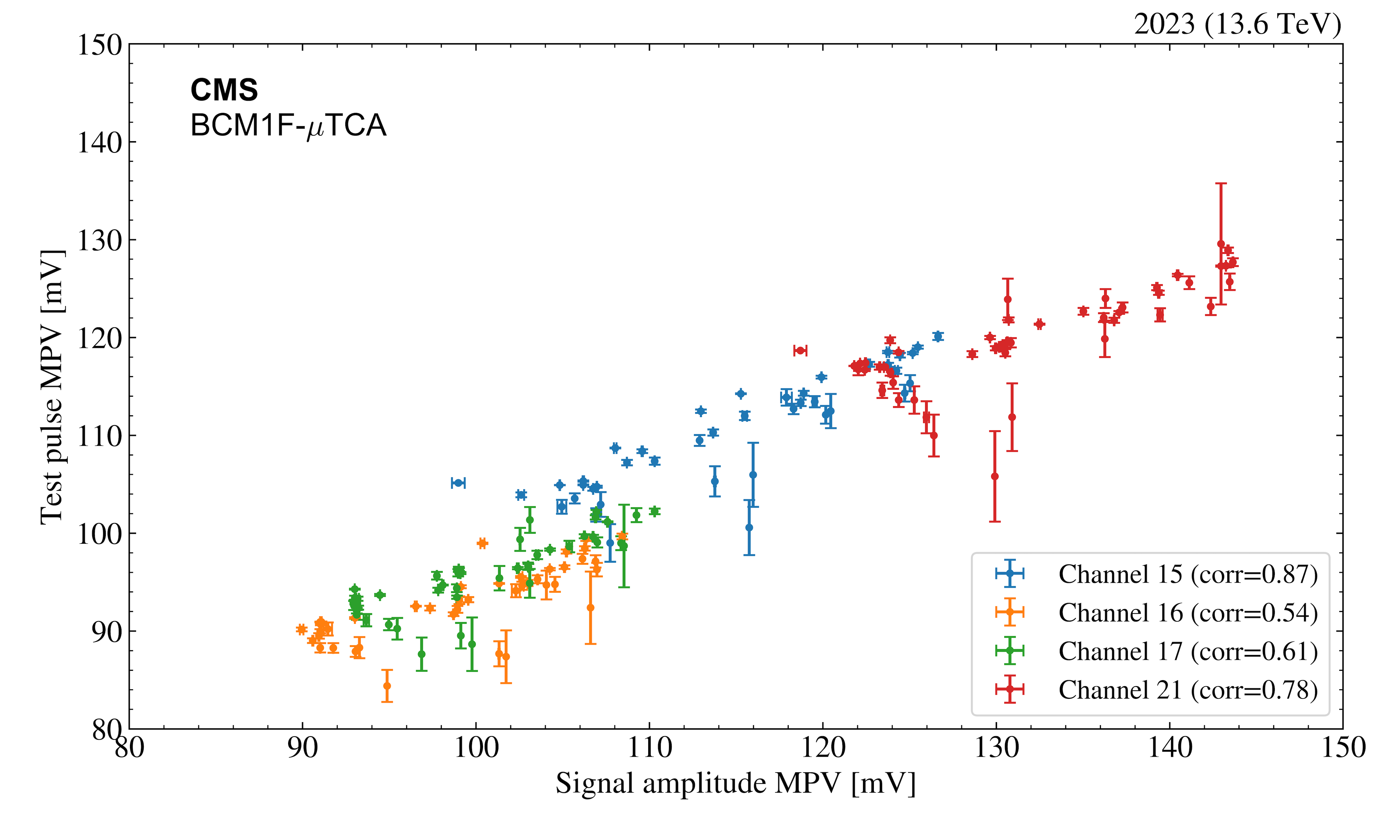}
\caption{%
    Correlation of the MPV of the signal amplitude spectra vs. the MPV of the test pulse spectra for four BCM1F-\utca channels during 2023 data-taking. The correlation (corr) per channel is given in the parentheses  
    }
\label{fig:bcm1f-TP_MPV-correlation}
\end{figure}

The shift of the MPVs of the signal amplitude and the test pulse with respect to the reference values measured close to the van der Meer calibration period (Sec.~\ref{sec:vdm_scans}) 
as a function of the delivered integrated luminosity is shown in 
Fig.~\ref{fig:bcm1f:mpv-change} for four BCM1F-\utca channels during the 2023 proton-proton data taking. The efficiency changes were not uniform across channels. These MPV values serve for end of the year reprocessing as additional information to efficiency degradation measured from emittance scans (Sec.~\ref{sec:emit_scans}). The signal amplitude and test pulse MPV values differ per channel, as shown on the correlation plot in Fig.~\ref{fig:bcm1f-TP_MPV-correlation}. Therefore, the threshold settings for each channel are defined separately, and their data analysed individually. 

\subsection{Channel stability and selection for end of year reprocessing} \label{sec:channel_selection}

\begin{figure}[!ht]
\centering
\includegraphics[width=\columnwidth]{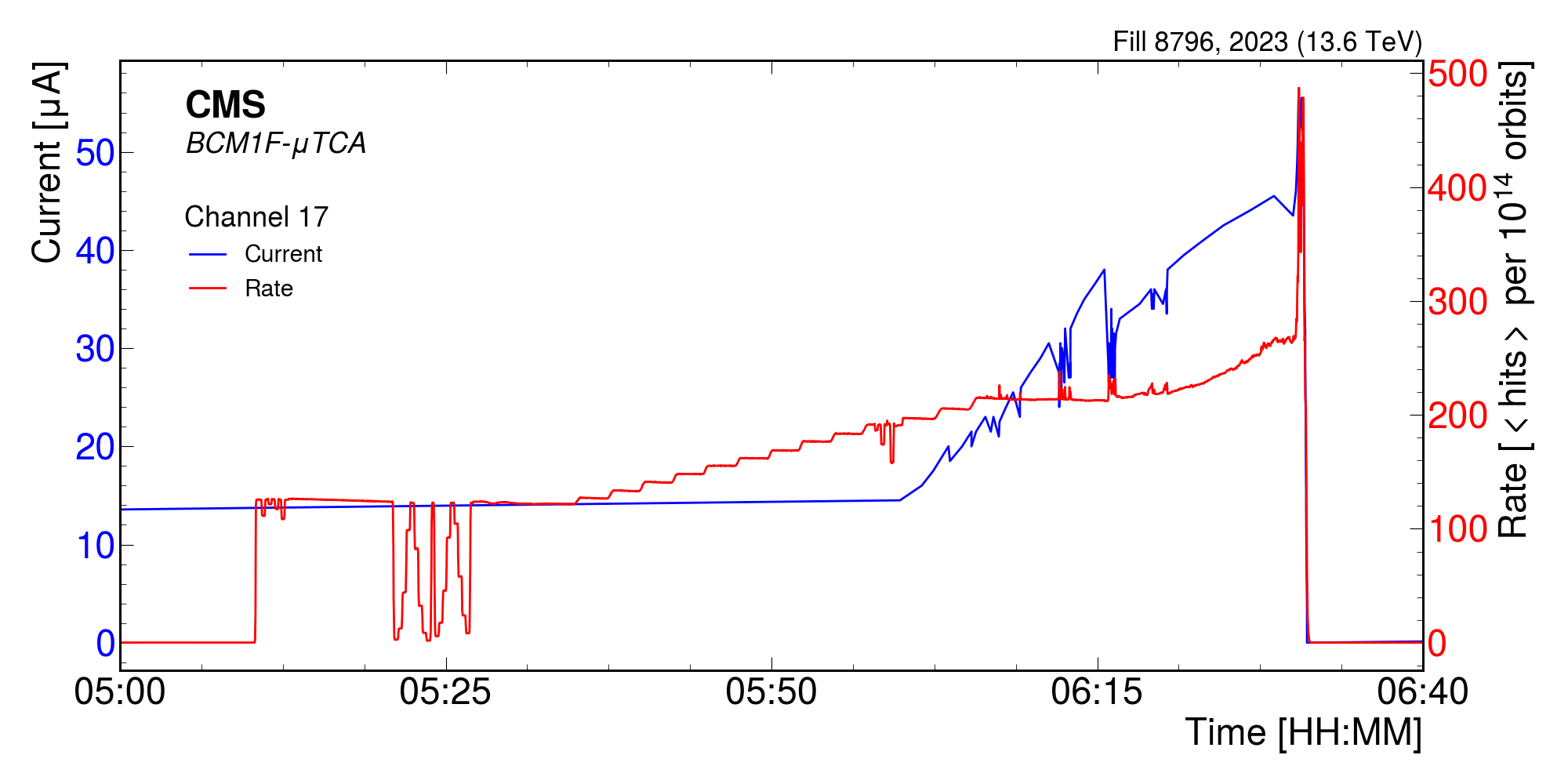}
\caption{%
    BCM1F-\utca channel 17 leakage current and rate as a function of time. A leakage current increase is observed after 06:00. A clear correlation between the spikes in the leakage current and the rate is visible at 6:12 and later. The safety current limit was reached around 06:22, and the channel was automatically switched off.  
    }
\label{fig:ch_17_trip_example}
\end{figure}

During the 2022 commissioning, several channels exhibited anomalous behaviour. It manifested as short-term fluctuations or else an abnormal increase in channel hit rate, driven by an unexpected increase in leakage current, as shown in Fig.~\ref{fig:ch_17_trip_example}. In most cases, leakage current increased until the safety current limit was reached, at which point the channel was switched off. This effect appeared after the first half-year of operation in 2022, remaining present until the end of Run 3 operation. After the increase in leakage current was observed, the channel was left off for at least 24 hours and then switched on again. No correlation between sensor position or production batch was found to explain this behaviour. Detailed sensor simulations and laboratory studies are presented in Sec.~\ref{sec:surface_charge}.

At the end of every year during Run 3 operations, per-channel BCM1F data reprocessing is performed, and all periods of instability due to detector issues or maintenance, such as HV scans, are excluded. 
The reprocessing of raw detector data indicated various causes for channel exclusion, such as intrinsic sensor quality, lower performance of reused optical and electrical lines from Run 1, and possible damage during transport or installation. Some datasets were corrected, while a few highly inefficient channels were completely excluded. For example, in the 2022 dataset, the 32 best-performing of 48 channels were retained for the end-of-year precision integrated luminosity measurement using both VME and \utca back ends. 

By the end of 2024, large integrated dose had affected detector performance to the point where less than half of the total channels were envisioned for use in the end-of-year precision re-analysis of the data. Enough spare parts were available to replace half of the detector in January 2025, with the 24 non-irradiated sensors helping ensure good performance of the detector through the end of LHC Run 3. Combined with results from other luminometers, these measurements are essential to provide the precision integrated luminosity for CMS Run 3.

\section{BCM1F sensor surface charge simulation and test under X-ray}
\label{sec:surface_charge}

To investigate the growth of leakage current which causes stability issues during BCM1F detector operation (described in Sec.~\ref{sec:channel_selection}), a series of simulation and laboratory studies were performed.
This section describes attempts to understand the issue via replication in simulated data samples (Sec.~\ref{sec:sensor_simulation}), as well as leakage current measurements performed under an X-ray lamp (Sec.~\ref{sec:x_ray_test}).

\subsection{Simulation results} \label{sec:sensor_simulation}

\begin{figure}[h]
\centering
\includegraphics[width=\columnwidth]{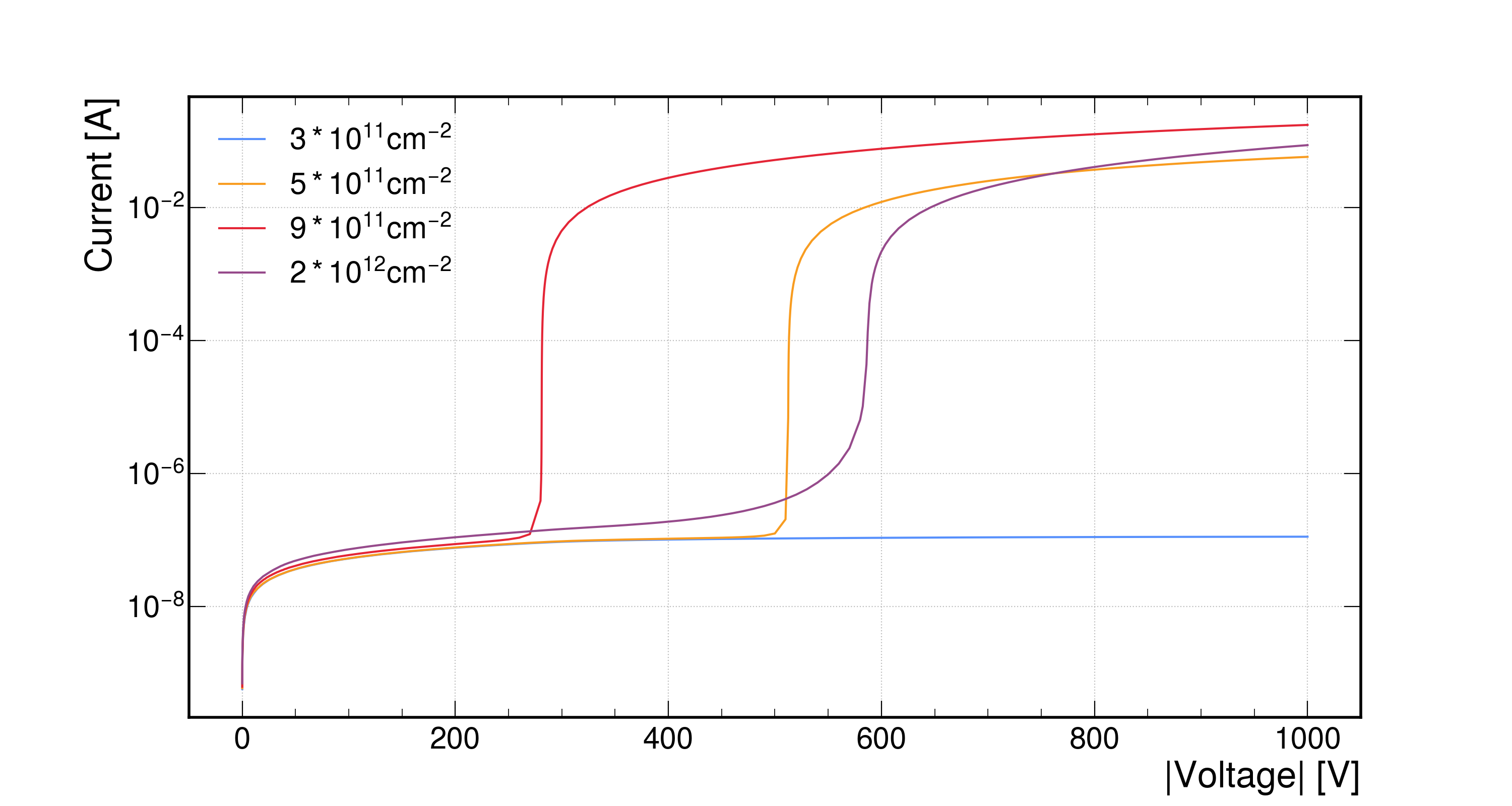}
\caption{Simulation of the sensor leakage current as function of bias voltage for different levels of surface charge in the transition area between silicon oxide and silicon }
\label{fig:TCAD2}
\end{figure}

Permanent irradiation creates stationary, predominantly positively charged defects in the silicon oxide layer~\cite{Hartmann}. To investigate the resulting phenomenological consequences for the leakage current, a 2D finite element simulation of the diode was created using the Sentaurus workbench~\cite{sentaurus_workbench}. In the simulation the defects in the oxide were simulated using a positive surface charge in the oxide-silicon boundary layer. This showed that, if the surface charge is too high, there is a sudden increase in the leakage current at the guard ring, as shown in Fig.~\ref{fig:TCAD2}. The positive surface charges lead to the formation of a conductive electron accumulation layer below the oxide layer, which short-circuits the edge implant and the guard ring. As the guard ring is grounded and the edge implant is at the same potential level as the backplane, the full potential difference drops over a short distance behind the edge implant. This creates a locally very strong electric field, which was shown in simulation to exceed the breakdown field of silicon ($2\,\times\,10^5\,$V/cm), and is thus capable of releasing large numbers of free charge carriers. As the oxide charge increases, the voltage at which the high leakage current occurs decreases  (Fig.~\ref{fig:TCAD2}, red line vs. yellow line). Eventually, the surface charge is high enough to displace the free charge carriers inside the edge implant and deplete it, leading to a reduced leakage current and an increase of the HV at which the breakdown sets in (violet line vs. red line).

\subsection{Sensor testing under X-Ray }
\label{sec:x_ray_test}


To validate the surface charge build up shown in the simulation in Sec.~\ref{sec:sensor_simulation}, laboratory tests were conducted using an X-ray source~\cite{X-ray} and a test-board with three irradiated BCM1F sensors. Unlike charged particles which penetrate the sensor bulk during LHC operation, X-ray irradiation generates only surface charges, making them suitable for studying the surface charge effect on the leakage current. The intensity of the X-ray tube was adjustable, allowing selection of the working point. For these measurements, the X-ray current was set to 2\,mA, resulting in sensor leakage current of about 16\,$\mu$A, comparable to what is observed during LHC operation.


\begin{figure}[ht!]
\centering
\includegraphics[width=\columnwidth]{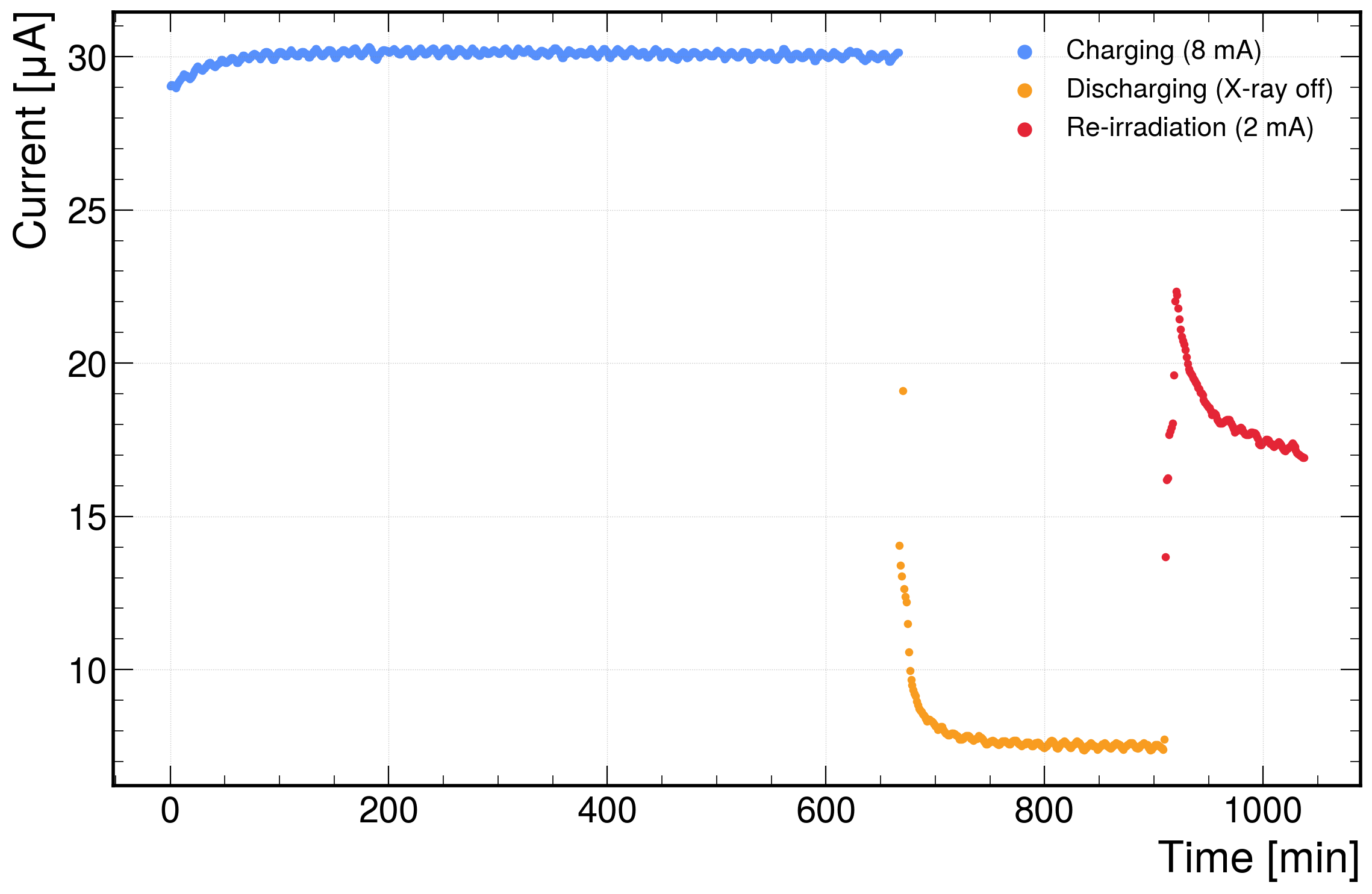}
\caption{ BCM1F leakage current under 8\,mA X-ray (blue), after turning off the X-ray (orange), and following re-irradiation at 2\,mA X-ray (red). There is a clear charging and discharging pattern visible during the first hour in blue, and in the orange and red data points (on a shorter timescale of minutes) correlating with the X-ray being turned on and off }
\label{fig:x_ray_charging}
\end{figure}

Extended overnight irradiation at a higher X-ray dose of 8\,mA showed current growth during the first hour, as shown in Fig.~\ref{fig:x_ray_charging}, consistent with the prediction of surface charge accumulation. When the X-ray source was switched off (at around 700\,min), a distinct discharge pattern was visible. Upon reactivation of the X-ray at 2\,mA (around 900\,min), the current initially reached the expected operation level, before rapidly growing over a ten-minute interval, exceeding the expected current levels established during a long calibration run. 


\begin{figure}[ht!]
\centering
\includegraphics[width=\columnwidth]{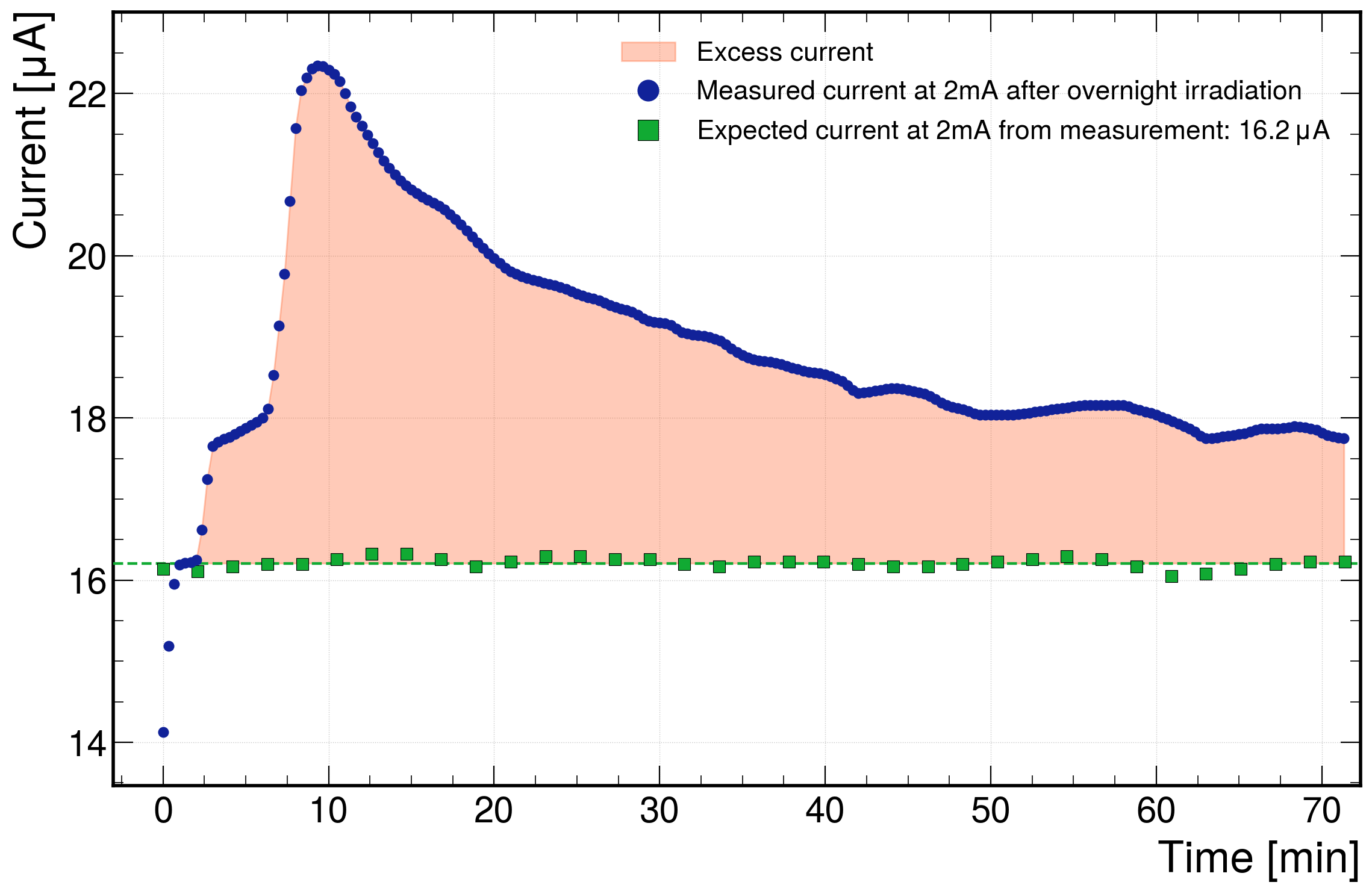}
\caption{ BCM1F leakage current (blue dots) at 2\,mA X-ray tube current after overnight irradiation at 8\,mA X-ray tube current, compared to the expected sensor current at 2\,mA (green squares) based on pre-irradiation data. The current is seen to reach the expected value before growing well beyond this value over the course of 10 minutes, and afterwards slowly discharging }
\label{fig:x_ray_growth}
\end{figure}

Fig.~\ref{fig:x_ray_growth} shows this growth in greater detail, with the sensor current after re-irradiation at 2\,mA (blue dots) exceeding the expected pre-irradiation sensor current from data (green squares) taken over a long calibration run of several hours. This current then discharges over the course of several hours, consistent with what is observed during LHC operation, where leakage current has also been seen to discharge in cases where the trip threshold is not reached.


\section{Data processing}
\label{sec:data_reprocessing}
The BCM1F data processing receives occupancy histograms from the BCM1F VME and \utca back ends, accumulated in the firmware over  2$^{12}$ and 2$^{14}$ LHC orbits respectively, synchronised to the Luminosity Nibble (NB) broadcast by the TCDS system as described in Sec.~\ref{sec:backend_and_opt}. The processing software synchronizes and aggregates histograms over time intervals of 4\,NB before broadcasting. 

\begin{figure}
    \centering
    \includegraphics[width=\columnwidth]{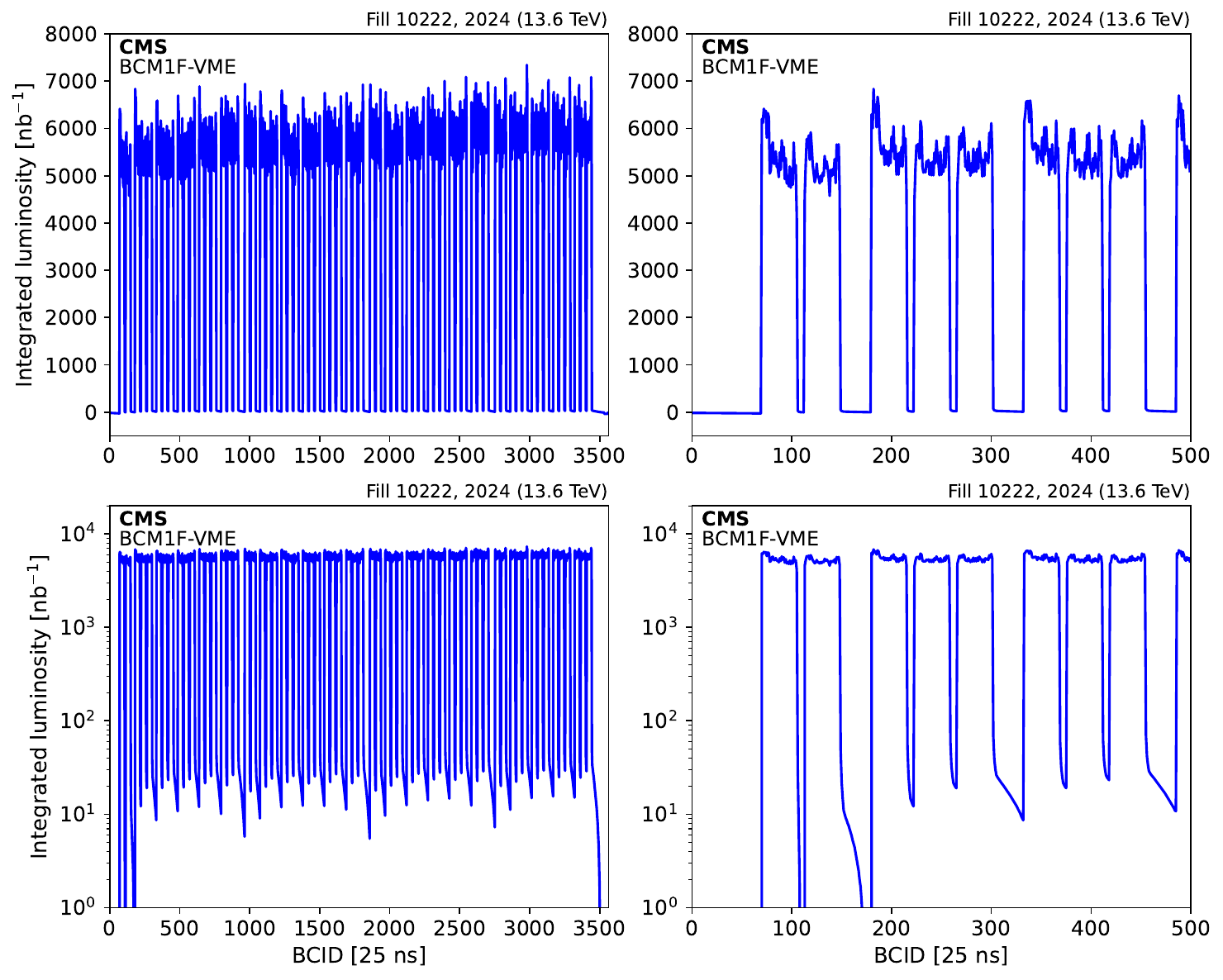}
    \caption{Example of the RHU luminosity histogram as a function of BCID in linear (top) and logarithmic (bottom) scales. The full LHC orbit is shown on the left, and a zoom into first 500 bunches on the right. Data was recorded during LHC fill 10222}
    \label{fig:bbb_histo}
\end{figure}

During Run 3, the LHC typically operates with up to 2400 bunches, although this depends on the chosen filling scheme. Bunch trains are separated by unfilled bunch slots to allow injection kicker ramping.  For beam extraction, a  3\,$\mu$s interval (the last 120 BXs of the orbit), contains no bunches  and is referred to as the \textit{abort gap}. Figure~\ref{fig:bbb_histo} shows examples of aggregated BCM1F hit rates converted to integrated luminosity (Sec.~\ref{sec:luminosity}) as a function of the BCID, using data from the RHU in the VME back-end system, illustrating this beam structure. The rate clearly visible between the trains from the so-called \textit{afterglow} hits is caused by imperfect timing, short term activation of material surrounding the detector, and slow neutrons scattering inside the experiment~\cite{Muller}. These hits are only visible in a logarithmic scale, where a lingering but slowly decaying background count rate tail is visible in each gap between bunch trains. As discussed in Sec.~\ref{sec:albedo}, this also affects the hit rate measured for colliding BCIDs in the bunch trains and leads to an over-counting which requires correction for both the BIB measurement (Sec.~\ref{sec:BIB}) and the luminosity measurement (Sec.~\ref{sec:luminosity}). 

\subsection{Afterglow correction}\label{sec:albedo}

The afterglow contributions are most clearly observed in data from a single colliding bunch pair. 
Afterglow signal arises from multiple sources which arrive at different times, and must be corrected to ensure reliable bunch-by-bunch luminosity reporting as well as stability with respect to the bunch train pattern. Short term afterglow ($<$100\,ns) can come from particles which either scatter off other detector components or loop in the CMS magnetic field, as well as from intrinsic detector timing misalignment or electronic time walk. A longer decay tail of afterglow hits on the timescale of $\mu$s is caused by activation of various materials in the environment and subsequent emissions.

\begin{figure}[ht]
    \centering
    \includegraphics[width=\columnwidth]{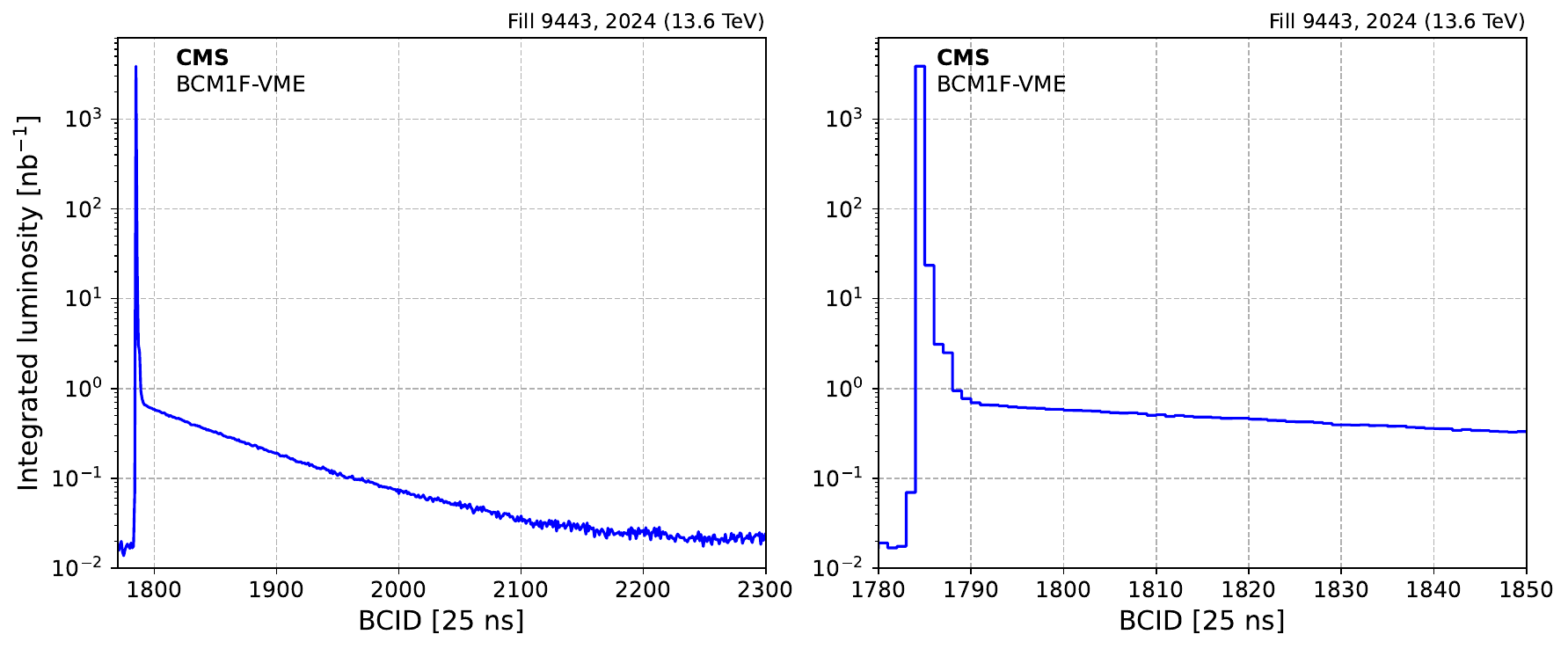}
    \caption{Single bunch luminosity histogram aggregated over one LHC fill. Data recorded by the RHU back end during fill 9443.}
    \label{fig:single_bunch_histo}
\end{figure}

Figure~\ref{fig:single_bunch_histo} measures these late albedo hits using luminosity data from LHC fill 9443, which contained only three maximally separated colliding bunch pairs. The peak in Fig.~\ref{fig:single_bunch_histo} is due to the in-time collision products, with the trailing albedo signal shown as a function of BCID. A very small increase above the noise level can also be seen in the preceding BCID due to out-of-time hits.

Due to their symmetric arrangement around the beam pipe at equal radius, all BCM1F sensors exist in a similar radiation environment, and therefore receive roughly the same contribution from most afterglow sources.

\begin{figure}
    \centering
    \includegraphics[width=1\columnwidth]{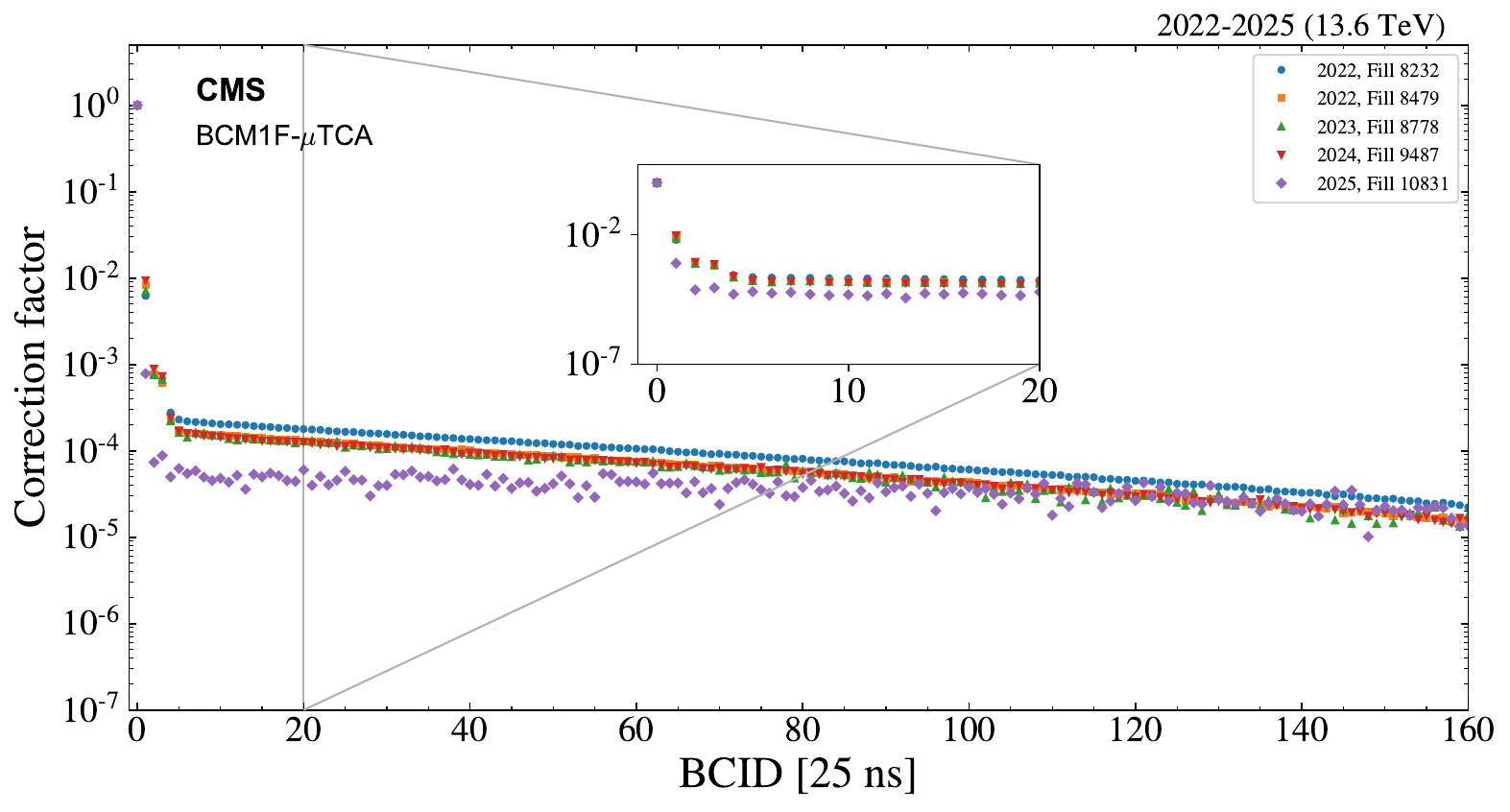}
    \caption{Run 3 afterglow correction model based on single colliding bunch data. BCM1F-\utca raw data was aggregated over many orbits in selected fills in the years 2022 to 2025 (with integration time between $\sim15$\,min and several hours in different data sets) and normalized to the collision counts such that the value for first BCID is set to 1. The resulting correction is applied to each per-bunch data set}
    \label{fig:utca_afterglow_2022_to_2025}
\end{figure}

To derive the afterglow correction, a single bunch detector response model is used. Dedicated fills with a single colliding bunch or alternatively fills with a minimal number of colliding bunches are selected for study. These must contain at least one sufficiently isolated bunch crossing to exclude contribution from the preceding afterglow tail to the model. 
The model is constructed by summing data over each fill (lasting several hours) and normalizing the resulting per-BCID histogram to the colliding bin signal. 
The resulting model entries represent the fractions of out-of-time afterglow hits in each bunch crossing after a collision. The model, applied to the \utca signal chain, and the data used to derive it are shown in Fig.~\ref{fig:utca_afterglow_2022_to_2025}.  
It has been measured that the afterglow correction can vary slightly from year to year and thus a dedicated single bunch fill is highly beneficial every year to derive an accurate afterglow model. 
A similar procedure is followed for the VME signal chain. 

The single bunch response model for afterglow is pileup-independent, and hence the amount of afterglow can be estimated by multiplying the afterglow model with the hits in the colliding bunch crossings.
Noise bias is removed by subtracting the noise level calculated from an empty orbit region where the afterglow had sufficiently decayed. In the fit, BCID locations are omitted where hits from beam-induced background due to the presence of unpaired bunches are expected.
The afterglow tail in the data can be separated into two distinct components: short-term ($<100$\,ns), which is extracted from averaged data, and the long-term decay tail ($\sim10\,\mu$s), modelled with an exponential function. 

The afterglow model describes the hit probability in the bunch crossings following a colliding bunch pair. It was validated across a broad single bunch instantaneous luminosity range ($2–8$\,Hz/$\mathrm{\mu}$b) and demonstrated stability over extended time periods ($\sim10$\,h in fill 7921). 

\begin{figure}
    \centering
    \includegraphics[width=\columnwidth]{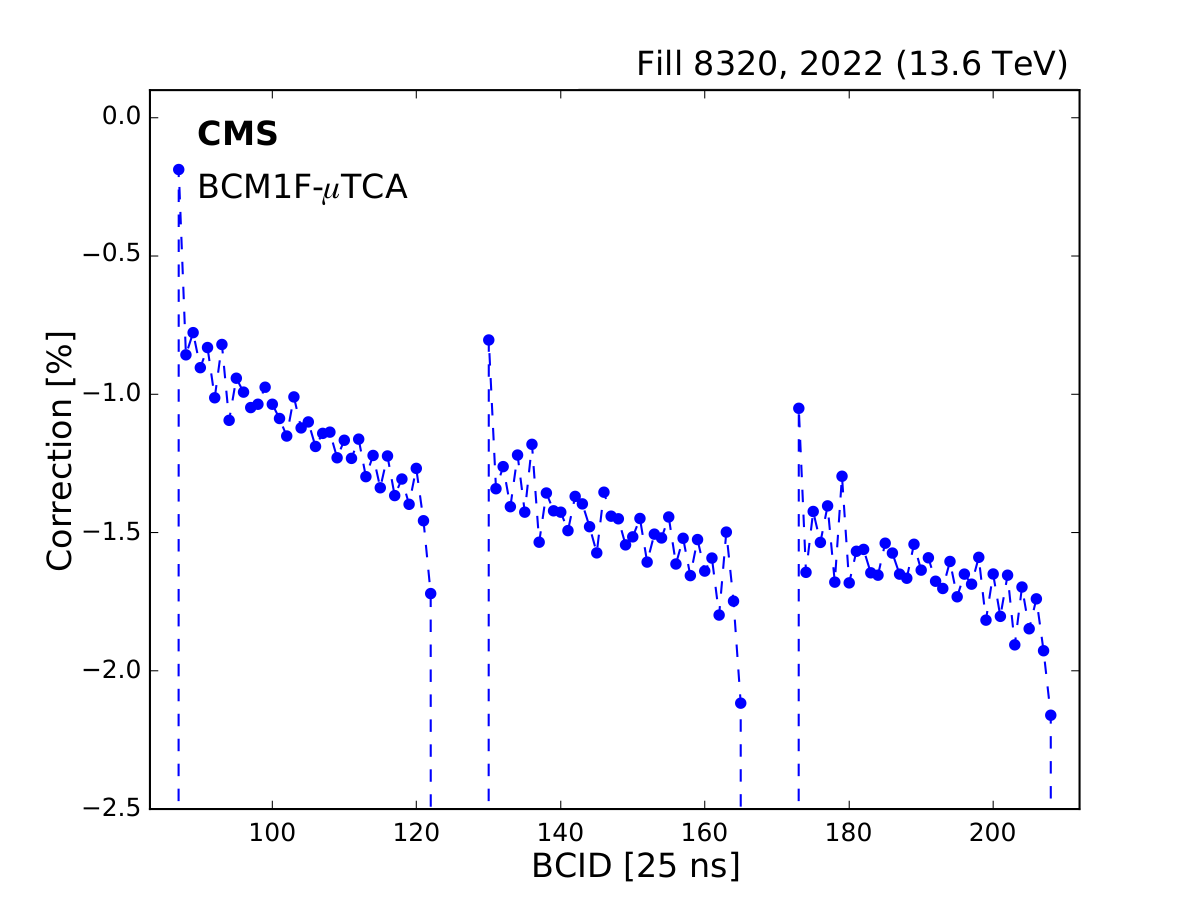}
    \caption{The BCM1F-\utca afterglow correction as a function of the position in a train for an example nominal physics fill (8320). The total correction is calculated by comparing the measured single bunch rate after applying the afterglow correction factor to all colliding BCIDs with reference to the non-corrected rates~\cite{Joanna_thesis}}
    \label{fig:albedo_vs_train}
\end{figure}

The short-term afterglow component is influenced by the sensitivity of each sensor, necessitating the reconstruction of the correction model for the specific subset of channels used in the final measurement. The timing misalignment of channels can result in a residual signal spillover. The correction algorithm subtracts the contamination of subsequent bunches based on the afterglow model, with the total relative correction depending on the train length and total number of bunches per orbit. An example of the relative correction dependence on the bunch position in the train is shown in Fig.~\ref{fig:albedo_vs_train}. The non-corrected data is used as a reference to calculate the correction percentage.

\begin{figure}
    \centering
    \includegraphics[width=\columnwidth]{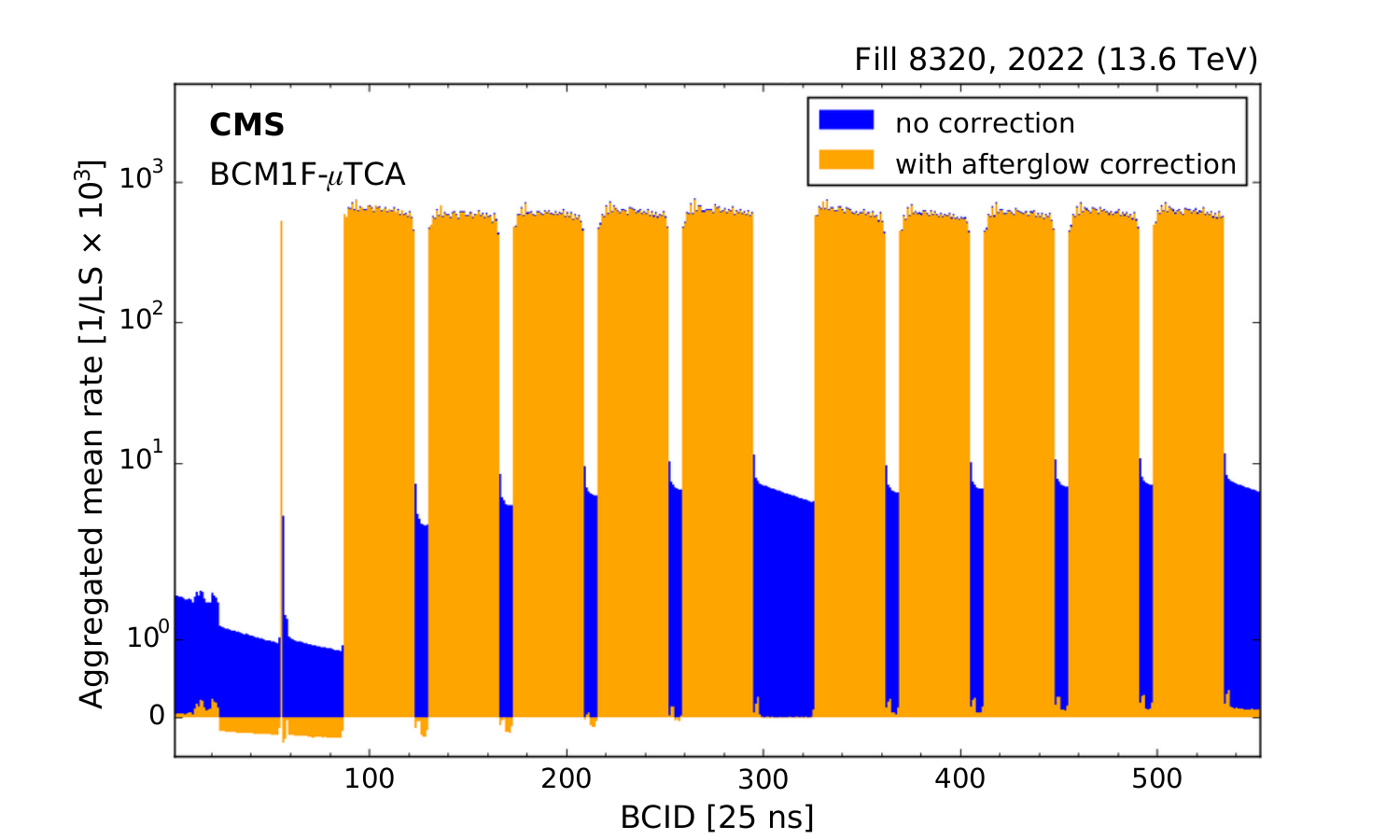}
    \caption{ The aggregated rates measured by BCM1F-\utca as a function of BCID in the LHC orbit. Rates are shown without (blue) and with the afterglow corrections (orange), aggregated over 200\,LS ($\approx$ 78\,min), normalized per LS and scaled by $10^3$~\cite{Joanna_thesis}}
    \label{fig:albedo_corrNocorr}
\end{figure}

An example of the per-BCID average rate histogram is shown in Fig.~\ref{fig:albedo_corrNocorr} for around 600 BCIDs from the start of the orbit  before and after the first step of the afterglow correction.
It is clear that this correction significantly reduces the afterglow contribution to the measured collision rate, as evidenced by near-zero rates after the correction in empty bunch slots between colliding trains. As a second step the baseline residual within the abort gap is calculated and used as an additional correction.

\begin{figure}
    \centering
    \includegraphics[width=\columnwidth]{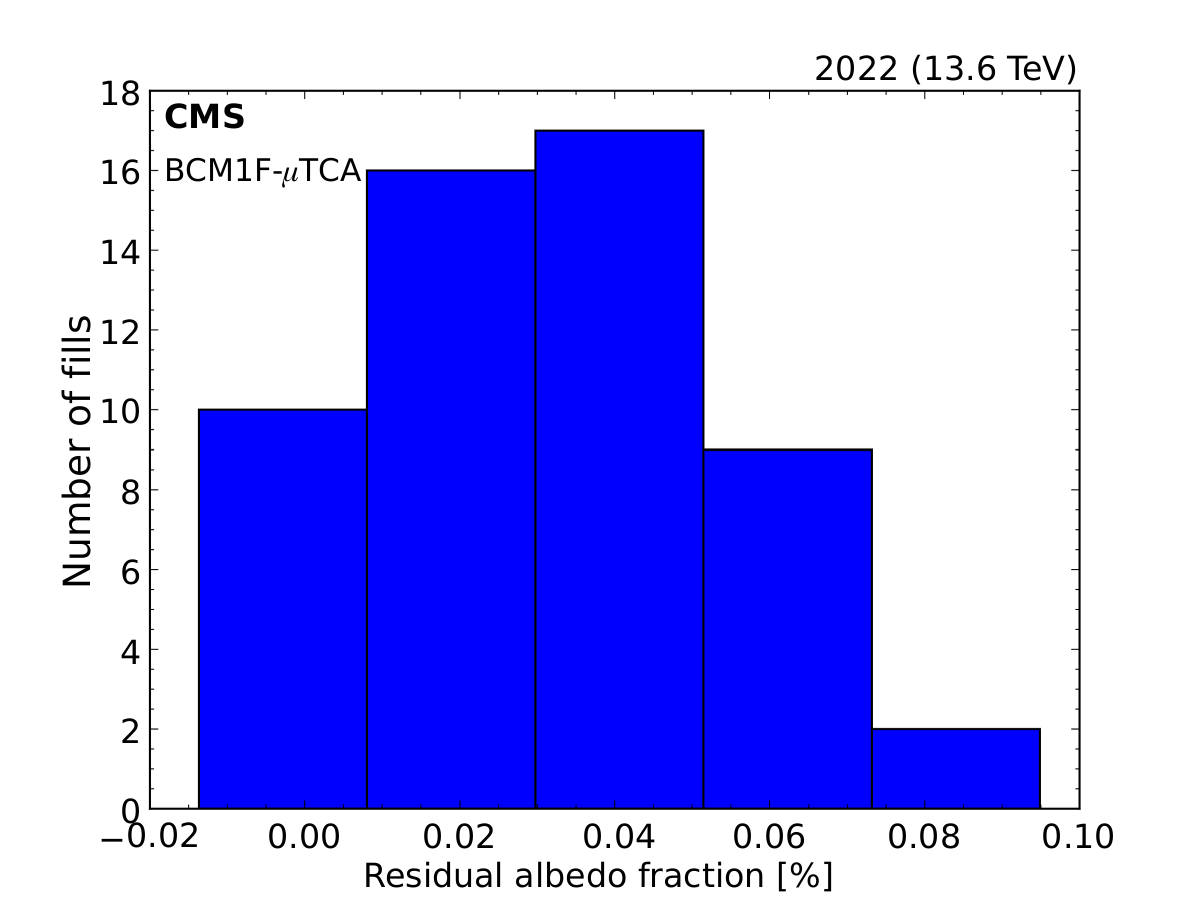}
    \caption{Afterglow correction residuals, calculated throughout 2022 fills, each entry includes 1\,h of integrated BCM1F-\utca rates~\cite{Joanna_thesis}}
    \label{fig:residual_hist}
\end{figure}

The afterglow correction applied is dependent on the train length and the total single-bunch instantaneous luminosity (SBIL), with values ranging from $-1\%$ to $-3\%$. Residuals of the applied correction were estimated through the analysis of remaining rate fractions in the bunch crossings subsequent to each train within the LHC orbit. An hour of integrated data was used to determine the average fractions per fill. The data for all year 2022 fills are histogrammed in Fig.~\ref{fig:residual_hist} after all corrections, indicating that the residual afterglow fractions are below 0.1\%. This value is used as the systematic uncertainty associated with the afterglow correction.

\subsection{Beam-Induced-Background} \label{sec:BIB}

Beam-induced background is originating from interactions of the beam with either the residual gas in the beam pipe or the collimators that define the beam aperture. The particle showers travel mostly parallel to the beam and are in time with the associated bunch. BIB rates can become problematically high in certain conditions, such as poor vacuum or beam loss due to so-called ``UFO" (Unidentified Falling Object) events~\cite{UFO}, where the beam interacts with micron-scale dust particles inside the beam pipe. Therefore, it is important to monitor the BIB, and in severe cases adjust the high voltage power system to protect the CMS tracker.  High BIB values can also increase the occupancy in the CMS inner tracker, increasing the event size and making the track finding algorithms more computationally intensive.

\begin{figure}[!htb]
   \centering
   \includegraphics[width=\columnwidth]{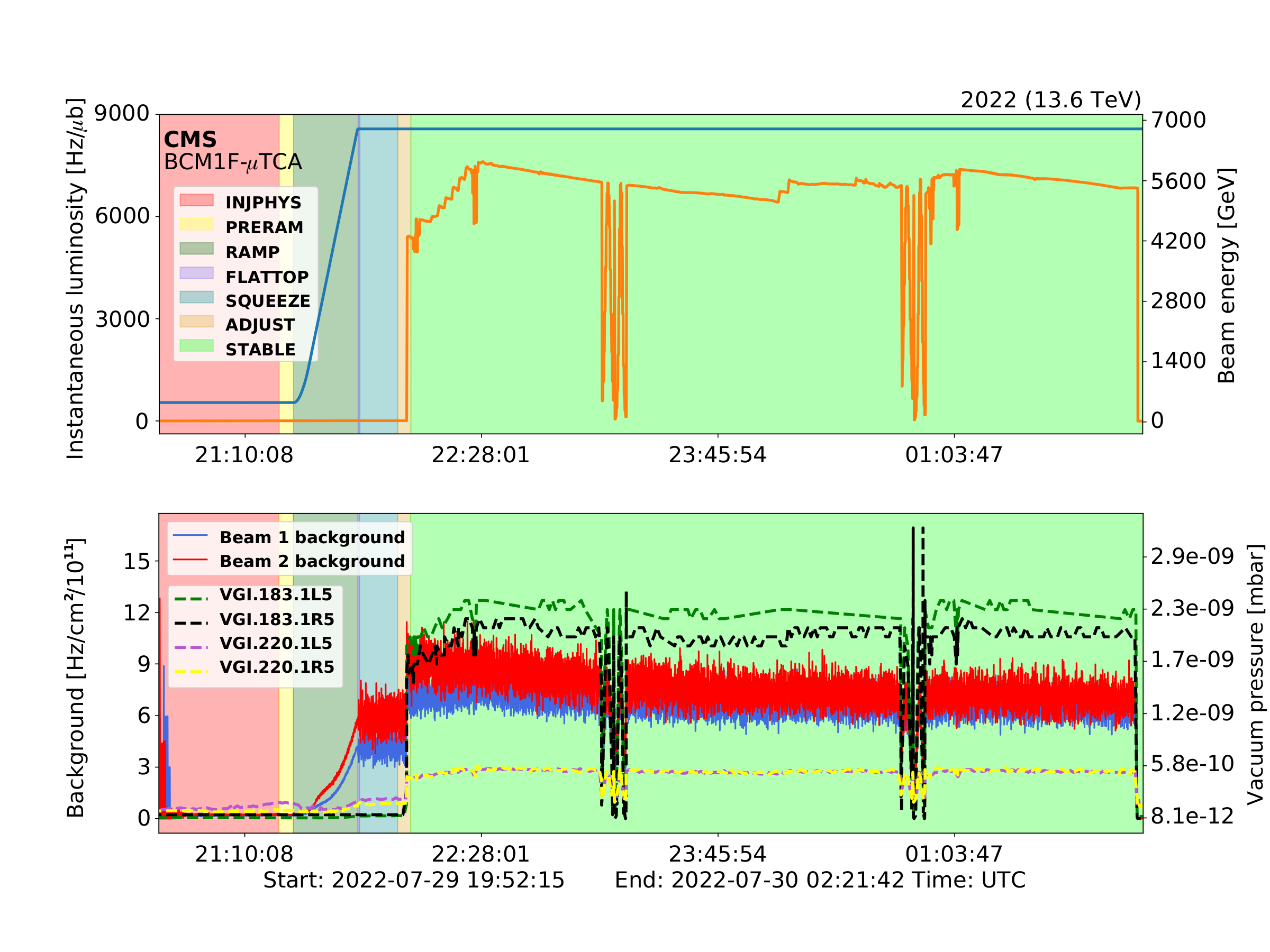}
   \caption{BCM1F measured Beam-1 and Beam-2 backgrounds (bottom) over the LHC beam mode cycle indicated with the different background shades (listed in the top figure legend). The measured values are affected by the beam conditions (top): beam energy (light blue curve) and BCM1F measured luminosity (orange curve). The readings from the vacuum pressure gauges (VGI) located close to the BCM1F detector at $\pm1.8$\,m and $\pm2.2$\,m from the interaction point are also shown (bottom)~\cite{Joanna_thesis}}
   \label{fig:bkgLHC}
\end{figure}

The BIB is measured as the particle rate per cm$^2$ per second per 10$^{11}$ protons in the beam, normalized to beam intensity to allow comparison between fills. BIB is determined separately for beam 1 and beam 2: channels on the +Z end measure incoming BIB for beam 1, and those on the –Z end for beam 2. Hits are included only if no paired bunches have occurred within the preceding 100 crossings, ensuring the measurement is performed for the first bunch of each train, where afterglow is minimal. Furthermore, only the individual bin corresponding to the BIB location (width 6.25\,ns for VME and 4.17\,ns for \utca) is used. Residual afterglow is corrected by subtracting the average of the five preceding crossings, taken as a beam-absent noise estimate.

The method was stable in Runs 2 and 3. At high intensities, a correlation with instantaneous luminosity was observed, likely due to the beam pipe vacuum degradation caused by collision products. Figure \ref{fig:bkgLHC} shows luminosity, BIB, and local vacuum at the start of a representative fill. During Run 3, BIB remained below the \mbox{20\,Hz/cm$^2$} per 10$^{11}$ protons threshold at the start of all fills, and CMS Tracker turn-on was never inhibited. Elevated BIB occurred during commissioning and machine development, but vacuum pressures—and thus BIB rates—remained low during nominal physics data taking.

\subsubsection{Beam–gas losses and increased beam-pipe vacuum pressure}

A dedicated machine development study was performed during LHC Run 2 (fill 5005) to quantify the response of background monitors to degraded vacuum conditions. At the time, the first iteration of the Run-2 BCM1F detector, based on single-crystal diamond sensors, was installed. Although this detector was near the end of its operational lifetime and exhibited reduced sensitivity, the study remains the only dedicated measurement of this kind.
During the test, vacuum pressure was deliberately increased at three locations on each side of CMS, positioned at 148\,m, 58\,m, and 22\,m from the interaction point. The pressure surges were generated simultaneously on both sides. BCM1F measured the beam-induced background separately for beam 1 and beam 2, while vacuum gauges (VPIAN and VGPB) recorded the local pressure.

\begin{figure}
     \centering
     \begin{subfigure}{}
         \includegraphics[width=\columnwidth]{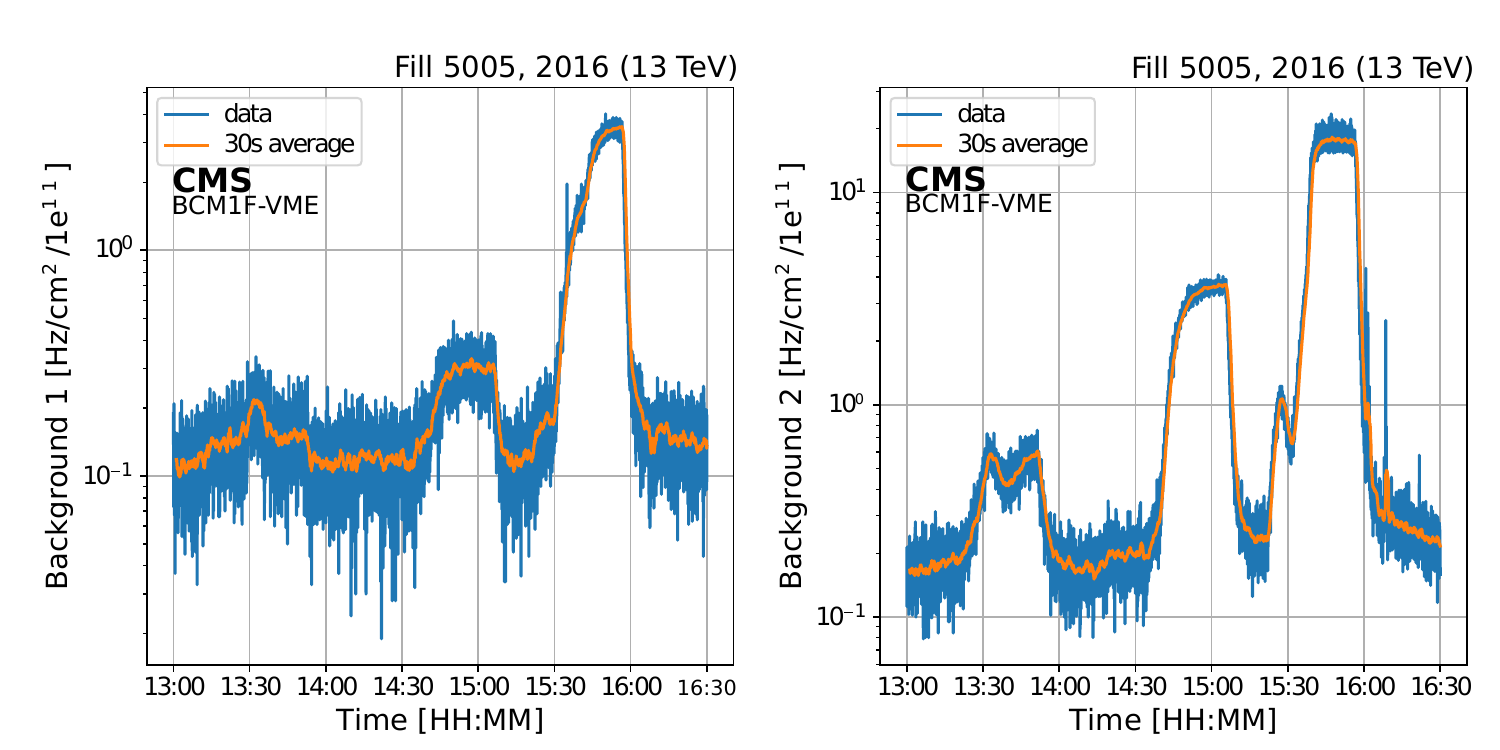}
         \label{fig:vactest_bkgd}
     \end{subfigure}
     \vspace{-0.5cm}
     \begin{subfigure}{}
         \includegraphics[width=\columnwidth]{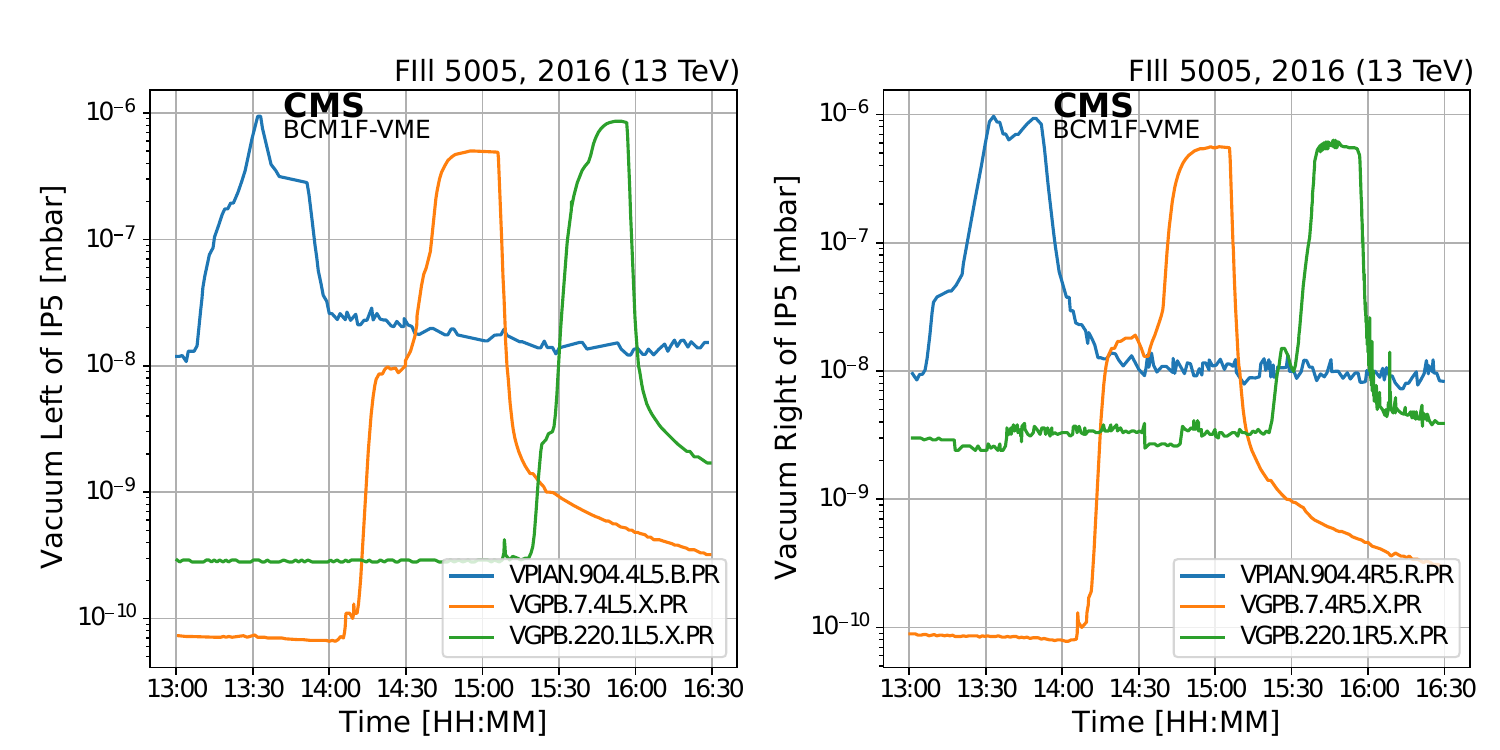}
         \label{fig:vactest_vac}
     \end{subfigure}
        \caption{
        The BIB rates (top) and the corresponding vacuum pressures (bottom) recorded by gauges at the excitation points on each side of CMS, positioned at 148\,m (blue), 58\,m (orange), and 22\,m (green) from the interaction point}
        \label{fig:vactest1}
\end{figure}

\begin{figure}
    \centering
        \includegraphics[width=0.99\columnwidth]{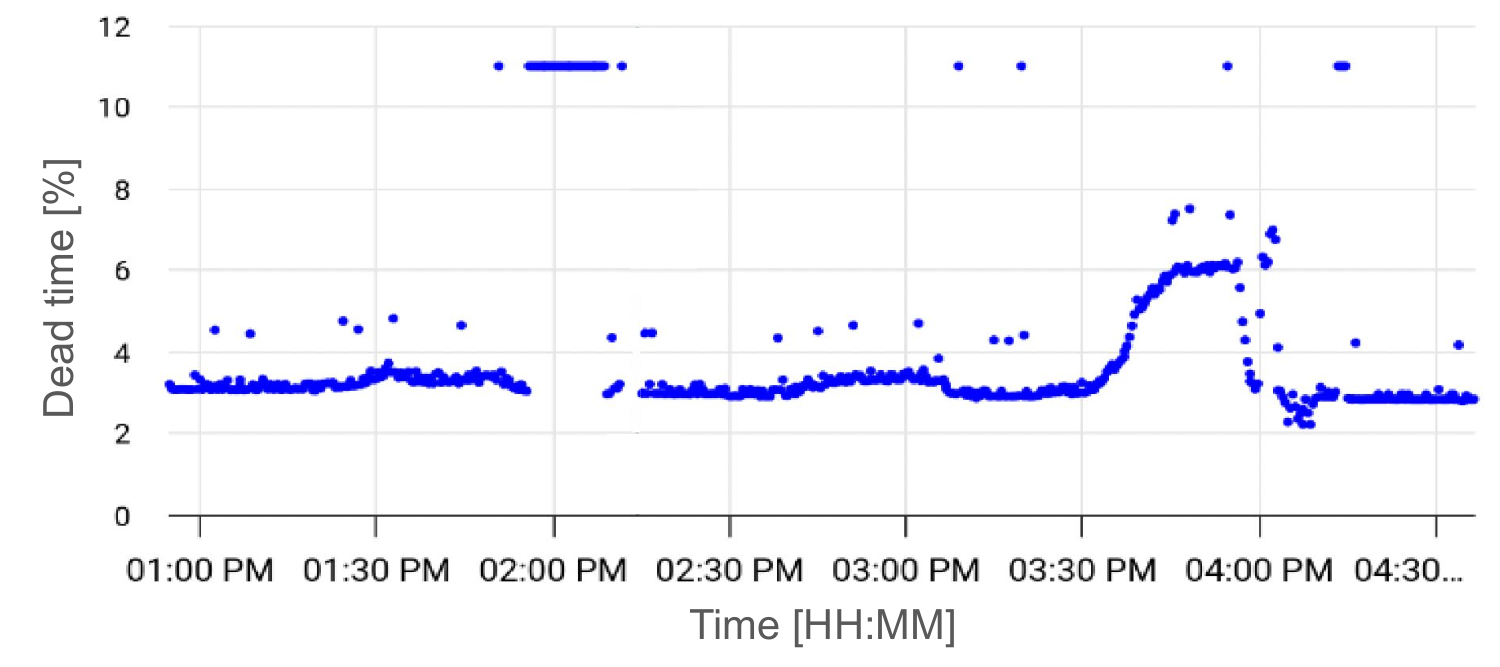}
       \caption{CMS dead time during the special machine development tests in LHC fill 5005. All values above 10\% are plotted as 11\%. The brief period of high dead time around 2:00 PM was not related to the pressure bump test, but the effect of the local pressure increase at 22\,m from the IP (between 3:30 and 4:00 PM) can clearly be seen when comparing with Fig.~\ref{fig:vactest1} }
       \label{fig:deadtime}
\end{figure}

Figure~\ref{fig:vactest1} shows the time evolution of BIB rates for each beam together with the corresponding vacuum pressures. The BCM1F background measurements closely followed the temporal profiles of the induced pressure bumps, with higher background rates observed for pressure increases closer to CMS. Differences between the left and right pressure signals were consistently reproduced in the corresponding beam-specific BIB measurements. These results demonstrate correlation between local beam–gas interactions and BIB rates at CMS, and the capability of BCM1F to provide fast, position-sensitive diagnostics of vacuum-related beam losses. 

To explicitly show the impact of increased BIB on typical CMS operations, the data-taking dead time is shown in Fig.~\ref{fig:deadtime} for the same time period as BIB and vacuum pressure increases in Fig.~\ref{fig:vactest1}. CMS dead time is affected differently by local and distant pressure variations: the former increases pixel detector occupancy and, consequently, the pixel system readout dead time, and is seen to cause a significantly increased overall CMS dead time; the latter effect is much smaller.


\subsection{Luminosity determination} \label{sec:luminosity}

The instantaneous luminosity is measured using an observable that scales linearly with it. In the BCM1F system, the particle hit count rate over 25\,ns (corresponding to the bunch spacing) is used. The single-bunch instantaneous luminosity value is derived from counts aggregated for a predefined number of orbits (2$^{12}$ for VME and 2$^{14}$ for \utca back end) in each colliding bunch crossing.   
The zero-counting method is used to estimate the mean hit count based on the Poisson distribution, as described in Sec.~\ref{sec:zero_counting}. 

The absolute luminosity scale needs to be obtained using a dedicated calibration procedure at the LHC, called the van der Meer (vdM) method, that uses transverse beam separation scans (Sec.~\ref{sec:vdm_scans}). 
Due to increased bunch spacing in vdM fills, afterglow corrections (Sec.~\ref{sec:albedo}) are negligible, however, background corrections  (Sec.~\ref{sec:vdm_super_separation}) must be applied. 
Raw rates are stored on disk for offline reprocessing. Each channel is calibrated independently, enabling its use as a separate luminometer. 

During routine LHC operations, shorter calibration scans are included in regular physics fills to monitor detector stability and linearity (Sec.~\ref{sec:emit_scans}). The analysis of these scans is discussed further in Sec.~\ref{sec:linearity_stability_data_quality}.

\subsubsection{Zero-counting method} \label{sec:zero_counting}

In the zero-counting method, the expected number of hits is derived from the probability of observing zero hits. Assuming that the number of hits within a bunch-crossing follows a Poisson distribution, the probability $p$ of observing $n$ hits is given by:
\begin{equation}\label{eq:poisson}
    p(n) = \frac{\mu^n e^{-\mu}}{n!},
\end{equation}
where $\mu$ is the average number of hits. The simplest way to express this is by counting instances where no hits are detected (i.e., zero hits), leading to:
\begin{equation}\label{eq:poisson_mu_zero}
    p(0) = e^{-\mu},\, \rightarrow \tab \mu = -\ln{[p(0)]}.
\end{equation}

The zero-probability $p(0)$ is calculated by subtracting from 1 the sum of probabilities for any number of hits greater than zero $p(n\neq 0)$:
\begin{equation}\label{eq:poisson_mu_nzero}
        \mu = -\ln{[p(0)]} = -\ln{[1-p(n\neq 0)]}. 
\end{equation}
Consequently, the average number of hits per bunch crossing is calculated as:
\begin{equation}\label{eq:poisson_mu}
        \mu = -\ln{\biggl[1-\frac{r}{N_{\text{orbits}}}\biggr]}, 
\end{equation}
 where $r$ is the hit count within a predefined number of orbits $N_{\text{orbits}}$, and the fraction of counts per-bunch is hereby referred to as $f_r = \frac{r}{N_{\text{orbits}}}$. The statistical uncertainty on the mean number of hits can be obtained as follows:
 
\begin{equation}\label{eq:err_mu}
    \sigma(\mu) = \frac{\sqrt{r}}{1-f_r}.
\end{equation}

\begin{figure}[!htb]
   \centering
   \includegraphics*[width=1.0\columnwidth]{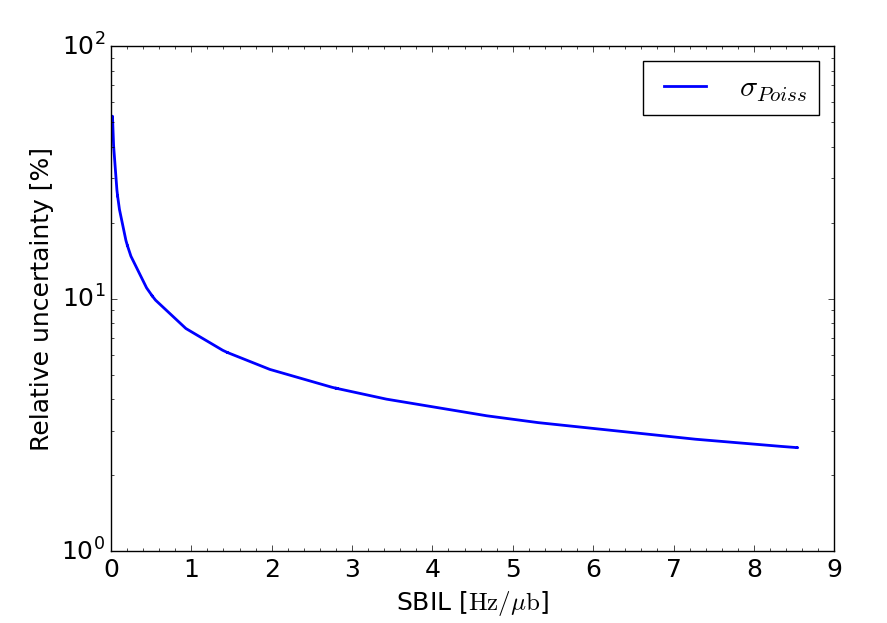}
   \caption{Relative uncertainty on the mean number of hits per channel as a function of SBIL, calculated using the Poisson estimator from Eq.~(\ref{eq:err_mu}). The smallest available aggregation-time granularity is used, corresponding to $t_{\mathrm{int}}=4\,\unit{NB}\simeq 1.46$\,s~\cite{Joanna_thesis} }
   \label{fig:relErrVsOcc}
\end{figure}

\begin{figure}[!htb]
   \centering
   \includegraphics[width=1.0\columnwidth]{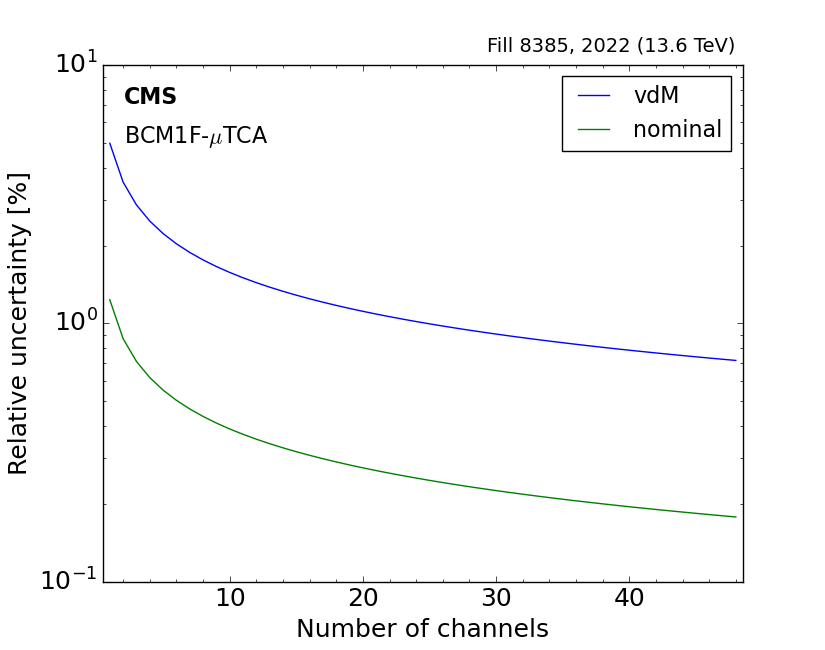}
   \caption{Statistical uncertainty on the BCM1F-\utca measured per-bunch luminosity as a function of the number of channels averaged for vdM conditions ($\mathrm{SBIL}\simeq 0.1\,\unit{Hz/\mu b},\ t_{\mathrm{int}} = 30$\,s, and for the nominal physics conditions (using average $\mathrm{SBIL}\simeq 5.3\,\unit{Hz/\mu b},\ t_{\mathrm{int}} = 10$\,s)~\cite{Joanna_thesis}}
   \label{fig:errVsCh}
\end{figure}

This uncertainty is shown in Fig.~\ref{fig:relErrVsOcc} as a function of the SBIL in the range representative of Run 3 conditions. 
The estimated per-channel statistical uncertainty in the figure is rather significant, but not representative of the final luminosity measurement, as it is further reduced with channel averaging and integration time. The final uncertainty for typical conditions during the special low pileup vdM calibration fills (Sec.~\ref{sec:vdm_scans}), as well as during the nominal data-taking as a function of the number of averaged channels, is shown in Fig.~\ref{fig:errVsCh}.
Under conditions during vdM scans, where the SBIL is very low, the typical integration time for the vdM scan steps is 30 seconds. This reduces the relative uncertainty from 3.5\% on a single channel to approximately 0.7\% in an ideal scenario where all of the 48 BCM1F channels are used. Under nominal physics conditions, the rate is significantly higher, which improves the statistical precision per unit time. The integration time can therefore be reduced to 10 seconds per step, as used in emittance scans, while still achieving a single-channel uncertainty of 1.25\%, which further drops to 0.2\% when averaging across all channels. 
Detailed comparison of beam conditions is given in~\ref{sec:emit_scans}. 
The above described method worked very well throughout Run 3 as the BCM1F location and sensor size was optimized in the design stage to avoid zero-starvation, i.e. conditions where the probability of no hit in the detector becomes small.  

\subsubsection{Van der Meer scans and calibration} \label{sec:vdm_scans}

The van der Meer calibration method for luminometers uses transverse beam-separation scans to vary and measure the overlap area of the colliding beams~\cite{vdm,Rubbia:1025746}. Each year, the LHC conducts a dedicated program of such scans to provide essential calibration data for the luminometers at CMS and other LHC experiments, for each beam energy and beam particle-type.
Details of the typical parameters and conditions of vdM scans can be found in Table~\ref{table:vdM_vs_emit}, where they are compared with shorter and more frequent emittance scans performed during standard LHC operations at high instantaneous luminosity (Sec.~\ref{sec:emit_scans}).

The primary purpose of the vdM scan program is to measure the detector-specific calibration constant that relates the observed rate to the absolute instantaneous luminosity~\cite{CMS-PAS-LUM-22-001}.
Under the assumption that the beam particle density distribution $\rho$ is factorizable into distributions in $x$ and $y$ directions ($\rho(x,y)=\rho_x(x) \times \rho_y(y)$), the transverse convolved bunch widths $\Sigma_x$ and $\Sigma_y$ can be extracted from the measured beam-separation $\Delta_x$ and $\Delta_y$ dependence of the collision rate $R(\Delta_x,\Delta_y)$: 
 \begin{equation}\label{eq:capsigmasFromLumi}
     \Sigma_x = \frac{1}{\sqrt{2\pi}} \frac{\int R(\Delta_x,0)d\Delta_x}{R(0,0)}, \\
     \Sigma_y = \frac{1}{\sqrt{2\pi}} \frac{\int R(0,\Delta_y)d\Delta_y}{R(0,0)}.
 \end{equation}
The measured convolved bunch widths from the scans in $x$ and $y$ directions from Eq.~(\ref{eq:capsigmasFromLumi}), the bunch intensities of the two proton beams $N_{1,2}$, and the rate at zero separation of the beams $R(0,0)$ are used to calculate the luminometer-specific calibration constant, the so-called visible cross section: 
\begin{equation}\label{eq:sigvisFromPar}
\sigma_\text{vis} = \frac{R(0,0)}{N_1 N_2}  2\pi\Sigma_x\Sigma_y \, .
\end{equation}

Subsequently, this calibration constant is used to translate the measured detector rate $R$ into the luminosity directly, at any beam conditions:
\begin{equation}\label{eq:lumiFromsigvis}
    \mathcal{L}_{\text{inst}}  = \frac{R}{\sigma_{\text{vis}}},
\end{equation}
where the detector rate is calculated as the product of the average number of hits $\mu$ (described in Sec.~\ref{sec:zero_counting}) and the LHC revolution frequency $ f_\text{rev}$ = 11.246\,kHz: 
\begin{equation}\label{eq:rate}
    R  = \mu f_\text{rev}.
\end{equation}

\subsubsection{Background estimation during the vdM scans} \label{sec:vdm_super_separation}

Another component of the vdM program is the data taken in dedicated super-separation periods, in which the beams are separated by a distance equivalent to five beam widths ($5\,\sigma$) in both transverse planes, resulting in a total separation of nearly $7.1\,\sigma$. 
This procedure ensures there is no transverse overlap between the two beams, thereby excluding collisions. 
As detailed in Sec.~\ref{sec:BIB}, the BCM1F detector exhibits high sensitivity to background signals and is also used to provide online background measurements for both CMS and the LHC.
The super-separation scans allow for the measurement of the BIB and detector-specific noise during the vdM program. 

\begin{figure}[!ht]
\centering
\includegraphics[width=1.0\columnwidth]{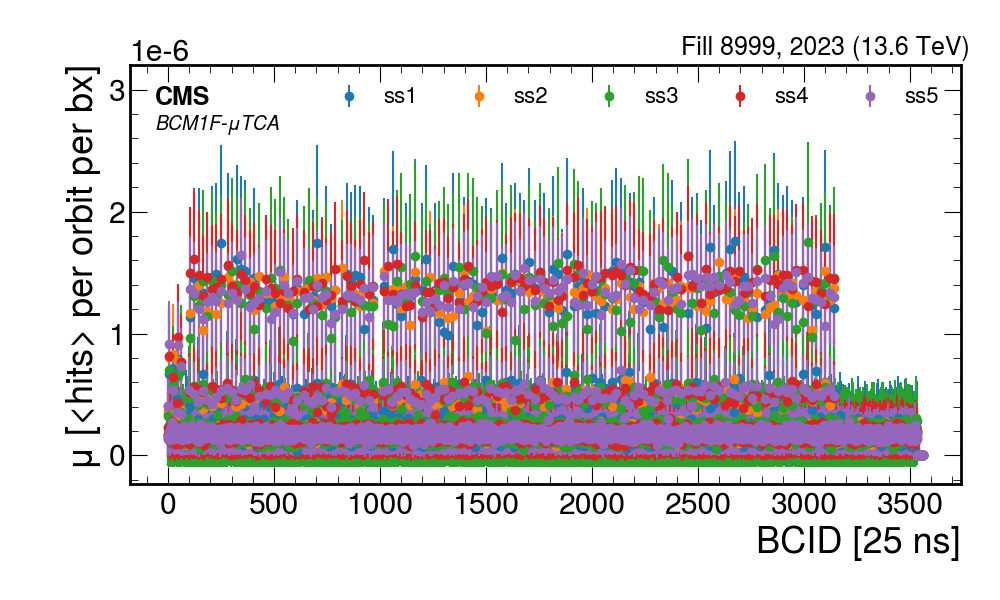}
\caption{
Per-bunch rates measured by BCM1F-\utca during five super-separation scans in a vdM fill. The upper band ($\mu \approx 1.4\cdot 10^{-6}$) corresponds to paired ``collidable" bunches used for collision-background estimation. Only four unpaired non-colliding bunches occur at the beginning of the orbit ($\mu \approx 0.8\cdot 10^{-6}$) . The lower band ($\mu \approx 0.5\cdot 10^{-6}$), slightly above the noise level ($\mu \approx 0.2\cdot 10^{-6}$), corresponds to incoming beam-induced background arriving 12.5\,ns before the collision signal} 
\label{fig:utca_bkg}
\end{figure}

The signal measured during five such periods in the 2023 calibration fill is shown in Fig.~\ref{fig:utca_bkg}. The statistical uncertainty of the measurement is minimized by extending the super-separation periods to 5 minutes. The average background of all BCIDs for which both beams are filled with bunches is subtracted directly from all BCM1F rates measured during vdM scans. As an example, in 2022 the total correction from the background measurement to \sigmavis was, on average, $-1.5 \pm 0.1\%$~\cite{Joanna_thesis}.

\subsubsection{Emittance scans} \label{sec:emit_scans}

\begin{figure}[!ht]
\centering
\includegraphics[width=1.0\columnwidth]{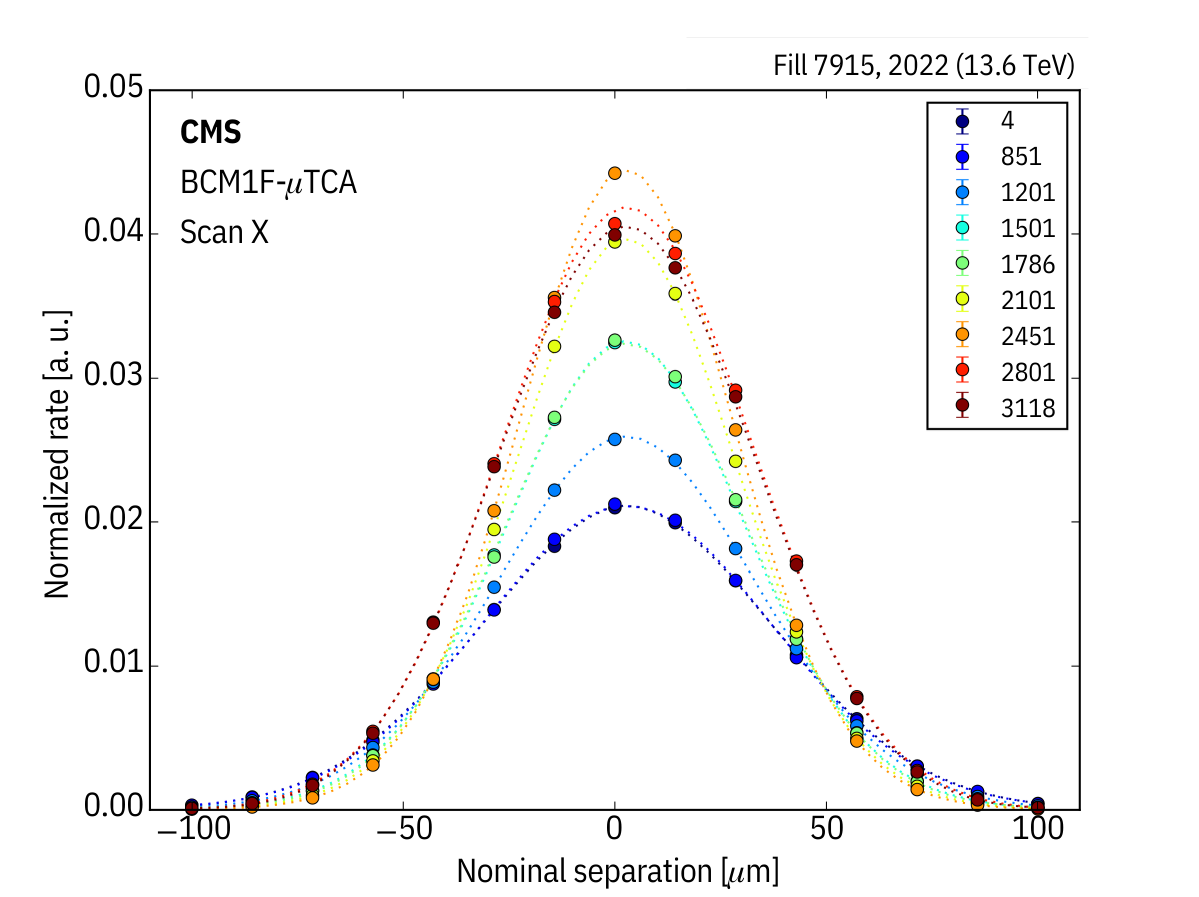} 
\caption{Beam overlap profiles obtained from an emittance scan in the $x$ direction for nine bunches with different emittances. The legend indicates the BCIDs corresponding to each set of measurement points, which are fitted with double-Gaussian functions~\cite{Joanna_thesis}} 
\label{fig:emittanceScan_fit}
\end{figure}

Short vdM-like emittance scans~\cite{Michi_Emit,Emit_scans_2017} were introduced in Run 2 by LHC for fast beam diagnostics during regular LHC operation. As emittance scans were proven to be also useful for the experiments to monitor the luminometers conditions (Sec.~\ref{sec:stability}), they continued to be performed on the regular bases since then. The emittance scan procedure follows one similar to a vdM scan (Sec.~\ref{sec:vdm_scans}), in which beams are first separated and moved in the $x$ and then in the $y$ direction. Emittance scans are performed in fewer steps, normally 9 or 15, and with less time (10\,s) spent at each step. An example measurement of the \utca rate as a function of separation in $x$ is shown in Fig.~\ref{fig:emittanceScan_fit}. 
Consequently, emittance scans allow for measuring the effective beam overlap and calculating a calibration coefficient $\sigma_\mathrm{vis}^\mathrm{em}$ per luminometer per bunch crossing~\cite{Emit_scans_2017} similar to the visible cross section $\sigma_\mathrm{vis}$ measured in vdM scans. However, due to the more challenging beam conditions, $\sigma_\mathrm{vis}^\mathrm{em}$ cannot be used as the absolute calibration $\sigma_\mathrm{vis}$. In particular, emittance scans are performed with various filling schemes which are used throughout the year for physics data collection. 
Physics fills differ from vdM fills in several key aspects: up to about 2500 bunches in total appear sequentially in trains, per-bunch intensity is approximately 20\% higher, pileup is up to two orders of magnitude larger, and collisions occur at a non-zero crossing angle and $\beta^*$.

These differences are summarized in Table~\ref{table:vdM_vs_emit}.

Comparison of effective beam overlap measurements from regular emittance scans across different detectors allows individual detector effects to be identified. As effective beam overlap is a property of the beams, all detector measurements must agree. 

\begin{table}[h!]
\centering
\renewcommand{\arraystretch}{1.2}
\begin{tabular}{P{2.1cm} | C{2.2cm} | C{2.4cm}}
\toprule
 & vdM scan & Emittance scan (nominal LHC operation) \\
\midrule
Number of bunches &  30--146  well-separated, isolated bunches& up to 2500 in bunch trains \\
Half crossing angle [\si{\micro\radian}] & 0 & in range $\sim 130$--160 \\
Transverse beam size [\si{\micro\metre}] & $\sim 100$ & $\sim 10$--$15$ \\
$\beta^{*}$ [m] & 19.2 & 0.3--1.2 \\
Number of scan steps & $\sim 25$ & 9 or 15 \\
Duration of scan step [s] & 30 & 10 \\
Peak pileup & $\sim 0.5$-0.8 & $\sim 60$ \\
SBIL [Hz/\si{\micro\barn}] & 0.1 & 5--9 \\
\bottomrule
\end{tabular}
\caption{Comparison of beam conditions and scan parameters between vdM calibration fill configuration and standard LHC operation when emittance scans are performed}
\label{table:vdM_vs_emit}
\end{table}

\begin{figure}[!htb]
   \centering
   \includegraphics*[width=\columnwidth]{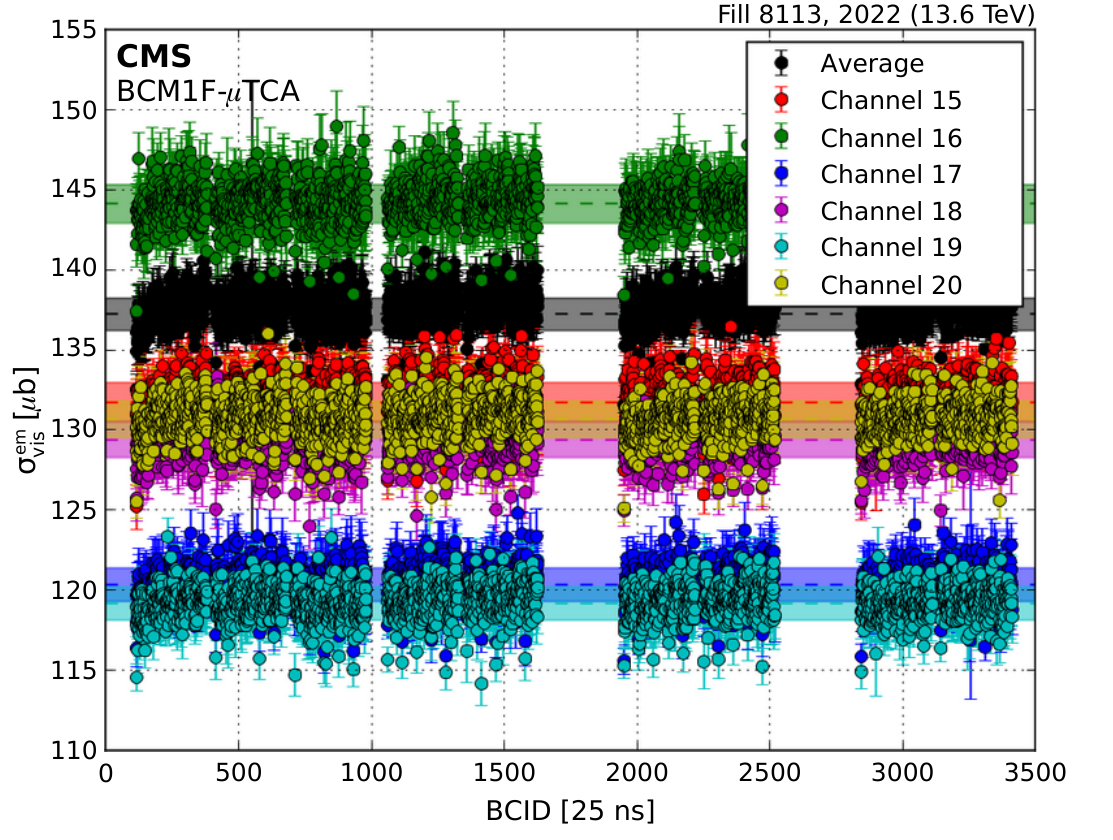}
   \caption{Calibration coefficient $\sigma_\mathrm{vis}^\mathrm{em}$ measured for each of the filled BCIDs, shown for six example channels and compared to the average of all BCM1F-\utca channels (black). The preliminary measurement from an emittance scan in fill 8113 is shown based on a single Gaussian fit without any corrections~\cite{Joanna_thesis} }
   \label{fig:bcm1futca_perChannelCalib}
\end{figure}

Differences in per-channel BCM1F rates result in measured $\sigma_\mathrm{vis}^\mathrm{em}$ differences, as presented in Fig~\ref{fig:bcm1futca_perChannelCalib}. The standard deviation across all channels was determined to be 8\%. Linear efficiency factors are calculated to scale individual per-channel measurements (shown in different colours) to the average value (shown in black) and applied to equalize the contributions of all channels. The average per-channel rate for $N$ channels is:

\begin{equation}
     \overline{\mu} =  \frac{\overline{\sigma}_\mathrm{vis}^\mathrm{em}}{N}  \sum^N_i \frac{\mu_i}{\sigma_{\mathrm{vis,}i}^\mathrm{em}}  \label{eq:mu_avg}   ,
 \end{equation}
where the ratio $\mu_i / \sigma^{\text{em}}_{\text{vis},i}$ removes the per-channel efficiency,  and the prefactor $\bar{\sigma}^{\text{em}}_{\text{vis}}$ restores the common scale, making Eq.~(\ref{eq:mu_avg}) equivalent to a simple arithmetic mean of the efficiency-corrected rates.

\begin{figure}[ht!]
    \centering
    \includegraphics[width=1.0\columnwidth]{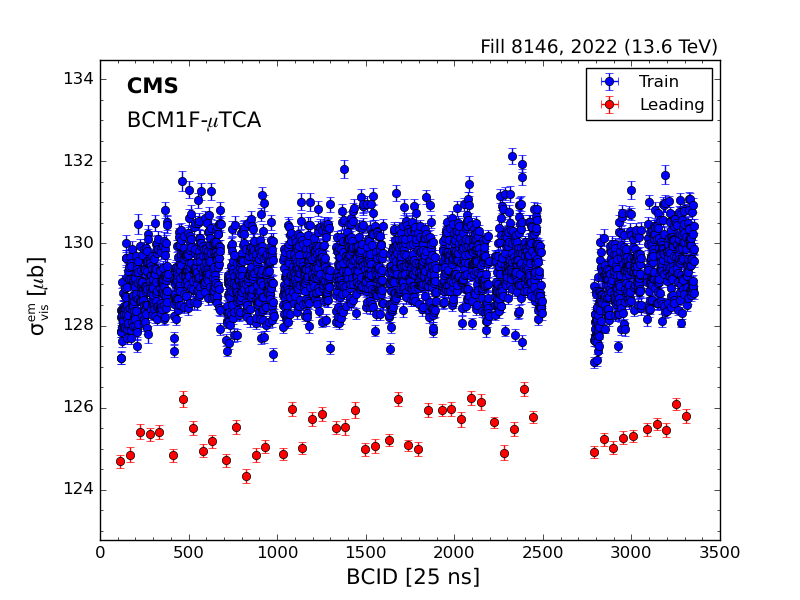}
    \includegraphics[width=1.0\columnwidth]{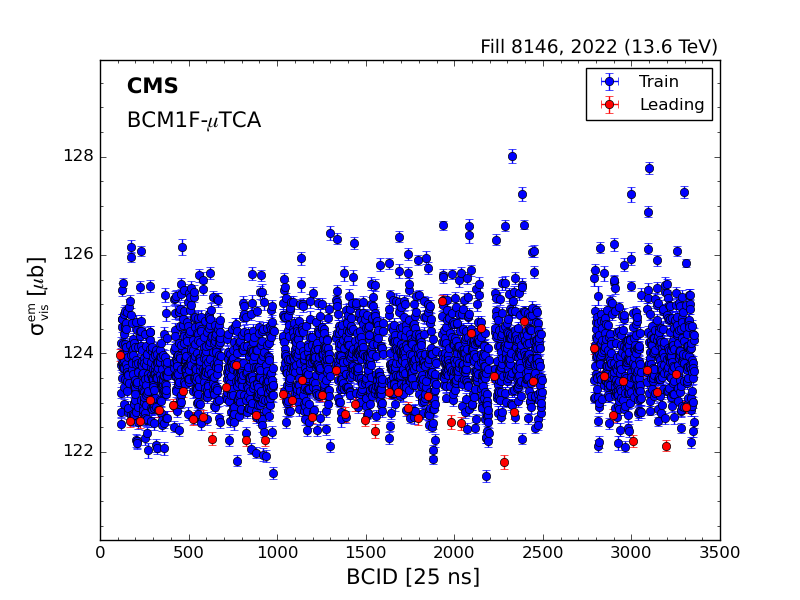}
    \caption{Per-bunch $\sigma_\mathrm{vis}^\mathrm{em}$ measured during the emittance scan in one of the 2022 fills before the afterglow correction (top) and after (bottom). The first bunch in each train is plotted in red~\cite{Joanna_thesis} }\label{fig:sigvis_vs_bcid}
\end{figure}

Calibration coefficient measurements per BCID from emittance scans also serve as a cross-check for the afterglow correction. An example from fill 8146 in 2022 is shown in Fig.~\ref{fig:sigvis_vs_bcid}. On the top plot, corresponding to the online data, a systematic discrepancy of about 3\% is observed for the first colliding bunches of each train (red), as these are the least affected by afterglow. By subtracting the activation fractions from each preceding colliding bunch during the reprocessing, this systematic bias is effectively removed, as seen on the bottom plot.

Finally, $\sigma_\mathrm{vis}^\mathrm{em}$ is a property of the detector and, under stable detector conditions, should remain constant. Variations of $\sigma_\mathrm{vis}^\mathrm{em}$ with changing beam conditions indicate a nonlinearity in the detector response, whereas variations over time point to changes in detector efficiency, for example due to ageing, as will be shown in Sec.~\ref{sec:stability}.

\section{Linearity, stability and data quality} \label{sec:linearity_stability_data_quality}

The absolute calibration constant $\sigma_\mathrm{vis}$ is obtained once every year in a special vdM calibration fill for each operation mode utilized for data taking, e.g. pp or PbPb collisions at a given energy (Sec.~\ref{sec:vdm_scans}). However the detector efficiency changes over the course of the year (Sec.~\ref{sec:commisioning_ops}). 
Multiple systematic effects must be considered and corrected for to achieve precise and accurate luminosity calibration.  
Details about the calibrations of BCM1F and other CMS luminometers in the first year of LHC Run 3 can be found in Ref.~\cite{CMS-PAS-LUM-22-001}. 
In addition to the beam-related systematic effects, a BCM1F-specific background correction (Sec.~\ref{sec:vdm_super_separation}) must be applied during vdM scan conditions.
Although afterglow corrections (Sec.~\ref{sec:albedo}) are negligible in vdM fills due to the large spacing between colliding bunches, they become essential during physics fills, where the bunch train structure leads to significant out-of-time contributions in the following BCIDs from activated detector material.
To enable the use of the vdM calibration constant throughout the year, the linearity of the observed rate with respect to pileup, as well as the detector performance stability over time must be monitored and corrected. These are discussed in Secs.~\ref{sec:linearity} and~\ref{sec:stability}, respectively.

\subsection{Linearity determination} \label{sec:linearity}

\begin{figure}[!ht]
\centering
\includegraphics[width=.8\columnwidth]{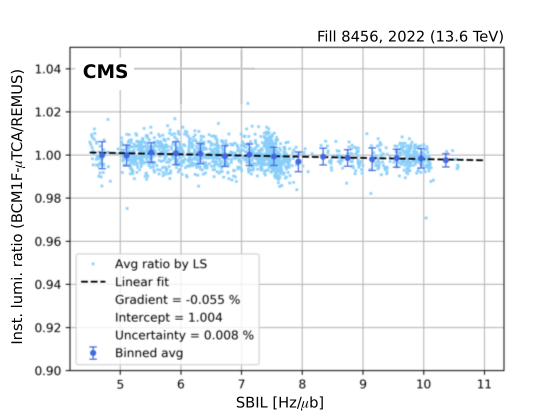} \\
\includegraphics[width=.8\columnwidth]{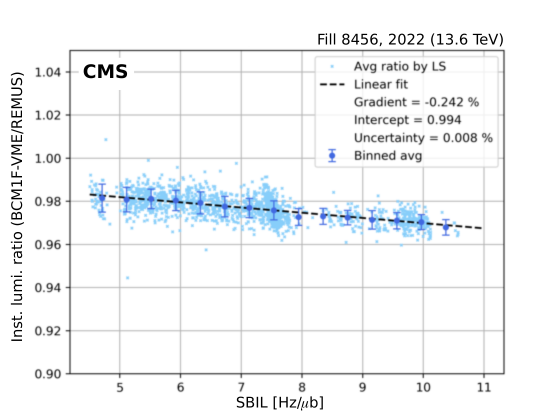}
\caption{Relative linearity measurement for both BCM1F-\utca (top) and BCM1F-VME (bottom) with respect to a reference detector with known linear behaviour (REMUS) for a typical fill in 2022}
\label{fig:sigvis_vs_sbil}
\end{figure}

To illustrate the BCM1F linearity response after the corrections discussed in Sec.~\ref{sec:data_reprocessing}, BCM1F results for $pp$ collision data collected during 2022 are presented in Ref.~\cite{CMS-PAS-LUM-22-001,CMS:2021xjt}. 
The linearity is assessed by comparing the luminosity measured by BCM1F to that of reference detectors with known linear behaviour as a function of the SBIL for each fill. Example measurements are shown for fill 8456 in Fig.~\ref{fig:sigvis_vs_sbil}. The top plot shows the ratio of BCM1F-\utca instantaneous luminosity with respect to one of the reference detectors REMUS~\cite{CMS-PAS-LUM-18-002}, and the bottom plot shows the same ratio for BCM1F-VME. A linear fit is applied to determine the slope, which quantifies relative BCM1F linearity. BCM1F-\utca linearity is shown to be comparable to the reference detector REMUS (-0.055\% relative slope). The BCM1F-VME linearity slope is -0.242\% per unit of SBIL, resulting in about a 1.2\% linearity change over 5 units of SBIL during a typical LHC fill; thus, a  linearity correction is required. The difference in linearity of the two BCM1F readouts has its origin in the different hit discrimination methods, with the BCM1F-\utca providing intrinsically better noise suppression and double--hit resolution. 

\begin{figure}[!ht]
\centering
\includegraphics[width=.8\columnwidth]{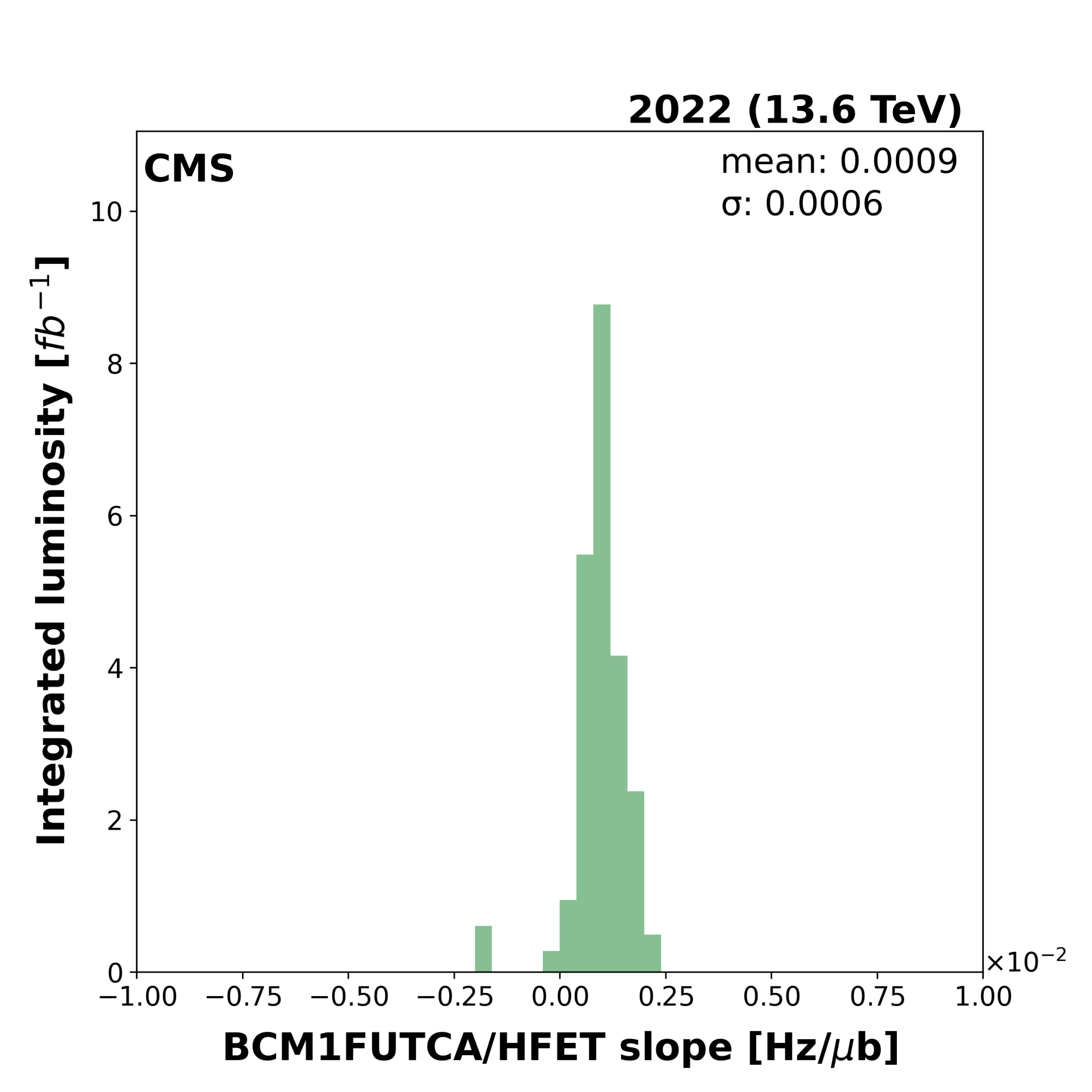}
\caption{%
Distribution of relative linearity slopes between BCM1F-\utca and HFET detectors. Each entry is weighted by the integrated luminosity of the corresponding fill~\cite{CMS-PAS-LUM-22-001}
}
\label{fig:bcm1futca_hfet}
\end{figure}

This linearity slope for each fill is then weighted by the integrated luminosity and aggregated in a histogram, showing linearity consistency for the entire 2022 data set.
Figure~\ref{fig:bcm1futca_hfet} shows the relative linearity slopes of BCM1F-$\mu$TCA  with respect to the CMS hadron calorimeter, a high-reliability luminometer employing a per-bunch transverse energy sum algorithm (HFET).  
The mean value (mean) and standard deviation ($\sigma$) of the distributions are provided in the legend.
Typically, the standard deviation of a histogram like Fig.~\ref{fig:bcm1futca_hfet} multiplied by the average SBIL ($\sigma\times\overline{SBIL}$) is assigned as systematic uncertainty on the luminosity measured for the inspected data-taking period due to linearity. In the case when the mean is larger than $\sigma$, meaning that the remaining residual nonlinearity is larger than the standard deviation, mean value is used for systematic uncertainty calculation. For 2022 $\overline{SBIL}$ = 5.32\,Hz/$\mu$b, yielding relative uncertainties of 0.48\% 
for BCM1F-\utca.

Emittance scans are also used to cross--check the linearity.  
At the beginning of the standard physics fill, the SBIL from all bunches covers a range of 5--9\,Hz/$\mu$b. Detector linearity is extracted by plotting per bunch crossing $\sigma_\mathrm{vis}^\mathrm{em}$ measurements in the emittance scans, and checking the slope of the linear fit to this distribution. 
However, since the $\sigma_\mathrm{vis}^\mathrm{em}$ measurement is affected by systematic effects arising from the beam dynamics of train bunches (for instance, beam–beam deflections due to long-range effects, x-y factorization), it is only used as a linearity cross-check and not as an absolute linearity measurement.


\subsection{Efficiency correction } \label{sec:stability}

\begin{figure}[ht!]
    \centering
    \includegraphics[width=\columnwidth]{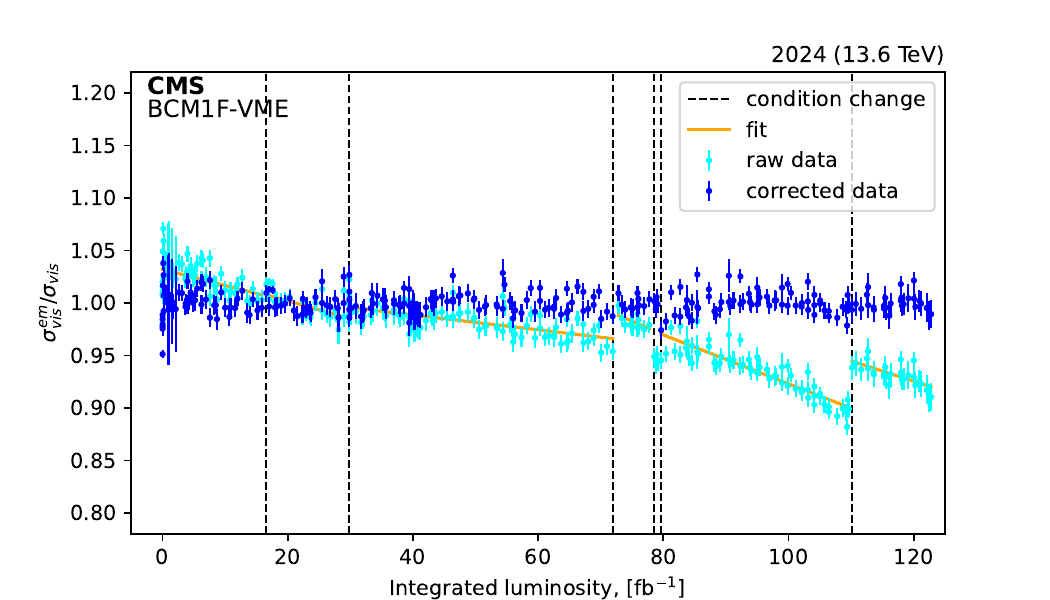}
     \includegraphics[width=\columnwidth]{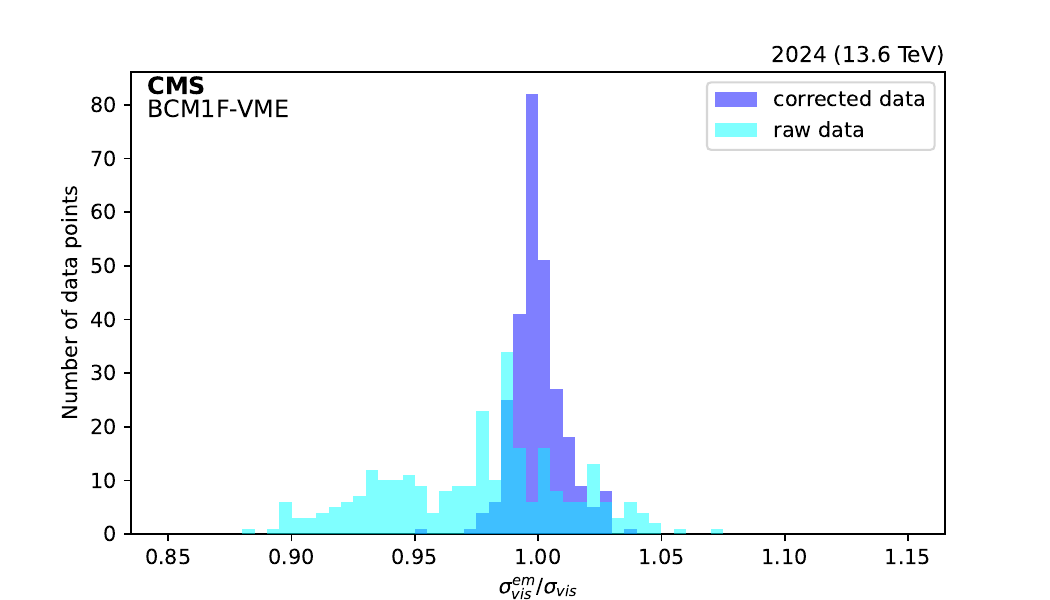}
    \caption{The ratio of $\sigma_\mathrm{vis}^\mathrm{em}$ measured during emittance scans to the reference $\sigma_\mathrm{vis}$ from the vdM scan for one channel in the VME readout chain, before and after the application of the efficiency corrections. The data points are shown as a function of integrated luminosity recorded by the detector (top), as well as histogrammed over the entire 2024 data-taking period (bottom)}
    \label{fig:vme_stability_2024}
\end{figure}

The long-term efficiency of the detector is influenced by accumulated radiation damage and annealing effects. These changes are analysed using the emittance scans (Sec.~\ref{sec:emit_scans}) throughout the operation period. The measured $\sigma_\mathrm{vis}^\mathrm{em}$ for the scans performed in the beginning of many fills using online rates are compared to the final precision $\sigma_\mathrm{vis}$ measured in the vdM fill.
This ratio, $\sigma_\mathrm{vis}^\mathrm{em}$/$\sigma_\mathrm{vis}$, is plotted in Fig.~\ref{fig:vme_stability_2024} as a function of the integrated luminosity for the 2024 pp dataset as an example. Light blue points show online (raw) data from which efficiency corrections are extracted, while dark blue points show results of the offline reprocessing after the efficiency corrections are applied (corrected data). 

The scan profiles are fit with a Q-Gaussian function to extract $\sigma_\mathrm{vis}^\mathrm{em}$.
No correction is applied for beam-related systematic effects~\cite{BB_DB}.

To derive the efficiency correction, raw data measurements in Fig.~\ref{fig:vme_stability_2024} (top) are treated independently across six distinct operational periods separated by vertical dashed lines.  
A significant efficiency degradation was observed. The optical converters configuration and high-voltage set point for the sensors were adjusted to recover efficiency between these six periods (Sec.~\ref{sec:commisioning_ops}).
A separate linear fit was used for each period (orange line) to define the efficiency correction, which is applied to scale the raw data on a per-fill basis to the reference efficiency measured during the vdM fill. As Fig.~\ref{fig:vme_stability_2024} (bottom) shows, the efficiency correction using the simple model of linear performance degradation over integrated luminosity accounts for much of the change in detector performance throughout the year. 

Other detector performance monitoring tools were also used during the data-taking periods to track fill-to-fill changes. For instance, baseline levels, the per-fill MPVs from the amplitude spectra and test pulse analysis (Sec.~\ref{sec:rhu_adc},~\ref{sec:uTCA_commissioning}) are regularly utilized.  

\begin{figure}[ht!]
    \centering
    \includegraphics[width=\columnwidth]{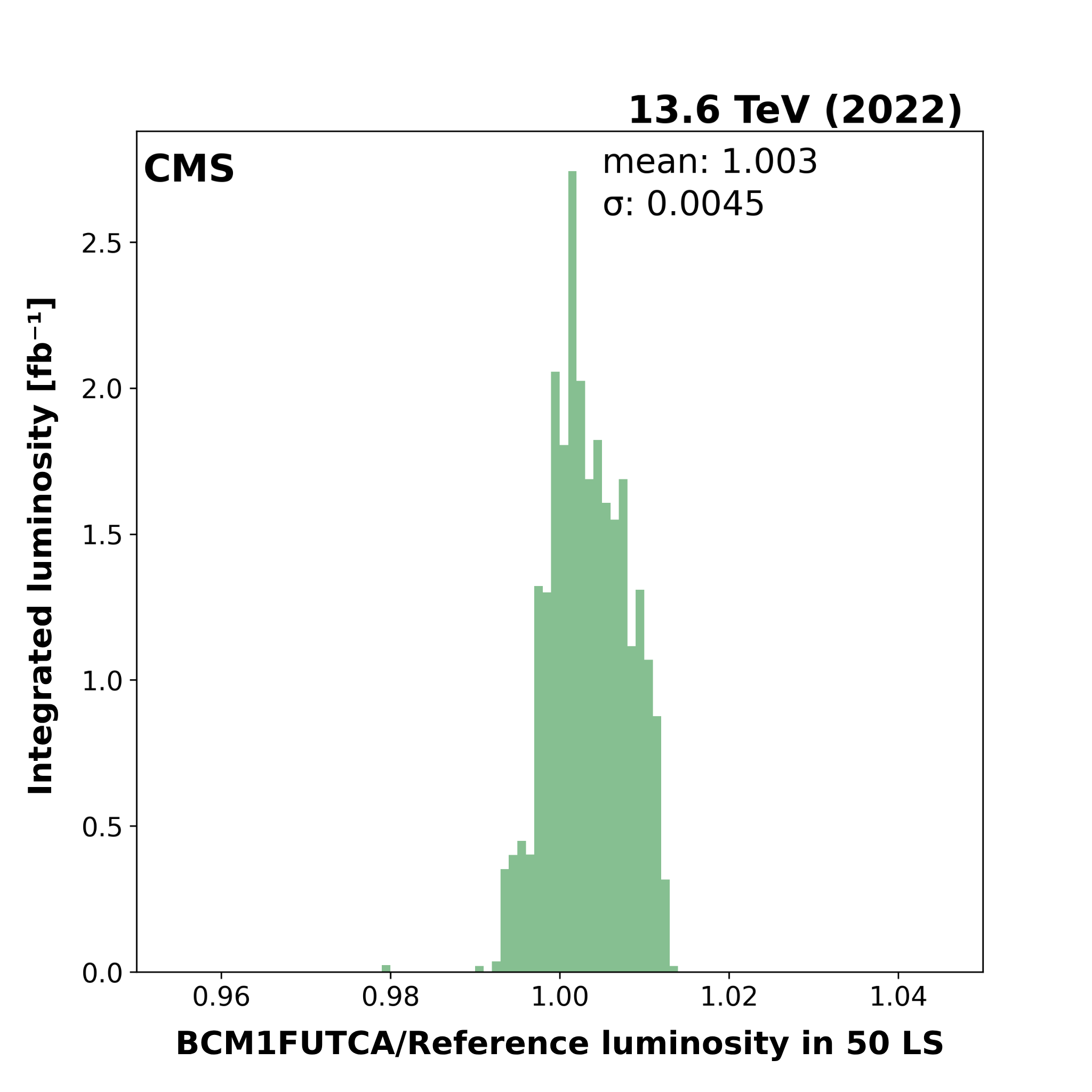}
    \caption{The distribution of the measured luminosity ratio in a subset of the 2022 data sample - BCM1F-\utca is compared to a reference CMS luminometer, chosen by availability during the integration period, with priority given to HF. Data is integrated for 50\,LS ($\sim$19\,min)~\cite{Joanna_thesis} }
    \label{fig:hfet_ratio}
\end{figure}

The systematic uncertainty associated with BCM1F stability can be estimated using comparisons with other stable luminometers. Taking as an example the 2022 dataset, the uncertainty is quantified as the standard deviation of the distribution of integrated luminosity ratios (BCM1F-\utca relative to HFET and muon drift tubes, depending on the availability), shown in Fig.~\ref{fig:hfet_ratio}, yielding a value of 0.45\%. 
Agreement in the absolute luminosity scale was further confirmed: the mean of the ratio distribution indicates consistency among detectors at the level of 0.3\% over the 2022 dataset~\cite{Joanna_thesis}.

\section{Impact of BCM1F operation experience on HL-LHC luminometer design}
\label{sec:fbcm}

\begin{figure}[!ht]
\centering
\includegraphics[width=\columnwidth]{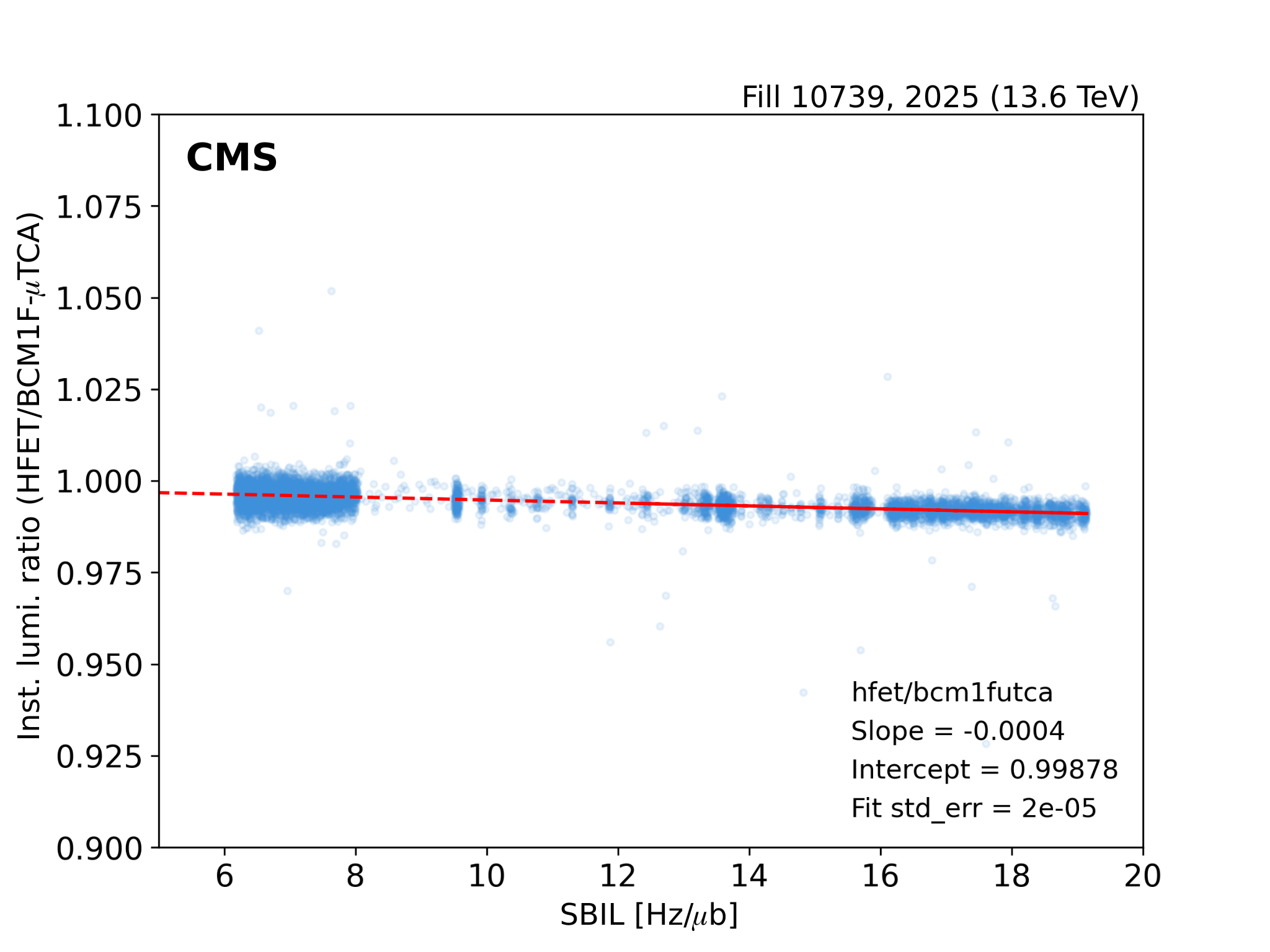}
\caption{Relative linearity of BCM1F-\utca luminosity response with respect to that of HFET in the high pileup machine development run, performed in fill 10739
}
\label{fig:bcm1f:utca_highpileup}
\end{figure}

In 2025, a dedicated high-pileup run took place, reaching pileup values up to 140 (corresponding to an SBIL of about 20\,Hz/$\mu$b). These conditions are the closest so far to those anticipated for the High-Luminosity LHC (HL-LHC). The dataset demonstrates excellent relative linearity of the BCM1F-\utca system with respect to HFET, as shown in Fig.~\ref{fig:bcm1f:utca_highpileup}. The latter is expected to exhibit highly linear behaviour based on simulation~\cite{BRIL-TDR}).

For the HL-LHC era, scheduled to start in 2030 with an average pileup of up to 200, a new standalone luminometer is in the final design stage: the Fast Beam Condition Monitor (FBCM)~\cite{BRIL-TDR,JINST_FBCM_2023,FBCM_TB2024_paper}. The BCM1F detector front-end architecture and mechanical solution, its back-end signal processing in FPGA, together with operation experience have provided a solid foundation for the design of this future per-bunch, triggerless luminosity and beam-induced background monitor. The operational insights from Run 2 and Run 3 and the CMS-wide electronics upgrade constraints have shaped the FBCM concept. The new luminometer will also increase the number of measurement channels by a factor of six.

Learning from experience with BCM1F, FBCM has a modular design enabling easy replacement of individual components in case of malfunction or production defects. The FBCM will employ silicon sensors, but uses a new 150\,$\mu$m-thick 6-pad structure with a double guard ring. A second guard ring was added to improve the stability of the response. The new sensor was tested under the same conditions as described in Sec.~\ref{sec:x_ray_test} for the BCM1F 290\,$\mu$m-thick 2-pad sensor. No leakage current increases or decreases were observed. The sensor choice has been recently validated via irradiation and beam tests. 

Signal processing will be performed using an FPGA, as it was done for the BCM1F $\mu$TCA readout chain. However, a different algorithm will be implemented for signal detection, as without the analogue ASIC signal, the derivative-based algorithm  cannot be utilized.
The FBCM detector will employ a novel custom-designed, radiation-hard fully asynchronous digital ASIC (FBCM23)~\cite{FBCM-ASIC}. 
The choice of a digital ASIC aligns with the broader CMS strategy for the HL-LHC, driven by the need to improve front-end timing and noise performance. The FBCM23 ASIC has an internal 4\,ns peaking time, with noise around 550 e$^-$ at 5\,pF input charge. In comparison, the BCM1F ASIC has a 8--10\,ns peaking time, with noise around 700 e$^-$ at 4\,pF input charge. 

Production of the FBCM luminometer is planned to begin in late 2026, followed by testing in 2027-2028 and installation in CMS at the end of LHC Long Shutdown 3 (LS3) in 2029.  
\section{Conclusions}

The BCM1F detector, which started from several diamond sensors with readout electronics installed around the beam pipe in 2008 in LHC Run 1, evolved  over years of operation. For LHC Run 2, a dedicated fast ASIC was designed to allow per-bunch luminosity and beam-induced background measurements. Multiple sensor types were tested in generations of BCM1F during LHC Run 2 and Run 3. 
The detector back end has also evolved, from the initial setup of dedicated histogramming units and separate ADC units to newer \utca cards with full data processing on FPGA. Both systems are still maintained and used for cross checks of the measurements.  

Precise luminosity measurement is important for the CMS physics program. 
The ability to study luminosity and beam-induced backgrounds per bunch crossing has opened a new avenue for understanding beam effects at the LHC, as well as afterglow and detector effects seen by luminometers. The BCM1F luminometer with its cooled silicon sensor and fast ASIC has played an important role in CMS in Run 3 to achieve $<1.5\%$ preliminary precision for luminosity measurements soon after data taking~\cite{CMS-PAS-LUM-22-001}. 
This represents an important milestone for the CMS Collaboration, which aims to reach the 1\% (2\%) target offline (online) luminosity precision for HL-LHC. The operational experience of the BCM1F detector has provided valuable lessons and impacted the design of the HL-LHC standalone luminometer FBCM, which is currently in the prototyping stage.

\begin{acknowledgements}
We congratulate our colleagues in the CERN accelerator departments for the excellent performance of the LHC and thank the technical and administrative staffs at CERN and at CMS institutes worldwide for their contributions to the success of the CMS effort. We thank the CERN bonding laboratory, in particular Florentina Manolescu and Marcin Tadeusz Poblocki; the LHC operations team, in particular Michi Hostettler and Jorg Wenninger; the CMS technical coordination team, for their help during operations; the CMS tracker project for their support with the services; the CMS cooling team for the maintenance of the cooling system and quality control on the cooling connections; and the CMS engineering and integration office for their work in routing and supervision of the installation of the cooling services. We acknowledge the enduring support for the construction and operation of the CMS BRIL detectors by the following institutes and funding agencies: 
CERN; 
the Secretariat for Higher Education, Science, Technology and Innovation (SENESCYT) (Ecuador);
the Helmholtz-Gemeinschaft Deutscher Forschungszentren (HGF) (Germany);
the Hungarian Academy of Sciences (MTA) and the National Research, Development and Innovation Office (NKFIH) (Hungary); 
the Mexican National Council for Science and Technology (CONACYT) (Mexico); 
the Ministry of Business, Innovation and Employment (New Zealand); 
the US CMS operations program, the US National Science Foundation (NSF), and the US Department of Energy (DOE) (USA).

Individuals have received support from 
the Estonian Research Council under grants CoE TK202 “Foundations of the Universe” and TARISTU24-TK10 (Estonia),
the NKFIH research grants K 143460, K 146913, K 146914 and TKP2021-NKTA-64 (Hungary), 
the Swiss Accelerator Research and Technology Institute (CHART) (Switzerland),
the US NSF research grants NSF-2121686, 
PHY-1945366,  
PHY-2111554, 
PHY-2209460, 
PHY-2411502, 
PHY-2512783 
and the US DOE Office of High Energy Physics awards 
DE-AC02-07CH11359, 
DE-SC0011845, 
DE-SC0015910, 
DE-SC0020267, 
and DE-SC0023908. 

\end{acknowledgements}

\bibliographystyle{auto_generated}
\bibliography{bibliography.bib}

\end{document}